# Data Justice in Practice:
# A Guide for Developers

Prepared by: **The Alan Turing Institute**

In collaboration with: 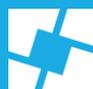 **ceimia** | International Centre of Expertise in Montreal on Artificial Intelligence

**GPAI** / THE GLOBAL PARTNERSHIP ON ARTIFICIAL INTELLIGENCE

# Data Justice in Practice: A Guide for Developers

By David Leslie, Michael Katell, Mhairi Aitken, Jatinder Singh, Morgan Briggs, Rosamund Powell, Cami Rincón, Antonella Perini, Smera Jayadeva, and Christopher Burr

**Acknowledgements**

This research was supported, in part, by a grant from ESRC (ES/T007354/1), Towards Turing 2.0 under the EPSRC Grant EP/W037211/1, and from the public funds that make the Turing's Public Policy Programme possible.

The creation of this material would not have been possible without the support and efforts of various partners and collaborators. The authors would like to acknowledge our 12 Policy Pilot Partners—AfroLeadership, CIPESA, CIPIT, WOUGNET, Gob_Lab UAI, ITS Rio, Internet Bolivia, Digital Empowerment Foundation, Digital Natives Academy, Digital Rights Foundation, Open Data China, and EngageMedia—for their extensive contributions and input. The research that each of these partners conducted has contributed so much to the advancement of data justice research and practice and to our understanding of this area. We would like to thank Thompson Chengeta, Noopur Raval, and Alicia Boyd, and our Advisory Board members, Nii Narku Quaynor, Araba Sey, Judith Okonkwo, Annette Braunack-Mayer, Mohan Dutta, Maru Mora Villalpando, Salima Bah, Os Keyes, Verónica Achá Alvarez, Oluwatoyin Sanni, and Nushin Isabelle Yazdani whose expertise, wisdom, and lived experiences have provided us with a wide range of insights that proved invaluable throughout this research. We would additionally like to express gratitude for the feedback and contributions of the Compliant & Accountable Systems Group at the University of Cambridge.

We would also like to thank those individuals and communities who engaged with our participatory platform on *decidim* and whose thoughts and opinions on data justice greatly informed the framing of this project. All of these contributions have demonstrated the pressing need for a relocation of data justice and we hope to have emphasised this throughout our research outputs. Finally, we would like to acknowledge the tireless efforts of our colleagues at the International Centre of Expertise in Montréal and GPAI's Data Governance Working Group. We are grateful, in particular, for the unbending support of Ed Teather, Sophie Fallaha, Jacques Rajotte, and Noémie Gervais from CEIMIA, and for the indefatigable dedication of Alison Gillwald, Dewey Murdick, Jeni Tennison, Maja Bogataj Jančič, and all other members of the Data Governance Working Group.

Cite this work as:
Leslie, D., Katell, M., Aitken, M., Singh, J., Briggs, M., Powell, R., Rincón, C., Perini, A., Jayadeva, S., and Burr, C. (2022). Data justice in practice: a guide for developers. The Alan Turing Institute in collaboration with The Global Partnership on AI.

# Contents

# Foreword to the Consultation Draft

Thank you for picking up *Data Justice in Practice: A Guide for Developers*. We are publishing this guide as a consultation draft in hopes of gathering feedback that will enable us to improve its content and presentation. This draft is therefore offered as a living document, and we appeal to you, the reader, for help in making it as useable, accessible, and actionable as possible. Please visit our project website at www.advancingdatajustice.org for details of how to submit feedback to our research team. Many thanks in advance!

# Introduction

The Advancing Data Justice Research and Practice project aims to broaden understanding of the social, historical, cultural, political, and economic forces that contribute to discrimination and inequity in contemporary ecologies of data collection, governance, and use. This guide for developers and organisations which are producing, procuring, or using data-intensive technologies, offers practical guidance to support responsible and equitable data innovation. As discussed in our *Integrated Literature Review and Annotated Bibliography*, the nascent field of data justice has, in its brief existence, done important work to illuminate how historically rooted conditions of power asymmetry, inequality, discrimination, and exploitation are drawn into processes of data production, extraction, and use. The Advancing Data Justice Research and Practice project offers conceptual framing and guidance to expand this area of scholarship and practice.

# What's in this Guide

This guide provides actionable information for developers who wish to implement the principles and priorities of data justice in their data practices and within their data innovation ecosystems. In this section we present the intended audience and the context of the Advancing Data Justice Research and Practice project. In the following section, we introduce the nascent field of data justice, from its early discussions to more recent intentions to relocate our understanding of what data justice means. This section includes an account of the outreach we conducted with stakeholders throughout the world in developing a nuanced and pluralistic conception of data justice and concludes with a description of the six pillars of data justice around which this guidance revolves. Following this section, we offer some examples of how these six pillars of data justice are being put into practice by organisations across the world.

Next, to support developers in designing, developing, and deploying responsible and equitable data-intensive and AI/ML systems, we outline the AI/ML project lifecycle through a sociotechnical lens, walking the reader through each phase and noting the ethics and governance considerations that should occur at each step of the way. This portion of the guide is intended to provide a background picture of the different stages of the lifecycle and to show how the data justice pillars can be woven into the stages and their respective sociotechnical considerations. We conclude by presenting some illustrative touchpoints between pillars and the lifecycle.

It is important to note here that, while much of our discussion focuses on projects that involve AI/ML to some degree, the questions and considerations raised are relevant for data-driven systems in general. This is crucial given that data justice issues are pertinent for and can manifest within and from technical systems which do not include models.

To support the operationalisation data justice throughout the entirety of the AI/ML lifecycle and within data innovation ecosystems, we then present five overarching principles of responsible, equitable, and trustworthy data research and innovation practices, the SAFE-D principles—Safety, Accountability, Fairness, Explainability, and Data Quality, Integrity, Protection, and Privacy. These principles support and underwrite the advancement of data justice within research and innovation practices. We elaborate upon them as high-level goals that are then followed by further specification through the presentation of additional properties, which are to be established in either the project or the system to ensure these goals are reached.

Depending on their contexts, potential impacts, and scale, data innovation activities should be carried out in a way that involves different degrees of stakeholder engagement. To facilitate this process, the next section provides an explainer of the Stakeholder Engagement Process and the steps it includes—preliminary horizon scanning, project scoping and stakeholder analysis, positionality reflection, and stakeholder engagement objectives and methods.

Finally, the last section presents guiding questions that will help developers both address data justice issues throughout the AI/ML lifecycle and engage in reflective innovation practices that ensure the design, development, and deployment of responsible and equitable data-intensive and AI/ML systems. This is done by presenting questions related to both the six pillars of data justice and the SAFE-D principles introduced previously.

There are four Annexes in this document. The first Annex outlines 12 Principles and Priorities of responsible innovation to provide developers and organisations producing, procuring, or using AI/ML or other data-driven technologies with a means of accessing and understanding some of the existing human rights, fundamental freedoms, and value priorities that could be impacted by the use of AI and data-driven technologies. This table draws on various charters, declarations, and conventions to help spur critical reflection on which salient rights, freedoms, and values could be affected by your project. The second Annex provides, for your reference, the list of Sustainable Development Goals (SDGs), as equitable implementation of data systems that furthers data justice should also serve to forward the SDGs (a set of general prompts about this is included in the Guiding Questions). The third Annex covers some of the insights we have gained about this project and the data justice pillars from the excellent reports that have been prepared by our Policy Pilot Partners. We have also included, as the fourth Annex, the positionality statement prepared by the Advancing Data Justice Research and Practice team as we started on our journey in this project.

## Intended Audience

This guide is designed for developers and organisations which are producing, procuring, or using AI/ML or other data-driven technologies in a variety of data innovation contexts. It is intended to support the responsible and equitable innovation practices of those who seek to integrate an understanding of data justice into their collection and use of data. Herein you will find practical guidance, background, and conceptual framings that are meant to help you appreciate and address many of the complex issues presented by contemporary networked societies. The concepts and activities in this guide are intended to support developers in promoting equitable, freedom-promoting, and rights-sustaining practices throughout the entirety of the AI/ML project lifecycle and throughout the data innovation ecosystem.

## Project Context

The Advancing Data Justice Research and Practice project seeks to initiate a new wave of data justice scholarship and practice. We utilise a decolonial lens that embraces a plurality of perspectives and situated knowledge, aiming to move beyond Anglo-European framings and recognising how existing relations of power among and within the world's societies are not inevitable. While recent, the data justice movement, and the transformative practices that are described in this guide, draw from an extensive history of critical insights and the energies of adjacent social justice movements from around the world. The application of an enlarged, inclusive, and decolonial approach to data justice research and practice is essential as we turn to address the manifold risks, harms, and opportunities presented by planetary scale datafication.

# Key Concepts: Data Justice

In this section we provide the reader with a portrayal of the emergent and evolving concept of data justice. We begin by describing the concept of data justice and present a brief history. We then expand on this concept with a set of "relocations" that shift our focus from exclusively Eurocentric framings and understandings of data justice to a more broadly inclusive concept. From there, we present six "pillars" of data justice that serve as the guiding priorities for this project, and which are informed by our efforts to connect with stakeholders from across the world. A goal for this section is to deepen the awareness of AI/ML developers about the role of data-driven technologies in the world's many economic and political dramas.

## What is data justice?

Before the advent of contemporary data justice research, prevailing approaches to data ethics and governance tended to frame issues surrounding the societal impacts of datafication and the increasing pervasiveness of data-intensive technologies in terms of data protection, individual rights, privacy, efficiency, and security.[1] They likewise tended to focus on building technical solutions to potential harms rather than on interrogating the social structures, human choices, and sociotechnical practices that lie behind the myriad predicaments arising out of an ever more "datafied society". The first wave of data justice scholarship sought to move beyond these limitations by situating the ethical challenges posed by datafication in the wider context of social justice concerns.

Beginning in 2014, several distinct strands of data justice research emerged in Western scholarship based in the varying but distinct implications of datafication.[2] In 2017, these strands were brought together by Linnet Taylor to create a data justice framework with three core pillars (Figure 1 below). Through these three pillars, data justice came to be understood as a conceptual framework based on 'fairness in the way people are made visible, represented, and treated as a result of their production of digital

**Key Term: Social Justice**

Social justice is a commitment to the achievement of a society that is equitable, fair, and capable of confronting the root causes of injustice. In an equitable and fair society, all individuals are recognised as worthy of equal moral standing and are able to realise the full assemblage of fundamental rights, opportunities, and positions.

In a socially just world, every person has access to the material means needed to participate fully in work life, social life, and creative life through the provision of proper education, adequate living and working conditions, general safety, social security, and other means of realising maximal health and well-being.

Social justice also entails the advancement of diversity and participatory parity and a pluralistically informed recognition of identity and cultural difference. Struggles for social justice typically include accounting for historical and structural injustice coupled to demands for reparations and other means of restoring rights, opportunities, and resources to those who have been denied them or otherwise harmed.

data'.[3] Taylor's work also calls for integrating elements of the 'capabilities approach' of social justice, borrowed from the work of Amartya Sen and Martha Nussbaum, which centres human flourishing and the creation of the material conditions necessary to enable people to realise their full potential and live freely.[4]

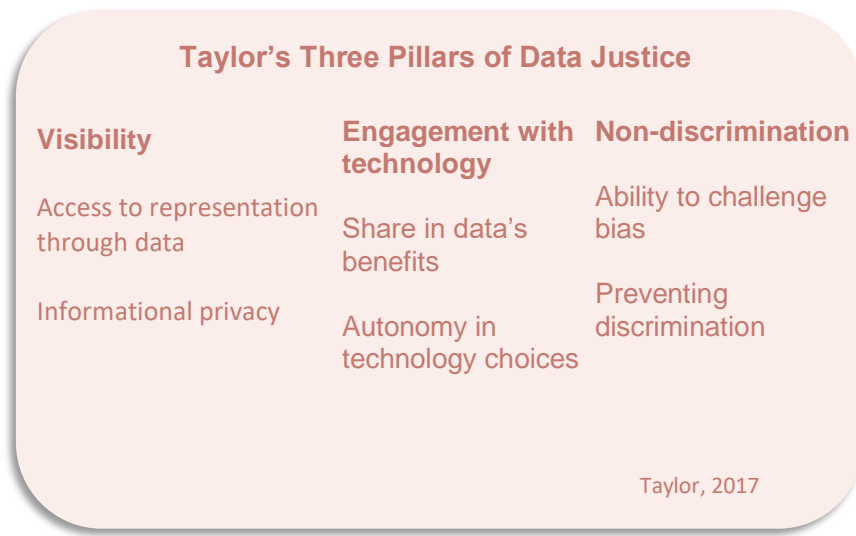

Figure 1: Taylor's Three Pillars of Data Justice

Since the publication of Taylor's 2017 data justice framework, the literature has expanded. Dedicated institutions including the Data Justice Lab at Cardiff University and the Global Data Justice Project at the Tilburg Institute for Law, Technology, and Society have been established.[5] The concept of data justice has been interrogated in a range of specific global contexts such as policing in Iran, activism in South Africa, indigenous agriculture in Africa, humanitarian work in post-earthquake Nepal, and more.[6] These academic understandings of data justice will continue to inform this work while additional perspectives, collected through our Policy Pilot Partners, our data justice survey, and our accompanying literature review broaden this definition even further.

## Key Term: Community

In this guide we frequently refer to "community", so it would be helpful to clarify what we mean by this. The term community relates to a group of people with some shared characteristics. This might be a "community of place"—a group of people who live or work in the same geographic area—or a "community of interest", which brings together people through shared activities, identities, interests, or concerns. As such, while some communities are located in a particular place, others are geographically dispersed (i.e., where people who share activities, identities, interests, or concerns live in different places).

It is also important to note that individuals typically belong to more than one community (e.g., someone might belong to a local community related to the place in which they live as well as communities formed around interests, identity characteristics, or hobbies). Moreover, communities are rarely homogeneous in their interests and experiences and so it is important to pay attention to power dynamics and inequalities within communities, noting that individual community members will have a range of experiences, interests, and perspectives.

---

[3] Taylor, 2017, p. 1
[4] Nussbaum, 2006; Sen, 1999; Taylor, 2019
[5] https://datajusticelab.org; https://globaldatajustice.org

[6] Akbari, 2019; Cinnamon, 2019; Dagne, 2020; Kennedy et al., 2019; Kidd, 2019; Mulder, 2020; Punathambekar & Mohan, 2019

# Timeline of Data Justice Literature 2014 to the Present

Johnson identifies power asymmetries in the governance and administrative functions of data which can lead to normatively coercive data structures and forms of extraction. He argues in favour of "information justice" in the context of open data as a framework to address these power dynamics.

Heeks and Renken propose that a framework of data justice is needed to account for local and global variations in how datafication impacts individuals and communities. While data justice needs to be applied differently in different contexts, human rights and fundamental freedoms are important guideposts. Heeks and Renken argue such a global approach is lacking.

Linnet Taylor defines Data Justice as 'fairness in the way people are made visible, represented and treated as a result of their production of digital data'.

Global Data Justice Project launched at Tilburg Institute for Law, Technology, and Society.

2020—Global Partnership on Artificial Intelligence (GPAI) is established. Its aim is 'to bridge the gap between theory and practice on AI by supporting cutting-edge research and applied activities on AI-related priorities'. GPAI's 15 founding members are Australia, Canada, France, Germany, India, Italy, Japan, Mexico, New Zealand, the Republic of Korea, Singapore, Slovenia, the United Kingdom, the United States, and the European Union. They were joined by Brazil, the Netherlands, Poland, and Spain in December 2020.

**2014**     **2015**     **2016**     **2017**     **2018**     **2020**     **2021**

World leaders adopt 17 Sustainable Development Goals (SDGs) at a UN Summit. These goals provide an important framing for the responsible adoption of AI.

Dencik et al. propose a data justice framework is needed to broaden the conversation around datafication to account for concerns beyond security, privacy, and data protection. They argue that the pursuit of data justice must include the involvement of activists and advocates in civil society.

Data Justice Lab officially launched at Cardiff University's School for Journalism, Media, and Cultural Studies.

Data Justice literature takes on increasingly globally oriented and intercultural approaches as authors explore local and contextual understandings of how social justice intersects with datafication.

# Relocating Data Justice

A central aim of this guide is to shift understandings of data justice away from the predominance of Eurocentric and "Global North" perspectives towards a more inclusive vision – one which may nudge and AI/ML developers beyond their existing viewpoints. This relocation operates among three dimensions: spatial, temporal, and vocational.

To relocate data justice *spatially* means to shift the 'where' of data justice away from practical approaches and research perspectives that emerge from current centres of social and economic power. This relocation attempts to account for meanings and values from outside the Global North as well as from marginalised voices within Global North societies. In so doing, data justice research and practice is enriched by frames of socio-cultural knowledge that are frequently overlooked by Western scholars and practitioners. Relocating data justice spatially is intended to promote greater cross-fertilisation of insights and experience in data justice research and practice, which are of particular importance in light of the ongoing failure of prevailing approaches to remediate the significant ecological and distributional challenges facing the world. Our goal here is to create conditions for participatory parity, so that crucial insights that have largely been excluded up to the present can now be centred.

The *temporal* relocation of data justice research and practice addresses the 'when' of data justice, accounting for its roots in social justice histories, including those whose relationship to data and digital infrastructures may not be immediately obvious. Data injustice is not an entirely new phenomena exclusively associated with the technological expansion of recent decades. Rather, it can be found in longstanding cultural, political, and socio-economic patterns of inequity and discrimination that find expression in contemporary networked society. These patterns are reflected in both the construction of data and its interpretation—given that the production of data is shaped by those with the power to collect it at scale and the degree of acceptance of the authority of the research products and practices informed by that data. A goal of this project is to urge researchers and practitioners to recognise the deep history of datafication and to bring an appropriately critical lens to the data innovation infrastructures and practices of the present.

To relocate data justice research and practice *vocationally* is to enlarge the "who" of data justice, transcending fixed notions of expertise to include and value the lived experience and "situated knowledge"[7] of impacted persons and communities, drawing from data advocacy and policymaking knowledge and from data justice adjacent activism (e.g., climate justice, global public health justice). This enlarged membership should be extended especially to those who have been historically discriminated against, disempowered, and marginalised. As such, this project embraces and promotes a constitutive plurality of knowledges to give an appropriate parity of voice to the academic articles and books, policymaking outputs, and activist papers, statements, and declarations that can contribute to conceptual and policy innovation.

For more information on the project, you can find further reading on the project website and our interim report.[8]

## Policy Pilot Partner Collaboration

A key element in our strategy to broaden our understanding of data justice is our ongoing partnership with twelve Policy Pilot Partner organisations recruited from across the world. These organisations were selected for their advocacy and activist work with local communities on topics related to media and technology adoption as well as experience researching topics surrounding datafication and human rights in distinct global contexts. From over 40 applicants across the globe, 12 partners across Africa, the Americas, Asia, and Oceania were selected and have provided invaluable local perspectives. Their critical assessments of the data justice pillars and of reflective questions for policymakers, developers, and impacted communities have shaped our work and will continue to guide subsequent editions of these guidelines. Please see Annex 3 for more information on the important insights of our PPPs about the project.

---

[7] Haraway, 1988
[8] https://advancingdatajustice.org; https://gpai.ai/projects/data-governance/data-justice/advancing-data-justice-research-and-practice-interim-report.pdf

## Decidim Analysis

As part of the research that informs this guide, we developed an online participatory engagement platform using the *decidim* digital interface[9] to enable individuals and communities to provide insights and ground our work in developing an inclusive and actionable conception of data justice. Our Policy Pilot Partners also contributed responses. Prompts and questions included prompts about defining and situating the concept of data justice.

Among the insights gained from this outreach, we identified gaps in existing portrayals of data justice that reveal tensions between individual and collective justice. Respondents highlighted the need to include the role of colonialism in entrenching historical inequalities between and within countries and entities. Additionally, we found that existing definitions of data justice adequately address neither the underlying historical, cultural, and economic patterns of discrimination that have cascading effects on data collection, processing, and use, nor how inequality and the exclusion of individuals and groups may be replicated, automated, or created through data-driven processes and tools. Respondents also indicated that data justice should include concepts of access, understanding, and consent to data collection processes.

## The Six Pillars of Data Justice Research and Practice

Taken together, our analysis of the *decidim* survey results, our critical exploration of the important conceptual work carried out in the first years of the academic data justice literature, our interactions with our Policy Pilot Partners, and our other desk-based research have led us to propose six pillars of data justice research and practice. These are the guiding priorities of power, equity, access, identity, participation, and knowledge.

While such pillars build on and expand previous attempts to specify the meaning of the term "data justice," they are not offered here as part of a definition *per se*. Key to the re-orientation of data justice undertaken in this guide is the idea that it is contextually determined. *It should be seen as a set of critical practices and procedures that respond to—and enable the transformation of—existing power asymmetries and inequitable or discriminatory social structures rather than as a collection of abstract principles or prescriptions.* Consequently, instead of answering the question "what is data justice" directly, the pillars are meant to be tools for orienting critical reflection and for generating constructive insights into how to transform data justice practice to redress the data inequities of the past and present in the ends of building more just societal and biospheric futures.

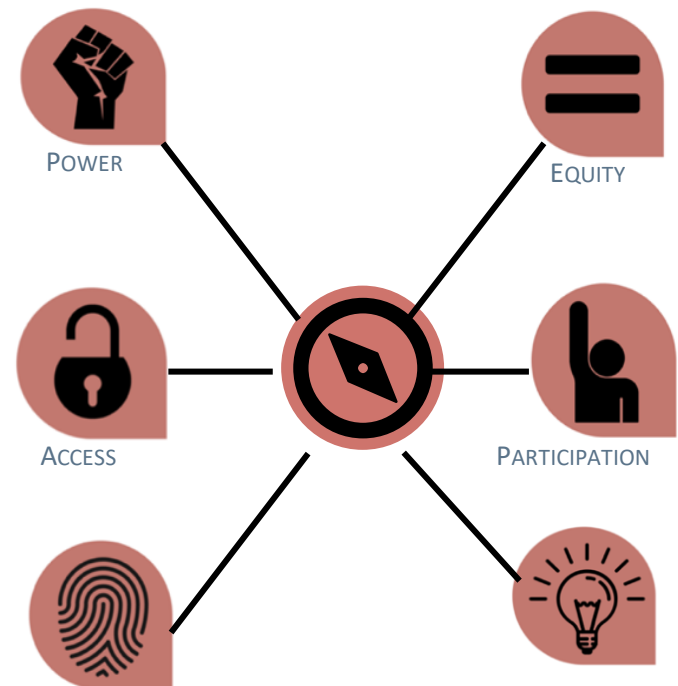

*Figure 2: The six pillars of data justice*

---

The six pillars shape this guide and our related research:

- **The pillar of power** demonstrates the importance of understanding the levels at which power operates and how power manifests in the collection and use of data in the world. The articulation of this pillar provides a basis from which to question power at its sources and to raise critical awareness of its presence and influence.
- **The pillar of equity** addresses the need to confront the root causes of data injustices as well as to interrogate choices about the acquisition and use of data, particularly where the goal or purpose is to target and intervene in the lives of historically marginalised or vulnerable populations.
- **The pillar of access** illuminates how a lack of access to the benefits of data processing is a starting point for reflection on the impacts and prospects of technological interventions. The beginning of any and all attempts to protect the interests of the vulnerable through the mobilization of data innovation should be anchored in reflection on the concrete, bottom-up circumstances of justice and the real-world problems at the roots of lived injustice.
- **The pillar of identity** addresses the social character of data and problematises its construction and categorisation, which is shaped by the sociocultural conditions and historical contexts from which it is derived.
- **The pillar of participation** promotes the democratisation of data scientific research and data innovation practices and the need to involve members of impacted communities, policymakers, practitioners, and developers together to collaboratively articulate shared visions for the direction that data innovation agendas should take.
- **The pillar of knowledge** involves recognising that diverse forms of knowledge and understanding can add valuable insights to the aspirations, purposes, and justifications of data use—including on the local or context-specific impacts of data-intensive innovation. Inclusion of diverse knowledges and ways of being can open unforeseen paths to societal and biospheric benefits and maximise the value and utility of data use across society in ways which take account of the needs, interests, and concerns of all affected communities.

# Data Justice Pillars in Focus

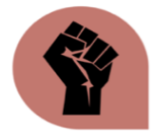

## Power

**1. Interrogate and critique power:** Power dynamics can be present in many different places and in several different ways. It is therefore important to:

***Understand where power operates in data innovation ecosystems.** This can include*

- *The **geopolitical level**. For example,* high-income nation-states and transnational corporate actors can control access to technological capabilities and pursue their own interests on the global stage. In doing this, they can exercise significant influence on which countries or regions are able to develop digital and data processing capacities.
- *The **level of economy and infrastructure**.* For example, large tech companies can decide which impacted communities, domestically and globally, are able to access the benefits of connectivity and data innovation, and they can control the provision of essential digital goods and services that directly affect the public interest.
- *The **legal, policy, and regulatory levels**.* For example, large international standards bodies, transnational corporations, trade associations, and nation states, can exercise disproportionate amounts of influence in setting international policies, standards, and regulation related to the governance of digital goods and services and data innovation.
- *The **organisational and political levels**.* For example, governments and companies can control data collection and use in intrusive and involuntary ways—especially where the public have no choice but to utilise the services they provide or must work in the environments they manage and administer.

- *The cultural level.* For example, power can operate through the way that large tech companies use relevance-ranking, popularity-sorting, and trend-predicting algorithms to sort users into different, and potentially polarising, digital publics or groups.
- The *psychological level.* For example, tech companies can use algorithmically personalised services to curate the desires of targeted data subjects. This can allow for the control or manipulation of consumer behaviour but also play an active and sometimes damaging role in identity formation, mental well-being, and personal development.

*Understand how power manifests and materialises in the collection and use of data in the world.* Power can surface in everyday life in several different ways. These include:

- *Decision-making power.* Here, an individual or organisational actor A has power over B to the extent that A can get B to do something that they would not otherwise do. Decision-making power is seen, for instance, in the way that government agencies collect and use data to build predictive risk models about citizens and data subjects or to allocate the provision of social services (and then act on the corresponding algorithmic outputs).
- *Agenda-setting power.* Here, an individual or organisational actor A has power over B to the extent that A sets the agenda that B then must fall in line with by virtue of A's control over the terms of engagement that set practical options within A's sphere of influence and interest. Agenda-setting power means that A can shoehorn the behaviour of B into a range of possibilities that is to A acceptable, tolerable, or desired. This kind of power is explicit, for example, in practices of regulatory capture, where large tech corporations secure light touch regulation through robust lobbying and legal intervention.
- *Ideological power.* This kind of power is exercised where people's perceptions, understandings, and preferences are shaped by a system of ideas or beliefs in a way which leads them—frequently against their own interests—to accept or even welcome their place in the existing social order and power hierarchy. For example, the priorities of "attention capture" and "screen-time maximisation", that are pursued by certain social media and internet platforms, can groom users within the growing ecosystem of compulsion-forming reputational platforms to embrace the algorithmically manufactured comforts of life-logging, status-updating, and influencer-watching all while avoiding confrontation with realities of expanding inequality and social stagnation.

- *Normalising power.* Normalising power manifests in the way that the ensemble of dominant knowledge structures, scientifically authoritative institutions, administrative techniques, and regulatory decisions work in tandem to maintain and 'make normal' the status quo of power relations. Where tools of data science and statistical expertise come to be used as techniques of knowledge production that claim to yield a scientific grasp on the inner states or properties of observed individuals, forms of normalising or disciplinary power can arise. Data subjects who are treated merely as objects of prediction or classification and who are therefore subjugated as objects of authoritative knowledge become sitting targets of disciplinary control and scientific management.

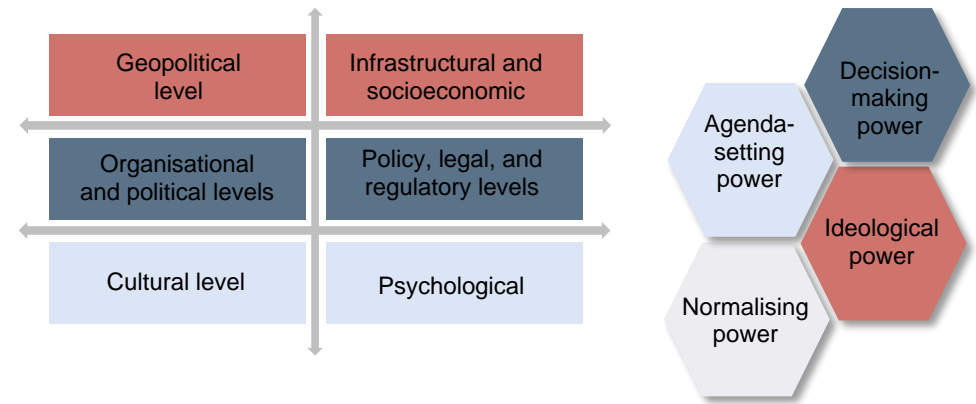

*Figure 3: Understanding the levels at which power operates in the collection and use of data, and how it manifests*

*Use this understanding to question power at its sources and to raise critical awareness of its presence and influence.* Interrogations of where and how power operates are first steps in a longer journey of questioning

and critical analysis. An active awareness of power dynamics in data innovation ecosystems should also lead to further questions:

- What are the interests of those who wield power or benefit from existing social hierarchy?
- How do these interests differ from other stakeholders who are impacted by or impact data practices and their governance?
- How do power imbalances shape the differing distribution of benefits and risks among different groups who possess varying levels of power?
- How do power imbalances result in potentially unjust outcomes for marginalised, vulnerable, or historically discriminated against groups?

**2. Challenge Power:** *Mobilise to push back against societally and historically entrenched power structures and to work toward more just and equitable futures.* While the questioning and critiquing of power are essential dimensions of data justice, its purpose of achieving a more just society demands that unequal power dynamics that harm or marginalise impacted individuals and communities must be challenged and transformed.

**3. Empower People:** *People must be empowered to draw on democratic agency and collective will to pursue social solidarity, political equity, and liberation.* When people and communities come together in the shared pursuit of social justice through mutually respectful practices of deliberation, collaboration, dialogue, and resistance, power becomes *empowerment*. It becomes constructive and opens transformative possibilities for the advancement of data justice, social solidarity, and political equity.

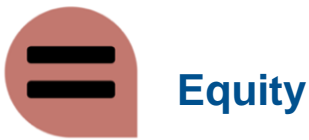 # Equity

**1. Consideration of equity issues should begin before any data are collected or used. Issues of equity should be confronted by developers and organisations at the earliest stage of project planning and should inform whether data innovation practices are engaged in**

**at all:** Data equity is only partially served by seeking to improve data and data practices, such as by pursuing data quality, or increasing its representativeness and accuracy. While errors and incompleteness are obstacles to data equity, the choice to acquire and use data can itself be a question of justice, particularly where the goal or purpose of a data practice is to target and intervene in the lives of historically marginalised or vulnerable populations. Here, the question may not be 'how can we repair an imperfect system or make it more effective', but rather 'does a particular use or appropriation of data enable or disable oppression?'; and 'does it preserve or combat harmful relations of power?' A perfectly engineered system employed by an oppressive regime (either governmental or commercial) can facilitate and potentially amplify data injustice.

**2. The purpose of the pursuit of data equity should be to transform historically rooted patterns of domination and entrenched power differentials:** Concerns with elements of data innovation practices like data security, data protection, algorithmic bias, and privacy are an important subset of data equity considerations, but the transformative potential of data equity to advance social justice comes in a step earlier and digs a layer deeper: It starts with questions of how longer-term patterns of inequality, coloniality, and discrimination penetrate data innovation practices and their governance. Data equity, in this deeper context, is about overhauling power imbalances and forms of oppression that manifest in harmful, unjust, or discriminatory data practices. To realise this sort of equity, those with power and privilege must be compelled to respond to and accommodate the claims of people and groups who have been marginalised by existing political and socioeconomic structures.

**3. Combat any discriminatory forms of data collection and use that centre on disadvantage and negative characterisation:** Data equity involves confronting and combating statistical representations of marginalised, vulnerable, and historically discriminated against social groups that focus mainly or entirely on measurements of 'disparity, deprivation, disadvantage, dysfunction, and difference', the '5 D's'. Approaches to statistical measurement and analysis that centre on disadvantage and negative characterisation produce feedforward effects which further entrench and amplify existing structures of inequity, discrimination, and domination.

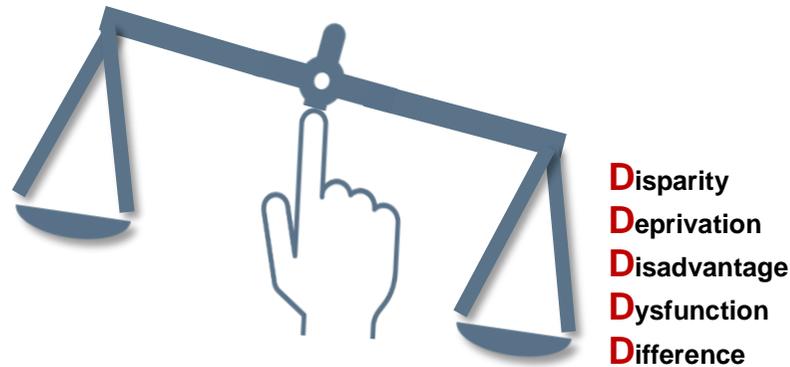

D**isparity**
D**eprivation**
D**isadvantage**
D**ysfunction**
D**ifference**

*Figure 4: Single axis modes of statistical representation; adopted from the 5 D's presented by Kukutai and Taylor (2016)*

**4. Pursue measurement justice and statistical equity:** Measurement justice and statistical equity involve focusing on collecting and using data about marginalised, vulnerable, and historically discriminated against communities in a way that:

- Advances social justice.
- Draws on their strengths rather than on perceived weaknesses.
- Approaches analytics constructively with community-defined goals that are positive and progressive rather than negative, regressive, and punitive.

This constructive approach necessitates a focus on socially licenced data collection and statistical analysis, on individual- and community-advancing outcomes, and strengths-based approaches.

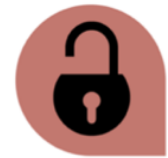

## Access

**1. Confronting questions of equitable access involves starting from real-world problems of material inequality and structural injustice. Access is about providing people tangible paths to data justice by addressing the root causes of social, political, and economic injustice:** Existing sociohistorical, economic, and political patterns of disadvantage must be taken as the starting point for reflection on the equitable access, because these create material conditions of injustice and a lack of access to the benefits of data processing. The beginning of any and all attempts to expand equitable access should be anchored in reflection on the concrete, bottom-up circumstances of justice, in its *historical and material preconditions*. Combatting the real-world problems at the roots of lived injustice should be a first priority.

**2. Equitably open access to data through responsible data sharing:** Calls for 'open data' sometimes run the risk of oversimplification and appropriation by market forces which could end up curtailing equitable access. The concept of 'open data' itself must be bounded and qualified. At all times, those who share data ought to remain critically aware of the moral claims and rights of the individuals and communities where the data came from, of the real-world impacts of data sharing on those individuals and communities, and of the practical barriers and enablers of equitable and inclusive research. There is also a need to consider the right of communities to access and benefit from the use of their data. Building on this, community-rights based approaches to data access and data sharing should include a strong participatory component. Here equitably opening access to community data entails the democratic governance of data collection and use as well as robust regimes of social license and public consent.

**3. Equitably advance access to research and innovation capacity:** Long-standing dynamics of global inequality may undermine reciprocal sharing between research collaborators from high-income countries (HICs) and those from low-/middle-income countries (LMICs). Given asymmetries in resources, infrastructure, and research capabilities, data sharing between LMICs and HICs, and the transnational opening of data, can lead to inequity and exploitation. Moreover, data originators from LMICs may generate valuable datasets that they are then unable to independently and expeditiously utilise for needed research, because they lack the aptitudes possessed by scientists from HICs, who are the beneficiaries of arbitrary asymmetries in education, training, and research capacitation. In redressing these access barriers, emphasis must be placed on 'the social and material conditions under which data can be made useable, and the multiplicity of conversion factors required for researchers to engage with data'. Equalising know-how and capability is a vital counterpart to equalising access to resources, and both together are necessary preconditions of just data sharing. Data scientists and developers engaging in international research collaborations should focus on forming substantively reciprocal partnerships where capacity-building and asymmetry-aware practices of cooperative innovation enable participatory parity and thus greater research access and equity.

**4. Equitably advance access to the capabilities of individuals, communities, and the biosphere to flourish:** This involves prioritising individual, social, and planetary well-being as well as an understanding that the attainment of well-being necessitates the stewardship of the human capabilities that are needed for all to freely realise a life well-lived. A capabilities- and flourishing-centred approach to just access demands that data collection and use be considered in terms of the affordances they provide for the ascertainment of well-being, flourishing, and the actualisation of individual and communal potential for these. It demands a starting point in ensuring that 'practices of living' enable the shared pursuit of the fullness, creativity, harmony, and flourishing of human and biospheric life (what Abya Yala Indigenous traditions of Bolivia and Ecuador have called 'living well' or *sumak kawsay* in Quechua, *suma qamaña* in Aymara, or *buen vivir* in Spanish).

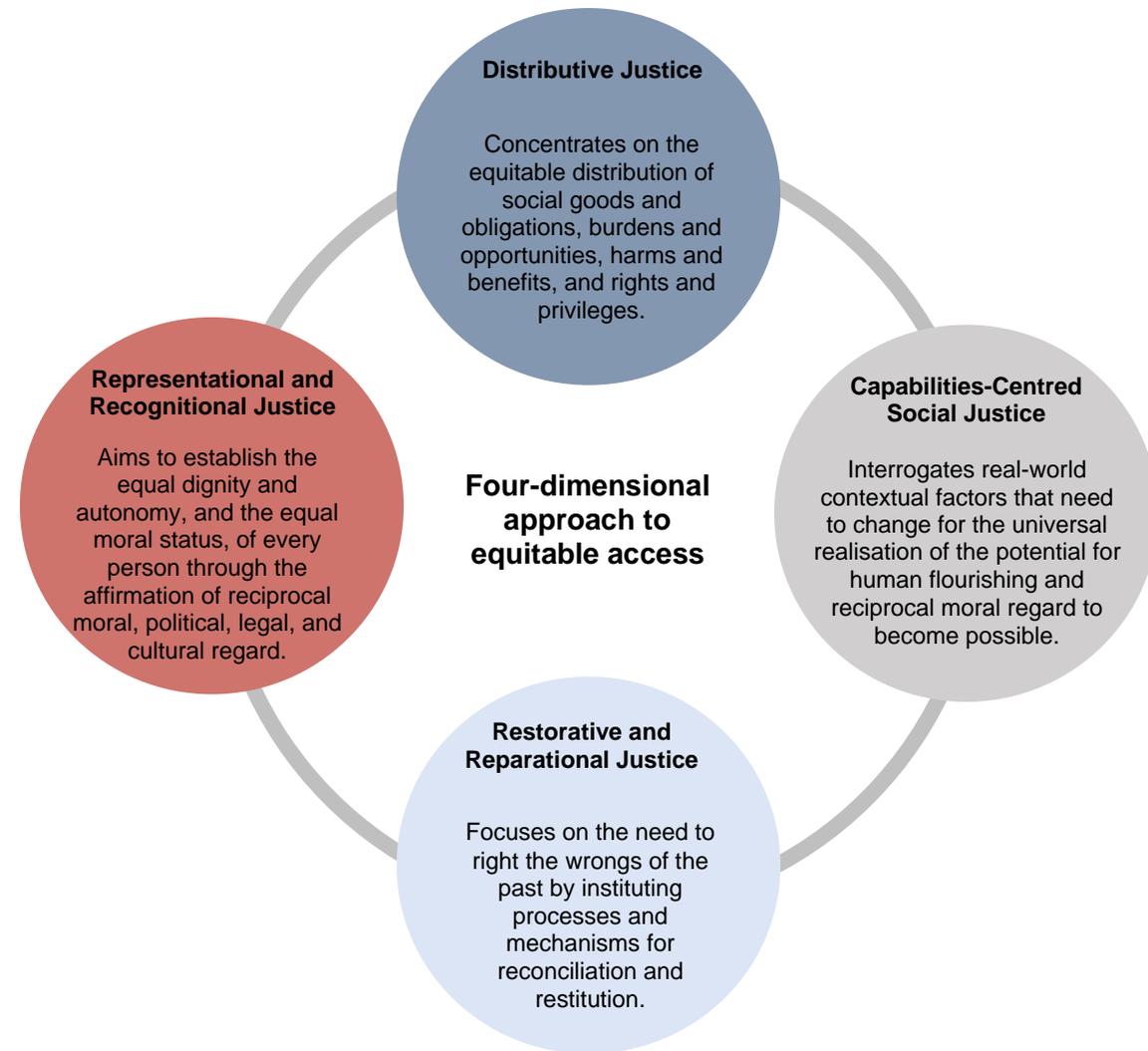

*Figure 5: Four-dimensional approach to equitable access*

**5. Confronting questions of equitable access involves *four dimensions of data justice:*** Concerns with equitable access should:

(1) Concentrate on the equitable distribution of the risks and benefits of data use. This is the dimension of *distributive justice*.
(2) Examine the material preconditions necessary for the universal realisation of justice. This is the dimension of *capabilities-centred social justice*.
(3) Rectify the identity claims of those who have faced representational injury. This is the dimension of *representational and recognitional justice*.
(4) Right the wrongs of the past so that justice can operate as a corrective dynamic in the present. This is the dimension of *restorative and reparational justice*.

This *four-dimensional approach to data justice* should use the ethical tools provided by the principles of social justice to assess the equity of existing social institutions, while also interrogating the real-world contextual factors that need to change for the universal realisation of the potential for human flourishing and reciprocal moral regard to become possible. It should likewise enable the reparation of historical injustices by instituting processes and mechanisms for reconciliation and restitution. While the first three of these facets remain integral to the advancement of access as it relates to data justice research and practice, they tend to focus primarily on addressing present harms and making course corrections oriented to a more just future. Restorative justice reorients this vision of the time horizons of justice. It takes aim at righting the wrongs of the past as a redeeming force in the present.

**6. Promote the airing and sharing of data injustices across communities through data witnessing:** Datafication makes possible the greater visibility of everyday life. Despite the ways increasing visibility may expose some to harm or exploitation, it can also be harnessed in positive ways to promote liberating transformation by exposing lived injustices, historical abuses, and moral harms. The growth of a networked and connected global society multiplies the transformative power of observation and communication. It enables the far-reaching airing and sharing of previously hidden inequities and mistreatment. This witnessing of injustice can occur both through the exposure of harms that are present in proximate data work and through the employment of digital media at-a-distance to observe harms that present in remote locations. Data witnessing should be marshalled as a force for change and as an opportunity to expand justice by means of transparency and voice.

**7. Promote the airing and sharing of data injustices across communities through transparency:** The role of transparency in the airing and sharing of potentially unjust data practices must also be centred. Transparency extends both to outcomes of the use of data systems and to the processes behind their design, development, and implementation.

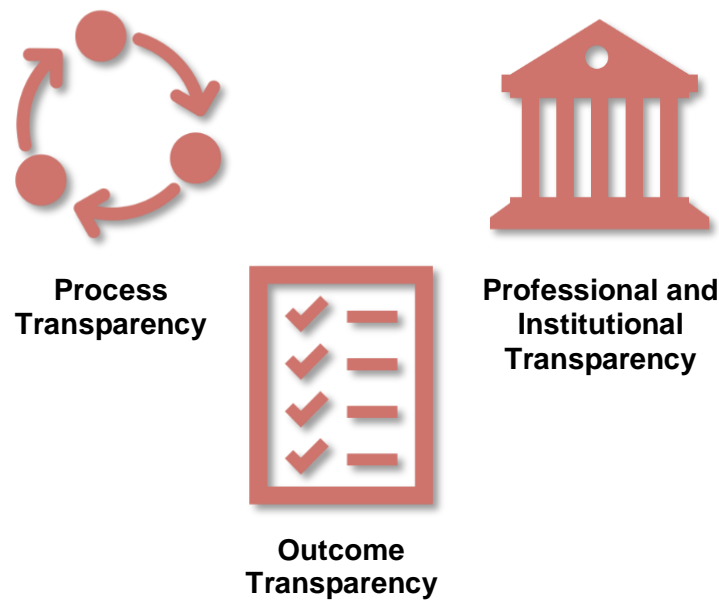

**Process Transparency**

**Professional and Institutional Transparency**

**Outcome Transparency**

*Figure 6: Different types of transparency*

- *Process transparency* requires that the design, development, and implementation processes underlying the decisions or behaviours of data systems are accessible for oversight and review so that justified public trust and public consent can be ascertained.

- *Professional and institutional transparency* requires that, at every stage of the design and implementation of a project, responsible team members should be identified and held to rigorous standards of conduct that secure and maintain professionalism and institutional transparency. These standards should include the core, justice-promoting values of integrity, honesty, and sincerity as well as positionality-aware modes of neutrality, objectivity, and impartiality. All professionals involved in the research, development, production, and implementation of data-intensive technologies are, first and foremost, acting as fiduciaries of the public interest and must, in keeping with these core justice-promoting values, put the obligations to serve that interest above any other concerns.

- *Outcome transparency* demands that stakeholders are informed of where data systems are being used and how and why such systems performed the way they did in specific contexts. Outcome transparency therefore requires that impacted individuals can understand the rationale behind the decisions or behaviours of these systems, so that they can contest objectionable results and seek effective remedy. Such information should be provided in a plain, understandable, non-specialist language and in a manner relevant and meaningful to those affected.

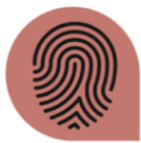

# Identity

**1. Interrogate, understand, and critique harmful categorisations:** The construction and categorisation of data, particularly when it is about people, is a fundamentally social activity that is undertaken by humans whose views of the world are, in part, the product of cultural contexts and historical contingencies. As such, the construction and categorisation of data is shaped by the sociocultural conditions and historical contexts from which it is derived. The social character of data coupled with the sorting and clustering that proceeds from its cleaning and pre-processing can lead to categorisations that are racialised, misgendered, or otherwise discriminatory. This can involve the employment of binary categorisations and constructions—for example, gender binaries (male/female) or racial binaries (white/non-white)—that are oriented to dominant groups and that ought to be critically scrutinised and questioned. Data justice calls for examining, exposing, and critiquing histories of racialisation and discriminatory systems of categorisation reflected in the way data is classified and the social contexts underlying the production of these classifications.

**2. Challenge the reification of identities by resisting the imposition of data categories as a convenience of computational sorting and optimisation:** In the construction and categorisation of data, system designers and developers can mistakenly treat socially constructed, contested, and negotiated categories of identity as fixed and natural classes. When this happens, the way that these designers and developers categorise identities can become naturalised and reified. This can lead to the inequitable imposition of fixed attributes to classify people who do ascribe to these categorisations or who view them as fluid and inapplicable to the way they identify or regard their themselves.

**3. Challenge the erasure of identities by contesting the deletion or omission of identity characteristics:** Where designers and developers miss, exclude, or group together categories or classes of data that pertain to self-ascribed identity characteristics (like race, gender, or religious affiliation), they run the risk of erasing or rendering invisible the identities of those who value or claim the identity characteristics that have been excluded or subsumed. For instance, the designers of a data system may group together a variety of non-majority racial identities under the category of "non-white" and thereby potentially erase a variety of distinctive identity claims, or they may record gender only in terms of binary classification (male/female) and, in turn, erase the identity claims of non-binary and trans people.

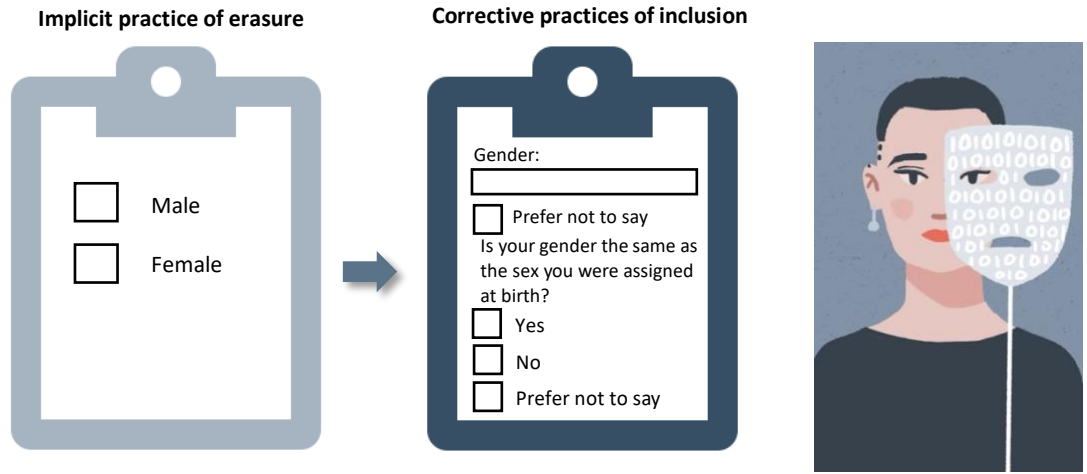

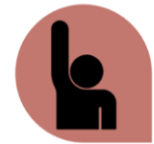

*Figure 7: Practices of erasure that take place during project lifecycle*

**4. Challenge the erasure of intersectional identity characteristics:** Intersectional discrimination occurs where protected characteristics like race and gender overlap in ways that compound or magnify discriminatory harms. Designers and developers can produce and use data systems that disparately injure people who possess unacknowledged intersectional characteristics of identity which render them vulnerable to harm, but which are not recognised in the bias mitigation and performance testing measures taken by development teams. For instance, a facial recognition system could be trained on a dataset that is primarily populated by images of white males, thereby causing the trained system to systematically perform poorly for darker skinned females. If the designers of this system have not taken into account the vulnerable intersectional identity (in this case, darker skinned females) in their bias mitigation and performance testing activities, this identity group becomes invisible and so too do injuries done to its members.

## Participation

**1. Democratise data and data work:** *Prioritise meaningful and representative stakeholder participation, engagement, and involvement from the earliest stages of the data innovation lifecycle to ensure social licence, public consent, and justified public trust.* The democratisation of data scientific research and data innovation practices involves bringing members of impacted communities, policymakers, practitioners, and developers together to collaboratively articulate shared visions for the direction that data innovation agendas should take. This entails the collective and democratically based determination of what acceptable and unacceptable uses of data research and innovation are, how data research and innovation should be governed, and how to integrate the priorities of social justice, non-discrimination, and equality into practices of data collection, processing, and use.

**2. Challenge existing, domination-preserving modes of participation:** Where current justifications and dynamics of data practices reinforce or institutionalise prevailing power structures and hierarchies, the choice to participate in such practices can be counterproductive or even harmful. When options for a community's participation in data innovation ecosystems and their governance operate to normalise or support existing power imbalances and the unjust data practices that could follow from them, these options for involvement should be approached critically. A critical refusal to participate is a form of critical participation and should remain a practical alternative where extant modes of participation normalise harmful data practices and the exploitation of vulnerability.

**3. Ensure transformational inclusiveness rather than power-preserving inclusion:** Incorporating the priority of inclusion into sociotechnical processes of data innovation can be detrimental where existing power hierarchies are sustained or left unaddressed. Where mechanisms of inclusion normalise or support existing power imbalances in ways that could perpetuate data injustices and fortify unequal relationships, these should be critically avoided. *Transformational inclusiveness demands participatory parity* so that the terms of engagement, modes of involvement, and communicative relationships between the includers and the included are equitable, symmetrical, egalitarian, and reciprocal.

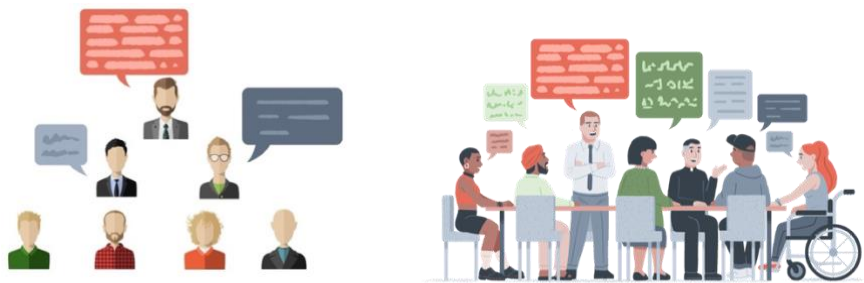

*Figure 8: Moving towards transformational inclusiveness*

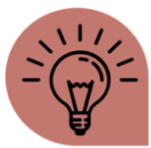

# Knowledge

**1. Embrace the pluralism of knowledges:** Different communities and sociocultural groups possess unique ways of seeing, understanding, and being in the world. This plurality of knowledges, and of lived experience, should inform and be respected in practices of data collection, processing, and use as well as in the policymaking practices surrounding the governance of data technologies. *Embracing the pluralism of knowledges involves recognising that diverse forms of knowledge, and ways of knowing and understanding, can add valuable insights to the aspirations, purposes, and justifications of data use—including on the local or context-specific impacts of data-intensive innovation.* Moreover, inclusion of diverse knowledges and ways of being can open unforeseen paths to societal and biospheric benefits and maximise the value and utility of data use across society in ways which take account of the needs, interests, and concerns of all affected communities.

**2. Challenge the assumed or unquestioned authority of technical, professional, or "expert" knowledge across scientific and political structures:** Processes of knowledge creation in data science and innovation are social processes which require scrutiny and wider public engagement to hold those with "expertise" to account and to ensure that data science and innovation progress in ways which align with wider societal values. This means that data technology producers and users have a responsibility to communicate plainly, equitably, and to as wide an audience as possible: *Clear and accessible public communication of research and innovation purposes/goals and data analytic and scientific results, should enable the public to interrogate the claims and arguments being put forward to justify data-driven decision-making and data innovation agendas.* This also means that members of the public have a corollary *responsibility to listen*—i.e., to pay attention to, engage with, and critically assess the scientifically authoritative knowledge claims and technological systems that impact them.

**3. Prioritise interdisciplinarity:** Approach the pursuit of understanding of data innovation environments—and the sociotechnical processes and practices behind them—through a holistically informed plurality of methods. This involves placing a wide range of academic disciplines and specialised knowledges conceptually on par, enabling an appreciation and integration of a wide range of insights, framings, and understandings. Ways of knowing that cannot (or are not willing to) accommodate a disciplinary plurality of knowledgeable voices that may contribute to richer comprehensions of any given problem cease to be knowledgeable *per se*.

**4. Pursue a reflexive and positionally aware objectivity that amplifies marginalised voices:** A robust approach to objectivity demands that knowers have positional self-awareness, which acknowledges the limits of everyone's personal, historical, and cultural standpoint. It also demands that knowers carry out critical and systematic self-interrogation to better understand these limitations. This launching point in *reflexive and positionally aware objectivity* can end up leading to *more objective and more universalistic understandings* than modes of scientific or technical objectivity which stake a claim to unobstructed neutrality and value-free knowledge that evades self-interrogation about the limits of standpoint and positionality. One reason for this has to do with power dynamics. Reflexive and positionally aware objectivity starts from a reflective recognition of how differential relations of power and social domination can skew the objectivity of deliberations by biasing the balance of voices that are represented in those deliberations. It then actively tries to include and amplify marginalised voices in the community of inquiry to transform situations of social disadvantage where important perspectives and insights are muted, silenced, and excluded into situations that are scientifically richer and *more advantaged*. Such richer and more inclusive ecologies of understanding end up producing more comprehensive knowledge and more just and coherent practical and societal outcomes. Reflexive and positionally aware objectivity amplifies the voices of the marginalised, vulnerable, and oppressed as a way to overcome claims of objectivity, impartiality, and neutrality that mask unquestioned privileges.

## Data Justice Pillars in Action

To help orient the reader to how the six data justice pillars could be applied in practice, we offer in this section concrete instances that illustrate the ways developers, producers of technologies, and civil society organisations with a focus on data-driven technologies have been able to engage in transformative practices that have advanced data justice. One example is offered for each pillar.

### POWER:

GobLab UAI, based at the School of Government at Adolfo Ibáñez University in Chile, is a public innovation lab that has worked with public agencies to develop decision-support algorithms that incorporate ethical standards. The lab provides research and assistance for developers to integrate ethical requirements like transparency, equity, privacy, explainability, and responsibility during the tender, procurement, and implementation processes while assisting public and private companies in the provision of social services. With their commitment to ethical data research and innovation, GobLab UAI challenges the tendencies of entrenched forms of political and administrative power to pursue technology development in purely strategic and instrumental ways.

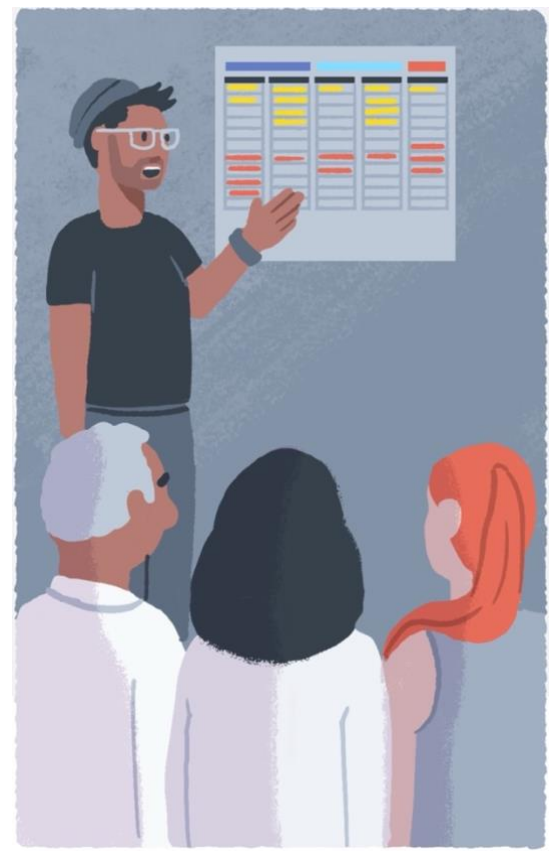

*Figure 9: Data justice is about social licence and democratic governance*

### ACCESS:

Founded as a potential solution to the 'biggest lie on the internet', Terms of Service; Didn't Read (ToS;DR) is an open-access website that provides brief overviews of complex terms of service and privacy policies of various organisations to individuals who may overlook the fine print of data processing and utilisation when they accept ToS's. Through peer reviews and multiple rounds of grading, the website allows users to share their perspectives on data privacy, processing, and utilisation policies that are often inundated with legal jargon that can obfuscate large-scale data extraction and potential privacy violations.

## EQUITY:

A collaboration between MIT's Senseable City Lab, BRTech 3D, and Rio's City Planning Commissioner, Favelas 4D is a project that aims to make visible the unmapped, unplanned, irregular, and complex sections of Rocinha, the largest favela (an urban area of informal settlements affected by socio-economic deprivation) in Rio de Janeiro. The project makes use of LiDAR (Light Detection and Ranging) for terrestrial scanning that attempts to fill gaps in spatial data. This is invaluable data, not present even in global mapping applications like Street View, that can assist urban designers while informing policies for public services.

## PARTICIPATION:

Women in Machine Learning and Data Science (WiMLDS) aims to overcome the long-standing gender gap in STEM fields through the organisation of the Scikit Open-Source Sprints that look to increase the contributions of women and gender minorities on GitHub. Currently gender minorities comprise only 11% of the contributions on the open-source community for software development. The sprints also serve as training platforms for participants to hone their skills in pull requests, virtual environments, and tests like flake8 and pytest. The website for WiMLDS provides networking options and posts employment opportunities for gender minorities to further their professional growth in STEM.

## KNOWLEDGE:

Following numerous instances of algorithmic bots working within coordinated disinformation campaigns across South America, the Institute for Technology & Society (ITS Rio) in collaboration with the National Democratic Institute for International Affairs launched Atrapabot, a tool providing a rating on the probability that an account is a bot. Complementary research and literature have also been released by ITS Rio on Atrapabot so that other researchers and organisations can strengthen their attempts to mitigate the spread of disinformation in the digital sphere. ITS Rio's interdisciplinary approach to addressing the automated spread of disinformation is an example of how enriched knowledge can contribute to the advancement of data justice.

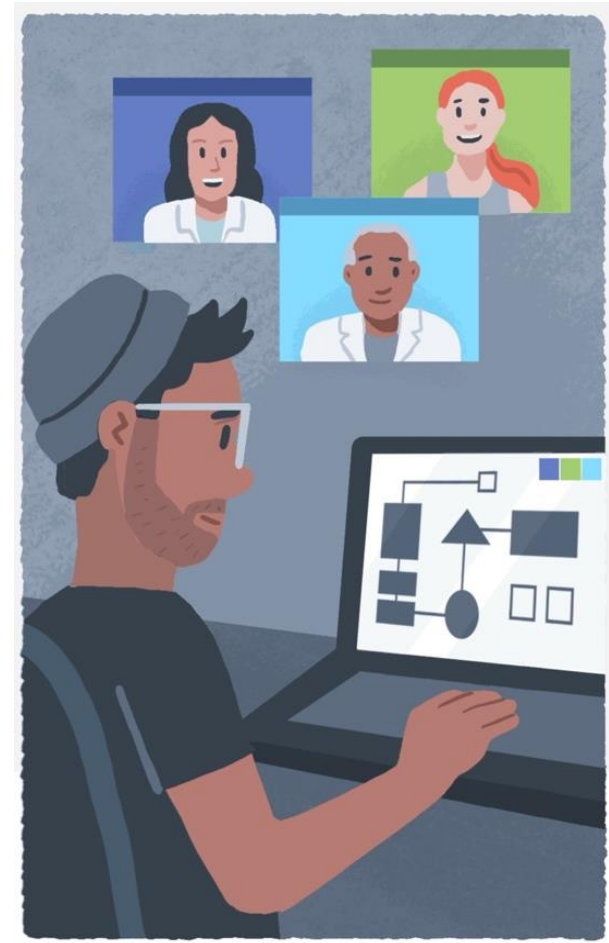

*Figure 10: Many voices should inform the design process*

## IDENTITY:

Coding Rights, an intersectional feminist organisation in Brazil, is developing notmy.ai, a feminist toolkit which can support anti-colonial and feminist movements in understanding and questioning algorithmic systems. The toolkit not only maps out those public projects in South America that may cause harms unique to the intersection of gender with race, class, sexuality, age, and territory but also subsequently conducts impact assessments. They have also proposed frameworks moulded by feminist theories of consent that can assist in negotiating or rejecting terms of service policies of digital platforms.

# Putting the Pillars into Practice I: Developing Shared Understandings of Data Justice

As our Policy Pilot Partner collaborations and research have shown, it is important to recognise that the idea of data justice is contextually bounded. It can mean something different to different people, depending on their varying histories, social and cultural backgrounds, needs, and circumstances. Variations in how communities understand data justice are rooted in differences in the shared values, languages, and lived experiences of the communities and groups who take it up and use it.

A durable concept of data justice should therefore be able to accommodate multiple understandings of justice and equity.[10] Moreover, it should remain open to revision. It should be able to evolve through continuous dialogue and re-evaluation so that it can stay responsive to diverse and changing realities of power, culture, and datafication.

It may be useful, along these lines, to carry out a reflective and deliberative process in developing the shared understandings of data justice that will animate the way you, and your community, approach putting the idea into practice. This will allow you to shape your data justice practices in accordance with your own values, goals, and purposes and, where helpful, to tailor the data justice pillars to your unique perspectives and vision.

**Here are some prompts to support this process of reflection:**

| Developing a Shared Understanding of Data Justice |
| --- |
| Reflection Questions |
| • *What comes to mind when you think of the words "justice" and "equity"? Do you understand these words as having to do with ethics or the legal sphere, or both? If you think of justice and equity as ethical or moral ideas, what are their main properties?* |
| • *Are there any other words that you see as equivalent to "justice" and "equity" or that you feel are better suited to your community's history, its social and cultural background, and the lived experience of its members?* |
| • *What comes to mind when you think of the words "injustice" and "inequity"? How, if at all, do these understandings enrich the way you think of the meanings of "justice" and "equity"?* |
| • *Before engaging in this guide, were you familiar with the idea of social justice? If so, what did this concept mean to you?* |
| • *Refer to the **Key Term: Social Justice** box above. How does this description of social justice align with your own understanding? How does it differ?* |
| • *How would you apply your understandings of justice, equity, and social justice in contexts of data collection and use? Do the data justice pillars outlined above (power, access, equity, participation, identity, and knowledge) align with these understandings?* |
| • *If the pillars differ from your understandings in significant ways, what, if any, resonance and harmonies do you feel are possible between your understandings and the pillars? What other pillars or guiding priorities can be included in your own approach to data justice?* |

---

[10] In undertaking this research, our team wanted to reflect on and recognise how our own positionality could shape the way we were approached our research on data justice. We have attached our positionality statement as . Details on the process of engaging in positionality reflection are explored below.

# Putting the Pillars into Practice Across a Project Lifecycle

In this section, we start to put the conceptual work discussed thus far into action by laying out the project lifecycle for data-driven technology (hereafter shortened to just 'project lifecycle') and then mapping the data justice pillars onto the specific stages of the lifecycle where they demand serious consideration and active intervention. Given the complexity of such projects, these stages are not necessarily linear and may occur simultaneously or in inverse order.

There are many ways of carving up a project lifecycle. For instance, Sweenor and colleagues break it into four stages: Build, Manage, Deploy and Integrate, Monitor.[11] Ashmore, Calinescu, and Paterson identify four stages, which have a more specific focus on data science: data management, model learning, model verification, and model deployment.[12] Furthermore, there are also well-established methods that seek to govern common tasks within a project lifecycle, such as data mining (e.g., CRISP-DM or SEMMA).

The multiplicity of approaches is likely a product of the evolution of diverse methods and approaches, particularly with regards to data mining/analytics, the significant impact of ML on research and innovation, and the specific practices and considerations inherent to each of the various domains where ML techniques are applied.[13] While there are many benefits of existing frameworks (e.g., carving up a complex process into smaller components that can be managed by a network of teams or organisations), they do not tend to focus on the wider social or ethical aspects that interweave throughout the various stages of a project lifecycle. Our team has developed a depiction of the AI/ML Project Lifecycle, which can be viewed on the following page. Note that while much of our discussion focuses on projects that involve AI/ML to some degree, the questions and considerations raised are relevant for data-driven systems in general. This is important given that data justice issues are relevant for and can manifest within and from technical systems which do not include models. After introducing the Project

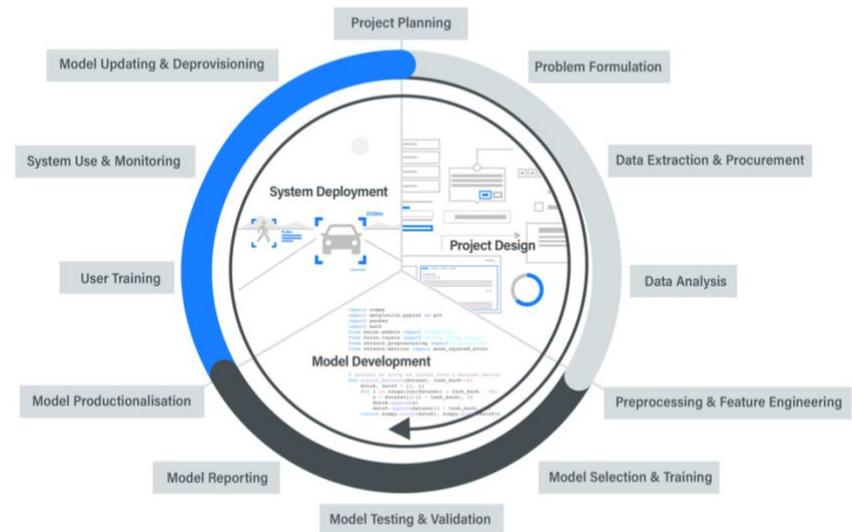

*Figure 11: Depiction of the AI/ML project lifecycle*

Lifecycle visualisation, we will provide an overview of how each of these phases can be viewed through a sociotechnical lens.

The Project Lifecycle—the overarching stages of design, development, and deployment (for a typical data-driven project)—can be split into indicative tasks and activities. In practice, both the stages and the tasks will overlap with their neighbours and may be revisited where a particular task requires an iterative approach.

To begin, the inner circle breaks the project lifecycle into a process of design, development, and deployment. These terms are intended to be maximally inclusive. For example, the design stage encompasses any project task or decision-making process that scaffolds or sets constraints on later project stages (i.e., design system constraints). Importantly, this includes ethical, social, and legal constraints, which we will discuss later.

---

[11] Sweenor et al., 2020
[12] Ashmore et al., 2019
[13] Ibid.

Each of the stages shades into its neighbours, as there is no clearly delineated boundary that differentiates certain project design activities (e.g., data extraction and exploratory analysis) from model design activities (e.g., pre-processing and feature engineering, model selection). As such, the design stage overlaps with the development stage, but the latter extends to include the actual process of training, testing, and validating what is built. Similarly, the process of productionising that is engineered within its runtime environment can be thought of as both a development and deployment activity. And, so, the deployment stage overlaps with the 'development' stage, and also overlaps with the 'design' stage, as the deployment of a system should be thought of as an ongoing process (e.g., where new data are used to continuously train the ML model, or the decision to de-provision a model may require the planning and design of a new model, if the older (legacy) system becomes outdated).

The three higher-level stages can be thought of as a useful heuristic for approaching the project lifecycle. However, each higher-level stage subsumes a wide variety of tasks and activities that are likely to be carried out by different individuals, teams, and organisations, depending on their specific roles and responsibilities. Therefore, it is important to break each of the three higher-level stages into their (typical) constituent parts, which are likely to vary to some extent between specific projects or within particular organisations. In doing so, we expose a wide range of diverse tasks, each of which give rise to a variety of ethical, social, and legal challenges. To fully understand the sociotechnical implications of each of the stages of the lifecycle, the next section will walk through an overview of each of the stages.

# Design Tasks and Processes

## Agenda Setting, Commissioning, and Project Planning

Rather than using data-driven technology as a "hammer" to go looking for nails, it is best to have a clear idea in mind of what the project's goals are at the outset. This can help to avoid a myopic focus on a narrow class of (AI/ML-based) technical "solutions", and also helps create space for a diversity of approaches—some of which may not require AI/ML, or indeed any technical intervention at all. Project planning, therefore, can comprise a wide variety of tasks, including, but not limited to:

- an assessment of whether such an undertaking is the right approach given availability and nature of resources and data, existing technologies and processes already in place, the complexity of the use-contexts involved, and the nature of the policy or social problem that needs to be solved[14];
- an analysis of user needs in relation to the prospective technology and whether a solution involving the latter provides appropriate affordances in keeping with user needs and related functional desiderata;
- mapping of key stages in the project to support governance and business tasks (e.g., scenario planning);
- an assessment of resources and capabilities within a team, which is necessary for identifying any skills gaps,
- a contextual assessment of the target domain and of the expectations, norms, and requirements that derive therefrom;
- stakeholder analysis and team positionality reflection to determine the appropriate level and scope of community engagement activities[15];
- stakeholder impact assessment, supported by affected people and communities, to identify and evaluate possible harms and benefits associated with the project (e.g., socioeconomic inequalities that may be exacerbated as a result of carrying out the project), to gain social license and public trust, and also feed into the process of problem formulation in the next stage;
- wider impact assessments—both where required by statute and done voluntarily for transparency and best practice (e.g., equality impact assessments, data protection impact assessments, human rights impact assessment, bias assessment).

[14] Leslie et al., 2022a

[15] Leslie et al., 2022b

## Problem Formulation

Here, 'problem' refers both to a well-defined computational process (or a higher-level abstraction of the process) that is carried out by the algorithmic system to map inputs to outputs and to the wider practical, social, or policy issue that will be addressed through the translation of that issue into the technical frame. For instance, on the computational side, a convolutional neural network carries out a series of successive transformations by taking (as input) an image, encoded as an array, in order to produce (as output) a decision about whether some object is present in the image. On the practical, social, and policy side, there will be a need to define the computational "problem" being solved in terms of the algorithmic system's embeddedness in the social environment and to explain how it contributes to (or affects) the wider sociotechnical issue being considered. In the convolutional neural network example, the system being produced may be a facial recognition technology that responds to a perceived need for the biometric identification of criminal suspects by matching face images in a police database. The social issue of wanting to identify suspects is, in this case, translated into the computational mechanism of the computer vision system. But, beyond this, diligent consideration of the practical, social, or policy issue being addressed by the system will also trigger, *inter alia*, reflection on the complex intersection of potential algorithmic bias, the cascading effects of sociohistorical patterns of racism and discrimination, wider societal and community impacts, and the potential effects of the use of the model on the actors in the criminal justice systems who will become implementers and subjects of the technology.

Sociotechnical considerations are also important for determining and evaluating the choice of target variables used by the algorithm, which may ultimately be implemented within a larger automated decision-making (or other) system. The task of formulating the problem allows the project team to get clear on what input data will be needed, for what purpose, and whether there exist any representational issues in, for example, how the target variables are defined. It also allows for a project team (and impacted stakeholders) to reflect on the reasonableness of the measurable proxy that is used as a mathematical expression of the target variable, for instance, whether being taken into care within six months of a visit from child protective services is a reasonable proxy for a child's being "at risk" in a predictive risk model for children's social care. The semantic openness and contestability of formulating problems and defining target variables in AI/ML and data-driven innovation lifecycles is why stakeholder engagement, which helps bring a diversity of perspectives to project design, is so vital, and why this stage is so closely connected with the interpretive burdens of the project planning stage (e.g., discussion about legal and ethical concerns regarding permissible uses of personal or sensitive information).

## Data Extraction and Procurement

Ideally, the project team should have a clear idea in mind (from the planning and problem formulation stages) of what data are needed prior to extracting or procuring them. This can help mitigate risks associated with over-collection of data (e.g., increased privacy or security concerns) and help align the project with values such as *data minimisation*.[16] Of course, this stage may need to be revisited after carrying out subsequent tasks (e.g., pre-processing, model testing) if it is clear that insufficient or imbalanced data were collected to achieve the project's goals. Where data is procured, questions about provenance arise (e.g., legal issues, concerns about informed consent of human data subjects). Generally, responsible data extraction and procurement require the incorporation of domain expertise into decision-making so that desiderata of data minimisation as well as of securing relevant and sufficient data can be integrated into design choices.

Procurement considerations also extend beyond the data. Many technical undertakings draw on supply chains that range beyond internal project teams, whereby various components and services that form parts of AI/ML systems might be commissioned or sourced from other actors, and where certain aspects of system design, development, deployment, operation, and management may be outsourced to various entities. Ethics and equity issues may arise at any point along the supply chain (for instance, where a procured component is produced through discriminatory or unjust labour practices or labour practices that are in violation human rights). It is

---

therefore important to diligently consider data justice concerns right throughout these supply chains, which includes amongst others the responsibilities, obligations, incentives, and positionalities of the suppliers involved.

## Preliminary Data Analysis

Exploratory data analysis is an important stage for hypothesis generation or uncovering possible limitations of the dataset that can arise from missing data, in turn identifying the need for any subsequent augmentation of the dataset to deal with possible class imbalances. However, there are also risks that stem from cognitive biases (e.g., confirmation bias) that can create cascading effects that effect downstream tasks (e.g., model reporting).

# Development Tasks and Processes

## Data Pre-Processing and Feature Engineering

Pre-processing and feature engineering is a vital but often lengthy process, which overlaps with the design tasks in the previous section and shares with them the potential for human choices to introduce biases and discriminatory patterns into the AI/ML workflow. Tasks at this stage include data cleaning, data wrangling or normalisation, and data reduction or augmentation. It is well understood that the methods employed for each of these tasks can have a significant impact on the model's performance (e.g., deletion of rows versus imputation methods for handling missing data). As Ashmore and colleagues note, there are also various desiderata that motivate the tasks, such as the need to ensure the dataset that will feed into the subsequent stages is relevant, complete, balanced, and accurate.[17] At this stage, human decisions about how to group or disaggregate input features (e.g., how to carve up categories of gender or ethnic groups) or about which input features to exclude altogether (e.g., leaving out deprivation indicators in a predictive model for clinical diagnostics) can have significant downstream influences on the fairness and equity of an AI/ML system.

## Model Selection and Training

This stage determines the model type and structure that will be produced in the next stages. In some projects, model selection will result in multiple models for the purpose of comparison based on some performance metric (e.g., accuracy). In other projects, there may be a need to first of all implement a pre-existing set of formal models into code. The class of relevant models is likely to have been highly constrained by many of the previous stages (e.g., available resources and skills, problem formulation), for instance, where the problem demands a supervised learning algorithm instead of an unsupervised learning algorithm; or where explainability considerations require a more interpretable model (e.g., a decision tree).

Prior to training the model, the dataset will need to be split into training and testing sets to avoid model overfitting. The *training set* is used to fit the ML model, whereas the *testing set* is a hold-out sample that is used to evaluate the fit of the ML model to the underlying data distribution. There are various methods for splitting a dataset into these components, which are widely available in popular package libraries (e.g., the scikit-learn library for the Python programming language). Again, human decision-making at this stage about the training-testing split and about how this shapes desiderata for external validation—a subsequent process where the model is validated in wholly new environments—can be very consequential for the trustworthiness and reasonableness of the development phase of an AI/ML system.

## Software Engineering

Technical systems are driven by software. In more model-centric systems, software might play a supportive role, by providing the 'plumbing' to enable the mapping from inputs to outputs in a model. However, many systems entail significant software undertakings, where a range of software components are built or used to deliver the system's functionality; define and provide the computational workflows; support the system's management, security, and governance; in addition to incorporating and supporting various models. It is therefore important that those involved in software engineering processes are informed about data justice principles and account for stakeholder and community perspectives as part of their development activities. Often there are opportunities for the software to be designed to provide functionality that better supports transparency, understanding, governance, agency, control, scrutiny, and other relevant aims.

---

[17] Ashmore et al., 2019

## Interface Design

Systems will generally entail some form of user interface. These interfaces form the basis for one to use and operate the system, present the outputs, results, and decisions from that system, and can be a mechanism for extracting data. User interfaces are important to consider, as the interface ultimately determines (and constrains) how a system may be used and managed, while working to influence a user's understanding about the nature of the system and its outputs. Clearly, designs that have the potential to be obfuscatory or manipulative (e.g., 'dark patterns') must be avoided; however, attention is also needed for ensuring that the controls, representations and means of presentation are aligned with the specific needs and desires from different user communities.

## Testing and Validation

Testing and validation processes must be undertaken to ensure that engineered systems meet their specifications, requirements, and intended purposes. This entails having the testing, validation, and quality assurance criteria, metrics, and evaluation cases that appropriately account for relevant data justice considerations. For example, as part of the model building process, the testing set is typically kept separate from the training set, in order to provide an unbiased evaluation of the final model fit on the training dataset. However, the training set can be further split to create a validation set, which can then be used to evaluate the model while also *tuning model hyperparameters*. This process can be performed repeatedly, in a technique known as (k-fold) cross-validation, where the training data are resampled ($k$-times) to compare models and estimate their performance in general when used to make predictions on unseen data. This type of validation is also known as 'internal validation', to distinguish it from external validation, and, in a similar way to choices made about the training-testing split, the manner in which it is approached can have critical consequences for how the performance of a system is measured against the real-world conditions that it will face when operating "in the wild".

## Reporting

Although the previous stages are likely to create a series of artefacts while undertaking the tasks themselves, reporting should also be handled as a separate stage to ensure that the project team reflect on the future needs of various stakeholders and end users. While this stage is likely to include information about the performance measures used for evaluating the various components that have been engineered, such as the model (e.g., decision thresholds for classifiers, accuracy metrics), it can (and should) include wider considerations, such as intended use of the model, details of the features used, training-testing distributions, and any ethical considerations that arise from these decisions (e.g., fairness constraints, use of politically sensitive demographic features).

# Deployment Processes

## System Productionalisation

This process, sometimes known as 'operationalisation', requires understanding how the engineered system will impact—and be impacted by—the functioning of the wider sociotechnical environment that the tool is embedded within (e.g., a decision support tool used in healthcare for patient triaging that may exacerbate existing health inequalities within the wider community). Ensuring a particular system or model works within the proximate systems and workflows can be a complex programming and software engineering task. But, more importantly, understanding how to ensure the system's sustainability given its embeddedness in complex and changing sociotechnical environments requires active and contextually-informed monitoring, situational awareness, and vigilant responsiveness.

## Implementer and User Training

Although the performance of the system is evaluated in earlier stages, the system's impact cannot be entirely evaluated without consideration of the human factors that affect its performance in real-world settings. The impact of human cognitive biases, such as algorithmic aversion must also be considered, as such biases can lead to over- and under-reliance on the model (or system), in turn negating any potential benefits that may arise from its use. Understanding the social and environmental context is also vital, as sociocultural norms may contribute to how training is received, and how the system itself is evaluated[18].

## System Use and Monitoring

Depending on the context of deployment, it is likely that the performance of the system could degrade over time. For example, *concept drift* is typically caused by increasing variation between how representative the training dataset was at the time of development and how representative it is at later stages, perhaps due to changing social norms (e.g., changing patterns of consumer spending, evolving linguistic norms that affect word embeddings). As such, mechanisms for monitoring the system's performance should be instantiated within the runtime protocols to track any divergence, and key thresholds should be determined at early stages of a project (e.g., during project planning or in initial impact assessment) and revised as necessary based on monitoring of the system's use. Also important are mechanisms that ensure the system is being used correctly and appropriately, and not being repurposed for tasks not envisaged nor accounted for in the previous design stages.

## Maintenance, Updating, or De-provisioning

A system (and its components) may require updating or modification, be it due to a degradation in real-world performance, a particular incident or failure, or a change in requirements. Updating the system requires either revisiting previous stages to make planned adjustments (e.g., model selection and training), or if more significant alterations are required the extant system may need to be entirely de-provisioned, necessitating an entirely new round of project planning.

---

[18] See Burton et al., 2020

# Examples of Pillar Touchpoints Across the AI/ML Lifecycle

| Data Collection and Use Lifecycle: | Design<br>• Agenda Setting, Commissioning, and Project Planning<br>• Problem Formulation<br>• Data Extraction and Procurement<br>• Preliminary Data Analysis | Develop<br>• Data Pre-Processing and Feature Engineering<br>• Model Selection and Training<br>• Software Engineering<br>• Interface Design<br>• Testing and Validation<br>• Reporting | Deploy<br>• Productionalisation<br>• Implementer and User Training<br>• System Use and Monitoring<br>• Maintenance, Updating, or Deprovisioning |
|---|---|---|---|
| **Power** | Consider whether the choice to build the system or the way its problem is formulated could reinforce or entrench harmful power relations | Interrogate how power imbalances could lead to the use of categorisations or discriminatory proxies in the model | Assess whether users are able to challenge aspects of system deployment that they believe to be reinforcing existing power dynamics |
| **Equity** | Assess whether the goals and purposes of the project will enable or disable oppression; Safeguard strengths-based, socially licensed, and community-involving data collection | Evaluate whether included features represent marginalised, vulnerable, and historically discriminated against social groups in negative ways that centre on the '5 D's' | Evaluate whether, in its operation and outcomes, the system is advancing social justice and combating rather than entrenching longer-term patterns of inequality and discrimination |
| **Access** | Interrogate how material inequalities could limit access for some to the benefits of the system | Make decisions about data annotation, labelling, and categorisation are publicly available to impacted communities | Pursue routes to responsible data sharing and research community-building that equitably advance access to research and innovation capacity |
| **Identity** | Ensure that data collection and procurement practices entail an active awareness of harmful categorisations that can misidentify, omit, or erase members of marginalised groups | Ensure that pre-processing and feature engineering activities do not involve harmful discriminatory categorisations and erasure practices that harm identity claims | Evaluate whether, in its operation and outcomes, the system's categorisations line up with the changing ways impacted people and groups self-identify and represent themselves |
| **Participation** | Empower impacted communities to get involved in project planning, problem formulation, and impact assessment; | Meaningfully involve impacted people in design and development processes to secure alignment with their understandings and goals | Prioritise meaningful and representative stakeholder participation in processes of impact monitoring and re-evaluation and in |

| | Pursue engagement objectives that secure transformational inclusion | and to ensure social licence, public consent, and justified public trust | the set-up of remedy and mitigation measures for harms |
|---|---|---|---|
| **Knowledge** | Include a diverse set of cultural views, practical knowledges, lived experiences, and disciplinary perspectives in project planning, problem formulation, and impact assessment | Ensure that sufficient domain knowledge and contextual understanding of social norms and expectations are present throughout all stages of development. | Consult members of impacted communities with lived experience of the operations and outcomes of the system (especially from marginalised and vulnerable groups) in the evaluation of known and unforeseen impacts |

# Key Concepts: The SAFE-D Principles

The SAFE-D principles (Safety, Accountability, Fairness, Explainability, and Data Quality, Integrity, Protection, and Privacy) are goals that assist with upholding responsible, equitable, and trustworthy data innovation practices throughout the entirety of the AI/ML lifecycle and within data innovation ecosystems. When these goals are met through responsible research and innovation practices, they can assist with advancing data justice within projects.

The specific meaning of the goals, however, will be delineated in different ways depending on the use context or domain of the AI/ML or data-driven system, and the processes and results of the Stakeholder Engagement Process (which will be detailed later on in this document). For example, depending on the system, SAFETY could relate to the *physical* safety of patients in the context of an AI/ML system used in healthcare, or *environmental* safety for an AI/ML system used in agriculture or forestry.

Each of these goals have implications for the advancement of data justice. For example, Accountability relates to ensuring transparency throughout the entirety of the AI/ML project lifecycle and data innovation ecosystem. This goal goes hand-in-hand with the Access Pillar of Data Justice which calls for the promotion of the airing and sharing of data injustices in communities through process, professional and institutional, and outcome transparency. To ensure the promotion of equitable access, transparency must be applied to both the outcomes of the use of data systems and the processes behind their design, development, and implementation.

We will first detail the top-level goals of SAFE-D followed by further specification through the presentation of additional properties, which are to be established in either the project or the system to ensure these goals are reached.

# Top-Level Goals

SAFETY is of paramount importance for ensuring the sustainable development, deployment, and use of an AI system. From a technical perspective, this requires the system to be secure, robust, and reliable. And from a social sustainability perspective, this requires the practices behind the system's production and use to be informed by ongoing consideration of the risk of exposing affected rights-holders to harms, continuous reflection on project context and impacts, ongoing stakeholder engagement and involvement, and change monitoring of the system from its deployment through to its retirement or deprovisioning.

ACCOUNTABILITY can include specific forms of process transparency (e.g., as enacted through process logs or external auditing) that may be necessary for mechanisms of redress, or broader processes of responsible governance that seek to establish clear roles of responsibility where transparency may be inappropriate (e.g., confidential projects).

**F**AIRNESS is inseparably connected with sociolegal conceptions of equity and justice, which may emphasise a variety of features such as non-discrimination, equitable outcomes, or procedural fairness through bias mitigation, but also social and economic equality, diversity, and inclusiveness.

**E**XPLAINABILITY is a key condition for autonomous and informed decision-making in situations where AI systems interact with or influence human judgement and decision-making. Explainability goes beyond the ability to merely interpret the outcomes of an AI system; it also depends on the ability to provide an accessible and relevant information base about the processes behind the outcome.

**D**ATA QUALITY, INTEGRITY, PROTECTION, AND PRIVACY must all be established to be confident that the (data-driven) AI system has been developed on secure grounds.

- 'DATA QUALITY' captures the *static* properties of data, such as whether they are (a) *relevant* to and *representative* of the domain and use context, (b) *balanced* and *complete* in terms of how well the dataset represents the underlying data generating process, and (c) *up-to-date* and *accurate* as required by the project.
- 'DATA INTEGRITY' refers to more *dynamic* properties of data stewardship, such as how a dataset evolves over the course of a project lifecycle. In this manner, data integrity requires (a) *contemporaneous* and *attributable* records from the start of a project (e.g., process logs; research statements), (b) ensuring *consistent* and *verifiable* means of data analysis or processing during development, and (c) taking steps to establish *findable*, *accessible*, *interoperable*, and *reusable* records towards the end of a project's lifecycle.
- 'DATA PROTECTION AND PRIVACY' reflect ongoing developments and priorities as set out in relevant legislation and regulation of data practices as they pertain to fundamental rights and freedoms, democracy, and the rule of law. For example, the right for data subjects to have inaccurate personal data rectified or erased.

# Properties of SAFE-D

Each of the SAFE-D goals has a variety of lower-level properties associated with them. These properties ought to be established in either the project or the system if the goal is said to have been obtained. Many of the lower-level properties link in closely with the assurance of the Six Pillars of Data Justice, as many of these properties have implications for the advancement of the data justice. For instance, when considering the SAFE-D goal of Fairness, one of the lower-level properties is Non-discrimination. Ensuring the model or system does not discriminate against any given group is critical to not only promoting data justice but also assuring the Equity and Identity Pillars are met.

## Safety

- **Sustainability:** The goal of safety must be achieved with an eye towards the sustainability of a safe system. This goes beyond environmental sustainability (e.g., the ecological footprint of the project and system). It also includes an understanding of the long-term use context and impact of the system, and the resources needed to ensure the system continues to operate safely over time. For instance, sustainability may depend upon sufficient *change monitoring* processes that establish whether there has been a substantive change in the underlying data distributions or social operating environment. Sustainability also involves engaging and involving impacted individuals and communities in the design and assessment of AI systems that could impact their human rights and fundamental freedoms.
- **Security:** Security encompasses the protection of several operational dimensions of an AI system when confronted with possible adversarial attack. A secure system is capable of maintaining the integrity of the information that constitutes it. This includes protecting its architecture from the unauthorised modification or damage of any of its component parts. A secure system also remains continuously functional and accessible to its authorised users and keeps confidential and private information secure even under hostile or adversarial conditions.

- **Robustness:** The objective of robustness can be thought of as the goal that an AI system functions reliably and accurately under harsh conditions. These conditions may include adversarial intervention, implementer error, or skewed goal-execution by an automated learner (in reinforcement learning applications). The measure of robustness is therefore the strength of a system's integrity the soundness of its operation in response to difficult conditions, adversarial attacks, perturbations, data poisoning, and undesirable reinforcement learning behaviour.

- **Reliability:** The objective of reliability is that an AI system behaves exactly as its designers intended and anticipated. A reliable system adheres to the specifications it was programmed to carry out. Reliability is therefore a measure of consistency and can establish confidence in the safety of a system based upon the dependability with which it operationally conforms to its intended functionality.

- **Accuracy and Performance Metrics:** In machine learning, the accuracy of a model is the proportion of examples for which it generates a correct output. This performance measure is also sometimes characterised conversely as an error rate or the fraction of cases for which the model produces an incorrect output. As a performance metric, accuracy should be a central component to establishing and nuancing the approach to safe AI. Specifying a reasonable performance level for the system may also often require refining or exchanging of the measure of accuracy. For instance, if certain errors are more significant or costly than others, a metric for total cost can be integrated into the model so that the cost of one class of errors can be weighed against that of another.

## Accountability

- **Traceability:** Traceability refers to the process by which all stages of the data lifecycle from collection to deployment to system updating or deprovisioning are documented in a way that is accessible and easily understood. This may include not only the parties within the organisation individuals who use the system.

- **Answerability:** Answerability depends upon a human chain of responsibility. Answerability responds to the question of *who is accountable* for an automation supported outcome.

- **Auditability:** Whereas the property of answerability responds to the question of *who is accountable* for an automation supported outcome, the notion of auditability answers the question of *how the designers and implementers of AI systems are to be held accountable*. This aspect of accountability has to do with *demonstrating and evidencing* both the responsibility of design and use practices and the justifiability of outcomes.

- **Clear Data Provenance and Data Lineage:** Clear provenance and data lineage consists of records that are accessible and simultaneously detail how data was collected and how it has been used and altered throughout the stages of pre-processing, modelling, training, testing, and deploying.

- **Accessibility:** Accessibility involves ensuring that information about the processes that took place to design, develop, and deploy an AI system are easily accessible by individuals. This not only refers to suitable means of explanation (clear, understandable, and accessible language) but also the mediums for delivery.

- **Reproducibility:** Related to and dependant on the above four properties, reproducibility refers to the ability for others to reproduce the steps you have taken throughout your project to achieve the desired outcomes and where necessary to replicate the same outcomes by following the same procedure.

- **Responsible Governance:** Responsible governance ensures accountability and responsibility for the processes that occur throughout the data lifecycle. This includes the identification and assignment of a data protection officer, as well as clearly identifying data controllers and processors. This may also include the creation of an independent oversight board to ensure these individuals are held accountable and the processes are well-documented.

## Fairness

- **Bias Mitigation:** It is not possible to eliminate bias entirely. However, effective bias mitigation processes can minimise the unwanted and undesirable impact of systematic deviations, distortions, or disparate outcomes that arise to a project governance problem, interfering factor, or from insufficient reflection on historical social or structural discrimination.

- **Diversity and Inclusiveness:** A significant component of fairness aware design is ensuring the inclusion of diverse voices and opinions in the design and development process through the participation of a more representative range of stakeholders. This includes considering whether values of civic participation, inclusion, and diversity been adequately considered in articulating the purpose and setting the goals of the project. Consulting with internal organisational stakeholders is also necessary to strengthen the openness, inclusiveness, and diversity of the project.
- **Non-Discrimination:** Your system should not create or contribute to circumstances whereby members of protected groups are treated differently or less favourably than other groups because of their respective protected characteristic.
- **Equality:** The outcome or impact of a system should either maintain or promote a state of affairs in which every individual has equal rights and liberties, and equal access or opportunities to whatever good or service the AI system brings about.

## Explainability

- **Interpretability:** Interpretability consists of the ability to understand how and why a model performed the way it did in a specific context and therefore to grasp the rationale behind its decision or behaviour.
- **Responsible Model Selection:** This involves meeting the normal expectations of intelligibility and accessibility that accompany the function the system will fulfil in the sector or domain in which it will operate. The availability of more interpretable algorithmic models or techniques in cases where the selection of an opaque model poses risks to the physical, psychological, or moral integrity of rights-holders or to their human rights and fundamental freedoms. The availability of the resources and capacity that will be needed to responsibly provide supplementary methods of explanation (e.g., simpler surrogate models, sensitivity analysis, or relative feature importance) in cases where an opaque model is deemed appropriate and selected.
- **Accessible Rationale Explanation:** The reasons that led to a decision—especially one that is automated—delivered in an accessible and non-technical way.
- **Responsible Implementation and User Training:** Training users to operate the AI system may include:
  - conveying basic knowledge about the nature of machine learning,
  - explaining the limitations of the system,
  - educating users about the risks of AI-related biases, such as decision-automation bias or automation-distrust bias, and
  - encouraging users to view the benefits and risks of deploying these systems in terms of their role in helping humans to come to judgements, rather than replacing that judgement.

## Data Quality

- **Source Integrity and Measurement Accuracy:** Effective bias mitigation begins at the very commencement of data extraction and collection processes. Both the sources and instruments of measurement may introduce discriminatory factors into a dataset. When incorporated as inputs in the training data, biased prior human decisions and judgments—such as prejudiced scoring, ranking, interview-data or evaluation—will become the 'ground truth' of the model and replicate the bias in the outputs of the system in order to secure discriminatory non-harm, as well as ensuring that the data sample has optimal source integrity. This involves securing or confirming that the data gathering processes involved suitable, reliable, and impartial sources of measurement and sound methods of collection.
- **Timeliness and Recency:** If datasets include outdated data, then changes in the underlying data distribution may adversely affect the generalisability of the trained model. Provided these distributional drifts reflect changing social relationship or group dynamics, this loss of accuracy with regard to the actual characteristics of the underlying population may introduce bias into an AI system. In preventing discriminatory outcomes, timeliness and recency of all elements of the data that constitute the datasets must be scrutinised.

- **Relevance, Appropriateness, and Domain Knowledge:** The understanding and utilisation of the most appropriate sources and types of data are crucial for building a robust and unbiased AI system. Solid domain knowledge of the underlying population distribution and of the predictive or classificatory goal of the project is instrumental for choosing optimally relevant measurement inputs that contribute to the reasonable determination of the defined solution. Domain experts should collaborate closely with the technical team to assist in the determination of the optimally appropriate categories and sources of measurement.
- **Adequacy of Quantity and Quality:** This property involves assessing whether the data available is comprehensive enough to address the problem set at hand, as determined by the use case, domain, function, and purpose of the system. Adequate quantity and quality should address sample size, representativeness, and availability of features relevant to problem.
- **Balance and Representativeness:** A balanced and representative dataset is one in which the distribution of features that are included, and the number of samples within each class is similar to the underlying distribution that exists in the overall population.

## Data Integrity

- **Attributable:** Data should clearly demonstrate who observed and recorded it, when it was observed and recorded, and who it is about.[19]
- **Consistent, Legible and Accurate:** Data should be easy to understand, recorded permanently, and original entities should be preserved. Data should be free from errors and conform with the protocol. Consistency includes ensuring data is chronological (e.g., has a date and time stamp that is in the expected sequence).

- **Complete:** All recorded data requires an audit trail to show nothing has been deleted or lost.
- **Contemporaneous:** Data should be recorded as it was observed, and at the time it was executed.
- **Responsible Data Management:** Responsible data management ensures that the team has been trained on how to manage data responsibly and securely, identify possible risks and threats to the system, and assign roles and responsibilities for how to deal with these risks if they were to occur. Policies on data storage and public dissemination of results should be discussed within the team and with stakeholders, as well as being clearly documented.
- **Data Traceability and Auditability:** Any changes or revisions to the dataset (e.g., additions, augmentations, normalisation) that occur after the original collection should be clearly traceable and well-documented to support any auditing.

## Data Protection and Privacy

- **Consent (or legitimate basis) for processing:** Each Party shall provide that data processing can be carried out on the basis of the free, specific, informed, and unambiguous consent of the data subject or of some other legitimate basis laid down by law. The data subject must be informed of risks that could arise in the absence of appropriate safeguards. Such consent must represent the free expression of an intentional choice, given either by a statement (which can be written, including by electronic means, or oral) or by a clear affirmative action and which clearly indicates in this specific context the acceptance of the proposed processing of personal data. Mere silence, inactivity or pre-validated forms or boxes should not, therefore, constitute consent. No undue influence or pressure (which can be of an economic or other nature) whether direct or indirect, may be exercised on the data subject and consent should not be regarded as freely given where the data subject has no genuine or free choice or is unable to refuse or withdraw consent without prejudice. The data subject has the right to withdraw the consent they gave at any time (which is to be distinguished from the separate right to object to processing). Full consideration should be given to the potential impact of any personal data processing on the fundamental rights and freedoms of the individuals involved.

---

[19] The properties of Data integrity have been adapted from SL Controls, n.d.

- **Data Security:** Each Party shall provide that the controller, and, where applicable the processor, takes appropriate security measures against risks such as accidental or unauthorised access to, destruction, loss, use, modification, or disclosure of personal data. Each Party shall provide that the controller notifies, without delay, at least the competent supervisory authority of those data breaches which may seriously interfere with the rights and fundamental freedoms of data subjects.

- **Data Minimisation:** Personal data being processed is adequate (sufficient to properly fulfil the stated purpose), relevant (has a rational link to that purpose), and limited to what is necessary (do not hold more data than needed for that purpose).

- **Transparency:** The transparency of AI systems can refer to several features, both of their inner workings and behaviours, as well as the systems and processes that support them. An AI system is transparent when it is possible to determine how it was designed, developed, and deployed. This can include, among other things, a record of the data that were used to train the system, or the parameters of the model that transforms the input (e.g., an image) into an output (e.g., a description of the objects in the image). However, it can also refer to wider processes, such as whether there are legal barriers that prevent individuals from accessing information that may be necessary to understand fully how the system functions (e.g., intellectual property restrictions).

- **Proportionality:** Proportionality in a broad sense encompasses both the necessity and the appropriateness (proportionality in a narrow sense) of a measure, that is, the extent to which there is a logical link between the measure and the (legitimate) objective pursued. The term proportionality is used as an evaluative notion, such as in the case of a data protection principle that states only personal data that are necessary and appropriate for the purposes of the task are collected.

- **Purpose Limitation:** The purposes for data processing must be outlined and documented from the beginning and made available to all individuals through privacy information. Personal data must adhere to the original purpose unless it is compatible with the original purpose, additional consent is received, or there is an obligation or function set out in law.

- **Accountability:** Appropriate measures and records must be in place to demonstrate compliance and responsibility for how data has been processed in alignment with the other principles.

- **Lawfulness, Fairness, and Transparency:** These three principles form part of the 'lawful basis' for the collection and use of personal data. Personal data must be used in a fair manner that is not unduly detrimental, unexpected, or misleading. Any processes in which data is used should not be in breach of any other laws, and teams must be clear, open, and honest with individuals about how their personal data is being used.

- **Respect for the rights of data subjects:** Respect for the rights of data subjects requires putting in place adequate mechanisms or undertaking necessary actions so as to ensure that the rights of data subjects as defined under the Council of Europe's Convention 108+ and the General Data Protection Regulation (GDPR) are upheld. Where necessary, this includes the responsible handling of sensitive data.

# Stakeholder Engagement Process

A valuable strategy for putting this data justice guide into practice is to engage with affected stakeholders to gain insights about proposed and ongoing data innovation projects. While the design, development, and deployment of AI/ML and other data-driven systems can be (and often are) conducted without active community engagement, innovation practices built around the inclusion of community-led participation and co-design from the earliest stages of stakeholder identification are more likely to support data justice goals. Stakeholder engagement is also a strategy for building better systems that are more likely to be accepted and adopted by affected individuals and communities.

Involving affected individuals and communities should, in all cases, be a significant consideration. Stakeholder involvement ensures that your project will possess an appropriate degree of public accountability and scaffolds trust. Stakeholder involvement can also help to strengthen the objectivity, reflexivity, reasonableness, and robustness of the choices your project team makes across the AI/ML project lifecycle and within the data innovation ecosystem. This is because the inclusion of a wider range of perspectives (especially of those who are most marginalised) can enlarge a project team's purview, expand its domain knowledge as well as its understanding of stakeholders' needs. It can likewise unearth potential biases that may arise from limiting the standpoints that inform decision-making to those of team members.

To facilitate proportionate stakeholder engagement and input when addressing guideline questions, developers must first gain a contextually informed understanding of the social environment and human factors that may be impacted by, or may impact, the tool or system you are planning to develop. This is the purpose of the Stakeholder Engagement Process, which is not a one-off activity, but rather should be occur each time the guide is used.

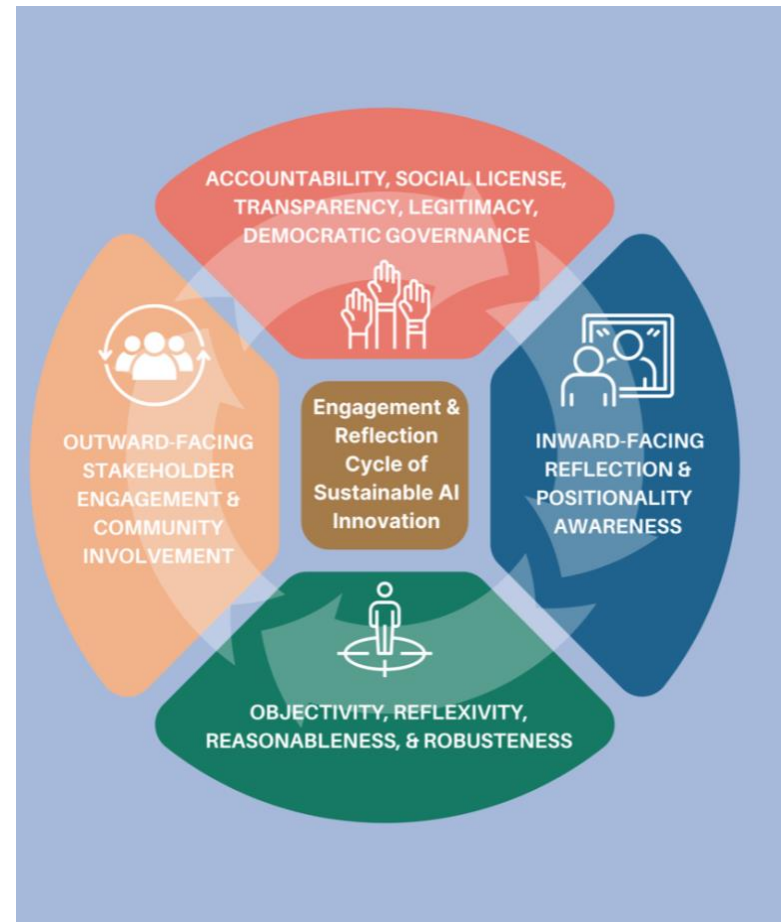

*Figure 12: Depiction of the engagement and reflection cycle of sustainable AI innovation*

The Stakeholder Engagement Process is comprised of three steps:

1. **Preliminary Project Scoping and Stakeholder Analysis:** Outline key project components, identify individuals or groups who may be affected by, or may affect, your innovation project, scope potential stakeholder impacts, and evaluate the salience and contextual characteristics of identified stakeholders.
2. **Positionality Reflection:** Evaluate team positionality as related to that of stakeholders. Consider strengths and limitations presented by team positionality.
3. **Stakeholder Engagement Objectives and Methods:** Establish engagement objectives that enable the appropriate degree of stakeholder engagement and co-production in project evaluation, and methods that support the achievement of defined objectives.

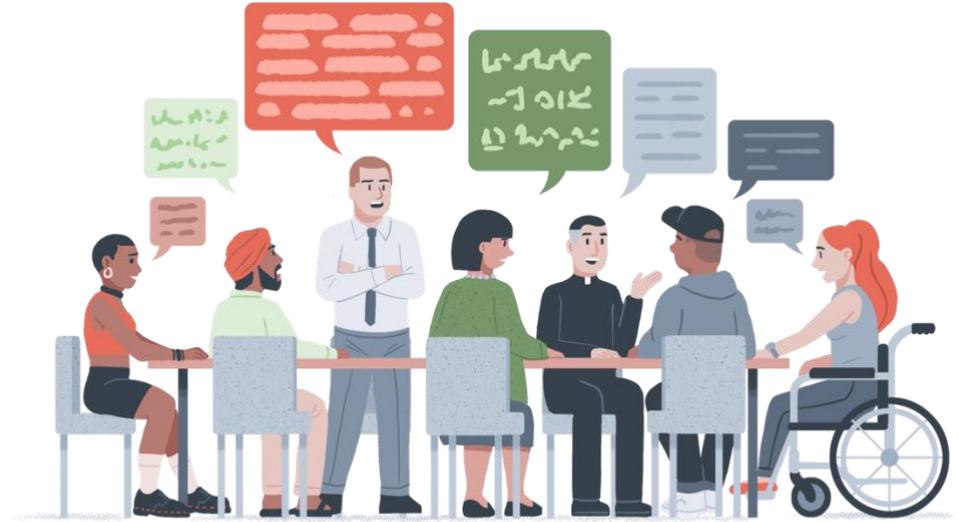

*Figure 13: Creating meaningfully inclusive dialogue*

**Key term: Stakeholder**

Scholars and practitioners from areas as diverse as public policy, land use, environmental and natural resource management, international development, and public health have offered many different definitions of "stakeholders" over the past several decades. Even so, these definitions have converged around a few common characteristics. Stakeholders are individuals or groups that (1) have interests or rights that may be affected by the past, present, and future decisions and activities of organisations; (2) may have the power or authority to influence the outcome of such decisions and activities; (3) possess relevant characteristics that put them in positions of advantage or vulnerability with regard to those decisions and activities.

# Preliminary Horizon-scanning, Policy Scoping, and Stakeholder Analysis

A proportional degree of stakeholder involvement will vary from project to project based upon a preliminary assessment of the potential risks and hazards of the model or tool under consideration.

Low-stakes AI/ML applications that are not safety-critical, do not directly impact the lives of people, and do not process potentially sensitive social and demographic data may need less proactive stakeholder engagement than higher-stakes projects.

You and your project team will need to carry out an initial evaluation of the scope of the possible risks that could arise from your project and of the potential hazards it poses to affected individuals and groups. This should include reasonable and context-based assessments of the dangers posed to human rights, fundamental freedoms, and priorities of AI ethics and data justice. The resulting evaluation will provide insight into a proportionate approach to stakeholder involvement.

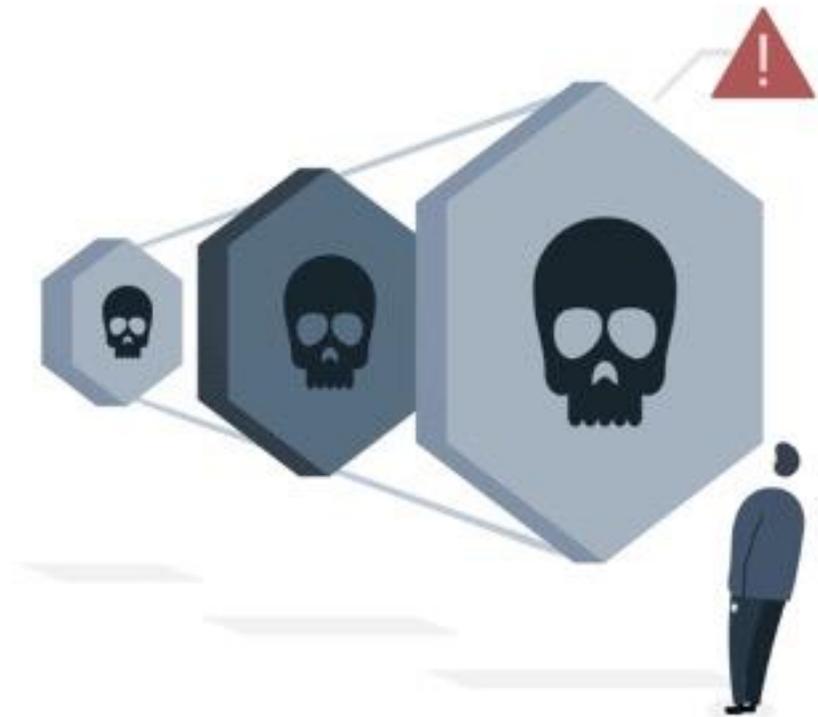

*Figure 14: Confronting the many layers of potentially harmful impacts of irresponsible data collection and use*

Preliminary Project Scoping and Stakeholder Analysis provides a structured approach to such assessment and is the first activity within the SEP process. It involves four sub-steps:

1. **Outlining project, use context, domain, and data:** Outline a high-level description of the prospective system, the domain in which it will operate, the contexts in which it will be used, and the data on which it will be trained. During this initial project scoping activity, you should draw on organisational documents (i.e., the project business case, proof of concept, or project charter), project team collaboration, and desk research (if necessary) to complete the description.
2. **Identifying stakeholders:** Building on this contextual understanding, identify who may be affected by, or may affect, your innovation project.
3. **Scoping potential stakeholder impacts:** Carry out a preliminary evaluation of the potential impacts of the prospective AI system on affected individuals and communities. At this initial stage of reflection, members of your project team should review *Annex 1: 12 Principles and Priorities of Responsible Data Innovation*, and then consider which of these principles and priorities could be impacted by the design, development and deployment of the prospective AI/ML or data-driven system and how.
4. **Analysing stakeholder salience:** Assess the relevance of each identified stakeholder group to your project and to its use contexts. Assess the relative interests, rights, vulnerabilities, and advantages of identified stakeholders as these interests, rights, vulnerabilities, and advantages may be impacted by, or may impact, the AI system your team is planning to develop and deploy. When identifying stakeholders, your team should also consider organisational stakeholders, whose input will likewise strengthen the openness, inclusivity, and diversity of your project.

**The following table presents a series of prompts and questions pertaining to each of the sub-steps, it is meant to help conduct the Preliminary Project Scoping and Stakeholder analysis step of the Stakeholder Engagement Process. Note that before you answer the questions pertaining to 'Scoping potential stakeholder impacts' sub-step you should first review *Annex 1: 12 Principles and Priorities of Responsible Data Innovation.***

| Preliminary Policy Scoping and Stakeholder Analysis | |
|---|---|
| Questions | Responses |
| **Outlining Project, Use Context, Domain, and Data** | |
| *What AI system is being built and what type of product or service will it offer?* | |
| *What benefits will the system bring to its users and customers, and will these benefits be widely accessible?* | |
| *Which parts or elements of the AI system, if any, will be procured from third-party vendors, suppliers, sub-contractors, or external developers? What are the responsibilities of these third-party vendors, suppliers, sub-contractors, or external developers?* | |
| *Which algorithms, techniques, and model types will be used in the AI system? (Provide links to technical papers where appropriate)* | |
| *In a scenario where your project optimally scales, how many people will it impact, for how long, and in what geographic range (local, national, global)? (Describe your rationale)* | |
| **USE CONTEXT** | |
| *What is the purpose of this AI system and in which contexts will it be used? (Briefly describe a use-case that illustrates primary intended use)* | |
| *Is the AI system's processing output to be used in a fully automated way or will there be some degree of human control, oversight, or input before use? (Describe)* | |
| *Will the AI system evolve or learn continuously in its use context, or will it be static?* | |
| *To what degree will the use of the AI system be time-critical, or will users be able to evaluate outputs comfortably over time?* | |
| *What sort of out-of-scope uses could users attempt to apply the AI system, and what dangers may arise from this?* | |
| **DOMAIN** | |
| *In what domain will this AI system operate?* | |
| *Which, if any, domain experts have been or will be consulted in designing and developing the AI system?* | |
| **DATA** | |
| *What datasets are being used to build this AI system?* | |
| *Will any data being used in the production of the AI system be acquired from a vendor or supplier? (Describe)* | |
| *Will the data being used in the production of the AI system be collected for that purpose, or will it be re-purposed from existing datasets? (Describe)* | |
| *What quality assurance and bias mitigation processes do you have in place for the data lifecycle—for both acquired and collected data?* | |
| **Identifying stakeholders** | |
| *Who are the stakeholders (both individuals and social groups) that may be impacted by, or may impact, the project?* | |
| *Do any of these stakeholders possess sensitive or protected characteristics that could increase their vulnerability to abuse, adverse impact, or discrimination, or for reason of which they may require additional protection or assistance with respect to the impacts of the project? If so, what characteristics?* | |
| *Could the outcomes of this project present significant concerns to specific groups of stakeholders given vulnerabilities caused or precipitated by their distinct circumstances?* | |
| *If so, what vulnerability characteristics expose them to being jeopardised by project outcomes?* | |
| **Scoping potential stakeholder impacts (Refer to <u>Annex 1: 12 Principles and Priorities of Responsible Data Innovation</u> for detailed descriptions)** | |
| *How, if at all, are each of the twelve following principles and priorities salient to the AI/ML or data-driven system I am planning to build, given its intended purposes and contexts?*<br>• Respect for and protection of human dignity<br>• Interconnectivity, solidarity, and intergenerational reciprocity<br>• Environmental flourishing, sustainability, and the rights of the biosphere<br>• Protection of human freedom and autonomy<br>• Prevention of harm and protection of the right to life and physical, psychological, and moral integrity<br>• Non-discrimination, fairness, and equality<br>• Rights of Indigenous peoples and Indigenous data sovereignty<br>• Data protection and the right to respect of private and family life<br>• Economic and social rights<br>• Accountability and effective remedy<br>• Democracy<br>• Rule of law | |
| *How could each of the twelve principles and priorities be impacted by the AI system we are planning to build?* | |
| *If things go wrong in the implementation of our AI system or if it is used out-of-the-scope of its intended purpose and function, what harms could be done to stakeholders in relation to each of the twelve principles and priorities?* | |
| *How, if at all could the AI/ML or data-driven system I am planning to build impact beneficence, safety, and non-harm?* | |
| **Analysing stakeholder salience** | |
| *Which affected stakeholder groups are most likely to be positively impacted by the deployment of the system or tool? Which affected stakeholder groups are most likely to be negatively impacted?* | |
| *Which affected stakeholder groups have the greatest needs in relation to potential benefits of the system/tool?* | |
| *How might different affected stakeholder groups be differentially impacted by the system?* | |
| *Are there any relevant power relations between these differentially impacted stakeholder groups that could affect the distribution of the prospective system's benefits and risks? Consider their relative advantages and disadvantages, and which affected stakeholders may have direct or indirect influence over the project and its outcomes?* | |
| *Which affected stakeholder groups have existing influence within relevant communities, political processes, or in relation to the domain in which the system will be deployed? How could these dynamics of influence impact the distribution of the prospective system's benefits and risks?* | |
| *Which affected stakeholder groups' influence is limited? How could these limitations impact the distribution of the prospective system's benefits and risks?* | |

## Engaging in Positionality Reflection

All individual human beings come from unique places, experiences, and life contexts that have shaped their thinking and perspectives. Reflecting on these is important insofar as it can help us understand how our viewpoints might differ from those around us and, more importantly, from those who have diverging cultural and socioeconomic backgrounds and life experiences. Identifying and probing these differences can enable developers to better understand how their own backgrounds, for better or worse, frame the way they see others, the way they approach and solve problems, and the way they carry out their respective data and development practices. By undertaking such efforts to recognise social position and differential privilege, they may gain a greater awareness of their own personal biases and unconscious assumptions. This then can enable them to better discern the origins of these biases and assumption and to confront and challenge them in turn.

**When taking positionality into account, developers are to reflect on their own positionality matrix:**

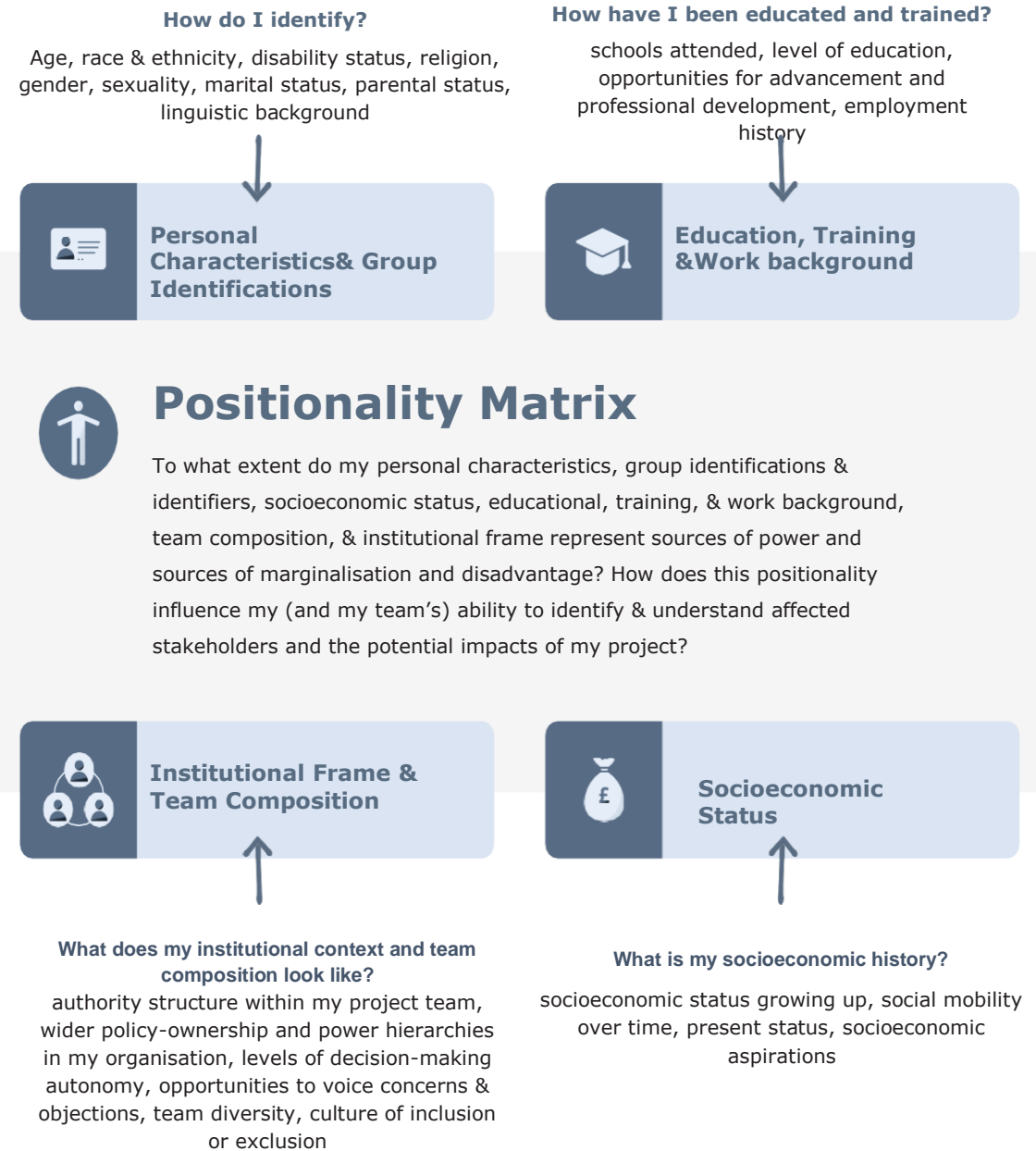

*Figure 15: Positionality Matrix*

**The following table presents a series of prompts and questions pertaining to positionality reflection. It is meant to help conduct this step of the Stakeholder Engagement Process.**

| Positionality Reflection | |
|---|---|
| Questions | Responses |
| *How does the positionality of team members relate to that of affected stakeholders?* | |
| *Are there any ways that your position as a team could influence your evaluation of the potential negative and positive impacts of this project?* | |
| *Are there any ways that your position as a team could limit your perspective when evaluating the impact of this project?* | |
| *Are there any ways that your position as a team could strengthen your perspective when evaluating the impact of this project? Consider overlapping identities and experience.* | |
| *What (if any) missing stakeholder viewpoints would strengthen your team's assessment of this system's potential impacts?* | |
| *How does the positionality of team members relate to that of affected stakeholders?* | |
| *Are there any ways that your position as a team could influence your evaluation of the potential negative and positive impacts of this project?* | |

# Stakeholder Engagement Objectives and Methods

**Determining Stakeholder Engagement Objectives**
All stakeholder engagement processes can run the risk either of being cosmetic tools employed to legitimate projects without substantial and meaningful participation or of being insufficiently participative, i.e., of being one-way information flows or nudging exercises that serve as public relations instruments. To avoid such hazards of superficiality, your team should shore up its proportionate approach to stakeholder engagement with deliberate and precise goal-setting. The objectives of engagement that your team chooses will depend on factors that divide into three categories, which are presented here with accompanying descriptions:

## Factors determining the objectives of engagement

| | |
|---|---|
| **Team-based assessments of risks of adverse impacts** | • Assessment of how to make stakeholder involvement proportionate to the scope of a project's potential risks and hazards |
| **Team-based assessments of positionality** | • Evaluation of team positionality—for instance, cases where the identity characteristics of team members do not sufficiently reflect or represent significantly impacted groups. How can the project team "fill the gaps" in knowledge, domain expertise, and lived experience through stakeholder participation? |
| **Establishment of stakeholder engagement goals** | • Determination of engagement objectives that enable the appropriate degree of stakeholder engagement and co-production in project evaluation and oversight processes<br>• Choosing participation goals from a spectrum of engagement options (informing, partnering, consulting, empowering) that equip your project with a level of engagement which meets team-based assessments of risk and positionality |

When weighing these three factors, you should use the results of your preliminary project scoping and stakeholder analysis to establish a **clear and explicit stakeholder engagement objective** and document this. **The following table outlines** *a range of engagement objectives, their means of participation, and the level of agency they support for stakeholders:*

| DEGREE OF PARTICIPATION | MEANS OF PARTICIPATION | LEVEL OF AGENCY |
|---|---|---|
| **INFORM** | | |
| Stakeholders are made aware of decisions and developments. | External input is not sought out. Information flows in one direction. This can be done through newsletters, the post, app notifications or community forums. | **LOW** Stakeholders are considered information subjects rather than active agents |
| **CONSULT** | | |
| Stakeholders can voice their views on pre-determined areas of focus, which are considered in decision-making. | Engagement occurs through online surveys or short phone interviews, door-to-door or in public spaces. Broader listening events can support consultations. | **LOW** Stakeholders are included as sources of information input under narrow, highly controlled conditions of participation. |
| **PARTNER** | | |
| Stakeholders and teams share agency over the determination of areas of focus and decision making. | External input is sought out for collaboration and co-production. Stakeholders are collaborators in projects. They are engaged trough focus groups. | **MODERATE** Stakeholders exercise a moderate level of agency in helping to set agendas through collaborative decision making. |
| **EMPOWER** | | |
| Stakeholders are engaged with as decision-makers and are expected to gather pertinent information and be proactive in co-operation. | Co-production exercises occur through citizens' juries, citizens' assemblies, and participatory co-design. Teams provide support for stakeholders' decision making. | **HIGH** Stakeholders exercise a high level of agency and control over agenda-setting and decision making. |

**Determining Stakeholder Engagement Methods**

Once you have established your engagement objective, you are in a better position to assess which method or methods of stakeholder involvement are most appropriate for carrying out your data practice.

Determining appropriate engagement methods for conducting this process necessitates that you (1) evaluate and accommodate of stakeholder needs, and (2) pay attention to practical considerations of resources, capacities, timeframes, and logistics that could enable or constrain the realisation of your objective:

| Factors determining engagement methods | |
| --- | --- |
| **Evaluation and accommodation of stakeholder needs** | • Identification of potential barriers to engagement such as constraints on the capacity of vulnerable stakeholder groups to participate, difficulties in reaching marginalised, isolated, or socially excluded groups, and challenges to participation that are presented by digital divides or information and communication gaps between public sector organisations and impacted communities<br>• Identification of strategies to accommodate stakeholder needs such as catering the location or media of engagement to difficult-to-reach groups, providing childcare, compensation, or transport to secure equitable participation, and tailoring the provision of information and educational materials to the needs of participants<br>• Consideration of engagement objectives |
| **Practical considerations of resources, capacities, timeframes, and logistics** | • The resources available for facilitating engagement activities<br>• The timeframes set for project completion<br>• The capacities of your organisation and team to properly facilitate public engagement<br>• The stages of project design, development, and implementation at which stakeholders will be engaged |

Developers should take a deliberate and reflective approach to deciding how to balance participation goals with practical considerations. You should also make explicit the rationale behind your choices and document this. **The following table outlines possible engagement methods along with their respective strengths, weaknesses, and relevant engagement objectives:**

| Mode of Engagement | Practical Strengths | Practical Weaknesses |
|---|---|---|
| 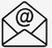 **newsletters** (email)<br><br>*Regular emails (e.g.: fortnightly or monthly) that contain updates, relevant news, and calls to action in an inviting format.*<br><br>**Degree of Engagement**<br>`INFORM` | Can reach many people; can contain large amount of relevant information; can be made accessible and visually engaging. | Might not reach certain portions of the population; can be demanding to design and produce with some periodicity; easily forwarded to spam/junk folders without project team knowing (leading to overinflated readership stats). |
| 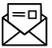 **Letters** (post)<br><br>*Regular letters (e.g.: monthly) that contain the latest updates, relevant news and calls to action.*<br><br>**Degree of Engagement**<br>`INFORM` | Can reach parts of the population with no internet or digital access; can contain large amount of relevant information; can be made accessible and visually engaging. | Might not engage certain portions of the population; Slow delivery and interaction times hampers the effective flow of information and the organisation of further engagement. |
| 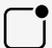 **App notifications**<br><br>*Projects can rely on the design of apps that are pitched to d stakeholders who are notified on their phone with relevant updates.*<br><br>**Degree of Engagement**<br>`INFORM` | Easy and cost-effective to distribute information to large numbers of people; Rapid information flows bolster the provision of relevant and timely news and updates. | More significant initial investment in developing an app; will not be available to people without smartphones. |

| Mode of Engagement | Practical Strengths | Practical Weaknesses |
|---|---|---|
| 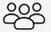 **Community fora**<br><br>*Events in which panels of experts share their knowledge on issues and then stakeholders can ask questions.*<br><br>**Degree of Engagement**<br>`INFORM` | Can inform people with more relevant information by providing them with the opportunity to ask questions; brings community together in a shared space of public communication. | More time-consuming and resource intensive to organise; might attract smaller numbers of people and self-selecting groups rather than representative subsets of the population; effectiveness is constrained by forum capacity. |
| 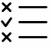 **Online surveys**<br><br>*Survey sent via email, embedded in a website, shared via social media...*<br><br>**Degree of Engagement**<br>`CONSULT` | Cost-effective; simple mass-distribution. | Risk of pre-emptive evaluative framework when designing questions; Does not reach those without internet connection or computer/smartphone access. |
| 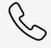 **Phone interviews**<br><br>*Structured or semi-structured interviews held over the phone.*<br><br>**Degree of Engagement**<br>`CONSULT`  `PARTNER` | Opportunity for stakeholders to voice concerns more openly. | Risk of pre-emptive evaluative framework when designing questions; Might exclude portions of the populations without phone access or with habits of infrequent phone use. |

| Mode of Engagement | Practical Strengths | Practical Weaknesses |
|---|---|---|
|  **Door-to-door interviews**<br><br>*Structured or semi-structured interviews held in-person at people's houses.*<br><br>**Degree of Engagement**<br>`CONSULT` `PARTNER` | Opportunity for stakeholders to voice concerns more openly; can allow participants the opportunity to form connections through empathy and face-to-face communication. | Potential for limited interest to engage with interviewers; time-consuming; can be seen by interviewees as intrusive or burdensome. |
|  **In-person interviews**<br><br>*Short interviews conducted in-person in public spaces.*<br><br>**Degree of Engagement**<br>`CONSULT` `PARTNER` | Can reach many people and a representative subset of the population if stakeholders are appropriately defind and sortition is used. | Less targeted; pertinent stakeholders must be identifie by area; little time/interest to engage with interviewer; can be viewed by interviewees as time-consuming and burdensome. |
|  **Focus groups**<br><br>*A group of stakeholders brought together and asked their opinions on a particular issue. Can be more or less formally structured.*<br><br>**Degree of Engagement**<br>`CONSULT` `PARTNER` | Can gather in-depth information; Can lead to new insights and directions that were not anticipated by the project team. | Subject to hazards of group think or peer pressure; complex to facilitate; can be steered by dynamics of differential power among participants. |
|  **Online workshops**<br><br>*Workshops using digital tools such as collaborative platforms.*<br><br>**Degree of Engagement**<br>`CONSULT` | Opportunity to reach stakeholders across regions, increased accessibility depending on digital access. | Potential barriers to accessing tools required for participation, potential for disengagement. |

| Mode of Engagement | Practical Strengths | Practical Weaknesses |
|---|---|---|
|  **Citizen panel or assembly**<br><br>*Large groups of people (dozens or even thousands) who are representative of a town/region.*<br><br>**Degree of Engagement**<br>`INFORM` `PARTNER`<br>`EMPOWER` | Provides an opportunity for co-production of outputs; can produce insights and directions that were not anticipated by the project team; can provide an information base for conducting further outreach (surveys, interviews, focus groups, etc.); can be broadly representative; can bolster a community's sense of democratic agency and solidarity. | Participant rolls must be continuously updated to ensure panels or assemblies remains representative of the population throughout their lifespan; resource-intensive for establishment and maintenance; subject to hazards of group think or peer pressure; complex to facilitate; can be steered by dynamics of differential power among participants. |
|  **Citizen jury**<br><br>*A small group of people (between 12 and 24), representative of the demographics of a given area, come together to deliberate on an issue (generally one clearly framed set of questions), over the period of 2 to 7 days (involve.org.uk).*<br><br>**Degree of Engagement**<br>`INFORM` `PARTNER`<br>`EMPOWER` | Can gather in-depth information; can produce insights and directions that were not anticipated by the project team; can bolster participants' sense of democratic agency and solidarity. | Subject to hazards of group think; complex to facilitate; Risk of pre-emptive evaluative framework; Small sample of citizens involved risks low representativeness of wider range of public opinions and beliefs; |

**The following table presents a series of prompts and questions pertaining to establishing stakeholder objectives and methods, it is meant to help conduct this step of the Stakeholder Engagement Process.**

| Stakeholder Objectives and Methods Questions | |
|---|---|
| *Questions* | *Responses* |
| **Engagement Objective** | |
| *Why are you engaging with stakeholders?* | |
| *What do you envision the ideal purpose and the expected outcomes of engagement activities to be?* | |
| *Ideally, how would stakeholders be able to influence the engagement process and the outcomes?* | |
| *What engagement objective do you believe would be appropriate for this project considering challenges or limitations to assessments related to positionality, and proportionality to the project's potential degree of impact?* | |
| *Considering answers to the above questions, what is your established engagement objective?* | |
| **Engagement Method** | |
| *What resources are available and what constraints will limit potential approaches?* | |
| *Which methods meet your team's engagement objective?* | |
| *What accessibility requirements might stakeholders have?* | |
| *Will online or in-person methods (or a combination of both) be most appropriate to engage salient stakeholders?* | |
| *Considering the above questions, what is your established engagement method answering the guideline questions?* | |
| *How will your team make sure that this chosen method accommodates different types of stakeholders?* | |
| *How will your team ensure that, where appropriate, content for answering the guideline questions is accessible to stakeholders?* | |

# Guiding Questions: Data Justice Pillars and SAFE-D Principles

This section will focus on providing guiding questions which draw from the six pillars of data justice and the SAFE-D Principles (Safety, Accountability, Fairness, Explainability, and Data Quality, Integrity, Protection, and Privacy). These questions are intended to support you and your organisation or firm in gaining a broader understanding of how to promote equitable, freedom-promoting, and rights-sustaining data collection, governance, and use as well as how to advance the 2030 Sustainable Development Goals.

It is important to note that these guiding questions are meant to be used as *reflective tools* to help make you and your organisation or firm aware of relevant elements of data justice and responsible and equitable data innovation practices and to prompt the reader to think differently and more critically about data practices by highlighting the data justice pillars and the SAFE-D principles in question form. The questions will therefore sometimes not assume or expect that you have a direct answer for the issue raised. Rather the questions are encouraging you to try to find a way to get that information or to pursue the initiative to improve equity, access, participation, etc. suggested in the question. For instance, a guide question might ask you to identify the interests of actors who control access to digital infrastructure (connectivity, computing resources, and data assets) and to think about the power imbalances that exist between these actors and your organisation or firm. Much of this information may be less-than-obvious, hidden, obscured, or even opaque. Raising these issues, however, is intended to provide a starting point for further examination and action—and, where this information is more ready-to-hand, to motivate the opening of critical paths towards challenging power and advancing data justice.

| Guiding Questions – Data Justice Pillars |
|---|
| |
| 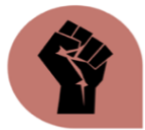 **Power** |
| |
| *Interrogate and critique power* |
| - As a designer, developer, or producer of data systems, what, if any, power imbalances exist between me (or my firm or organisation) and the communities impacted by the data innovation agendas I pursue?<br>    o Do the data innovation agendas I currently pursue reinforce or challenge these imbalances?<br>    o How, if at all, do these imbalances result in unjust exercises of power? Are my current activities entrenching or combating such exercises of power? |

- As a designer, developer, or producer of data systems, what are my interests (or those of my firm or organisation) in collecting or procuring data and in building and deploying data applications?
    - How, if at all, are these interests similar to or different from the interests of those in the communities that my data work impacts?
    - How, if at all, do any power imbalances that exist between me (or my firm or organisation) and impacted communities influence the pursuit of these interests in my (or firm's or organisation's) data innovation agendas?
    - How, if at all, do I (or my firm or organisation) exploit power imbalances to pursue these interests?
- What other actors hold power and influence over the data innovation agendas I pursue and the ways I collect or procure data and build and implement data applications?
    - How reliant am I on the data, tools, models, and digital infrastructure (connectivity, computing resources, and data assets) provided by other actors?
    - What are the interests of these actors? How are they similar to or different from my interests and from those of the members of the communities impacted by my data work?
    - What, if any, power imbalances exist between these actors and me (and my firm or organisation)?
    - What is the history of these power imbalances? Are current policies and available resources reinforcing or contesting these imbalances?
    - How, if at all, do these imbalances result in unjust exercises of power? Are current policies and available resources enabling or combating such exercises of power?
- What does the institutional context of my firm or organisation look like (taking into account the authority structure within my project team(s), wider policy-ownership and power hierarchies in my organisation, levels of decision-making autonomy, and opportunities to voice concerns)? Does this institutional context enable my innovation practices to safeguard the public interest and ensure that standards and governance regimes in the data innovation ecosystem are working towards just and societally beneficial outcomes?
- What is my relationship to the policymakers who control or influence standards and policies that govern the data I collect or procure and the data systems I develop or implement?
    - What are their interests (both stated/manifest and implicit) in controlling or influencing these regimes? How, if at all, do these interests promote societal benefit and accord with the public interest? How, if at all, do these interests run counter to the public interest?
    - What, if any, forms of power and influence do I hold in relation to these policymakers? What, if any, forms of power and influence do they hold in relation to me and my firm or organisation?
- How does my specific data work currently instantiate or break down existing power structures?
- How, if at all, does the product or service I contribute to instantiate or reproduce unequal power relations?

### Challenge Power and Empower People

- How would my development practices need to change to address the current power imbalances influencing data collection or procurement and the development of data systems?

- How could I use my development work to redress current power structures and to tackle power imbalances between my firm or organisation and impacted communities?

- How could I redress current power structures and tackle power imbalances within my firm or organisation?

- How could I redress current power structures and tackle power imbalances between my firm or organisation and other more powerful firms, organisations, standards setters, and policymakers?How could I use my work to support impacted communities to mobilize against unequal power structures?

- How could I use my work to support network-building between communities mobilising against these power structures?

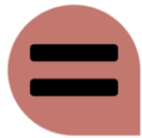 **Equity**

### Issues of equity should be confronted by developers and organisations at the earliest stage of project planning and should inform whether data innovation practices are engaged in at all

- *Consider the initial decision-making processes behind the choices made by you and your firm or organisation to engage in the collection or procurement of data and in the development and implementation of data-driven algorithms and applications.* Ask the following questions:

  o Who is involved in these processes? Are they inclusive, deliberative, and democratic? Do they involve the incorporation of a diversity of perspectives—especially from those stakeholders who will be most impacted by the data collection and use and those who are vulnerable or marginalised?

  o Are evaluations carried out to determine whether a technological solution such as AI (model-building) is necessary and appropriate for the given problem at hand, taking account of and given available resources, existing technologies, current user needs, available data, and nature of the problem?

  o Have these choices been made with considerations in mind of the equity and ethical justifiability of our collection and use of data?

  o Have these choices been made with considerations in mind of how to combat or disable the harmful relations of power that may be embodied in our practices of data collection and use (be it during system design or operation)?

- o Have these choices been made with considerations in mind of the potential effects of our data collection and use on the well-being of impacted communities and their members?

- o Have these choices been made with considerations in mind of the potential effects of our data collection and use on individual dignity and autonomy?

- o Have these choices been made with considerations in mind of the potential effects of our data collection and use on social solidarity, interpersonal connection, and democratic agency?

- o Have these choices been made with considerations in mind of the potential effects of our data collection and use on human rights and fundamental freedoms?

- o Have these choices been made with considerations in mind of possible discriminatory impacts?

- o Have these choices been made with considerations in mind of how our data practices may foster or impede a more equitable and just society?

- o Have transparent processes occurred, on the part of my firm or organisation, to air the rationale behind our choices to build and use data systems?

- o Have we undertaken, and made public, assessments of the potential adverse or beneficial social and ethical impacts of our choices to acquire and use data?

- o If such impact assessments have occurred, to what extent have affected individuals and communities been engaged and involved in them?

- o Where such evaluative processes and impact assessments have either not occurred or not been made public, how can we increase transparency and remedy these deficits?

- *Consider the role that your practices of data collection and use play in impacted communities.* Ask the following questions:

- o How have we introduced our data-practices into these communities?

- o Has this been done with public consent, community involvement, and social license?

- o Were these communities able to debate, contest, or challenge the implementation of these data systems?

- o To what extent do decision-making processes which currently determine our choices to engage in data collection and use enable contestation and revision? How can we introduce policies and controls that enable both internal and external, community-led practices of contestation and revision in our decision-making processes?

## Focus on the transformative power of data equity

- In what ways can we shape our data innovation agendas and practices to redress and transform the patterns of domination and entrenched power differentials that produce data injustices?

- How can we shape our data innovation agendas and practices to respond to the demands for rectification of those who have been harmed or marginalised by existing socioeconomic structures?

- How can we shape our data innovation agendas and practices to further economic equity and justice?

- How can we shape our data innovation agendas and practices to ensure effective interventions are held across the data pipeline which safeguard dataset representativeness and feature equity in data systems?

- How can we shape our data innovation agendas and practices to ensure that the long/short term, direct/indirect consequences of data systems on impacted communities are monitored after deployment to assess for improvements or deteriorations in their quality of life?

- How can we shape our data innovation agendas and practices to ensure that post-deployment assessments evaluate the equity of outcomes among affected groups?

- When undertaking machine learning:

  o Could our categorisation, annotation, or labelling practices serve to discriminate against certain groups?

  o Do we explore whether a model contains any lurking proxies or correlations that are discriminatory or inequitable? What are the processes we have in place to safeguard against these?

- If the dataset is publicly beneficial, has it been made available for equitable access across domains, levels of expertise, and roles (data subjects, aggregators, analysts)?

## *Pursue measurement justice and statistical equity*

- Are our decisions about data collection, labelling, and categorisation made publicly available to impacted communities? If not, how can we set up processes to provide this information?

- How can we initiate and organise community-involvement in the planning and implementation of our data systems, so that:

  o Statistical measurement and automation processes are equitable and help promote public interests?

  o Impacted communities are safeguarded against data over-collection and negative and discriminatory categorisation?

  o We consider and support community member's developmental, physical, cognitive, social, and emotional needs?

  o We focus on using data about marginalised, vulnerable, and historically discriminated against groups in a way that advances social justice, draws on their strengths rather than primarily on perceived weaknesses, and approaches analytics constructively with community-defined goals that are positive and progressive rather than negative, regressive, and punitive?

- How can we promote opportunities for community-involvement in the planning and implementation of data systems so that these are informed by community-led objective setting, problem formulation, and outcome definition as well as multi-stakeholder and interdisciplinary approaches to model planning and implementation?

**Combat any discriminatory forms of data collection and use that centre on disadvantage and negative characterisation**

- In what ways, if at all, are representations of the communities (or groups within them) in the data systems we produce and deploy focused on negative characteristics like disparity, deprivation, disadvantage, dysfunction, and difference (the '5 D's')?

    o Do these systems reinforce or enable existing social hierarchies and power dynamics that marginalise groups who are negatively characterised?

    o How can we adjust our approach to data collection and use to redress this kind of data injustice, where present or possible? How can we adjust our designs to better reflect these concerns?

- In what ways, if at all, are representations of these communities (or groups within them) in the data systems we produce and implement focused on single characteristics (like race, socioeconomic status, or gender) that are associated with relative disadvantages and negative characterisations—or proxies of these characteristics?

    o How, if at all, do such narrow representations detract from our focus on broader goals of advancing public good equitably?

    o How, if at all, do such representations obscure important intragroup differences (for instance differences between different genders within specific racial groups)?

    o How can we redress this kind of data injustice, where present or possible?

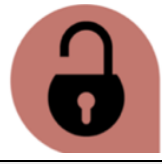 **Access**

**Confronting questions of equitable access involves starting from real-world problems of material inequality and structural injustice**

- What are the existing sociohistorical, economic, and political conditions of injustice experienced by the communities impacted by my firm's or organisation's data collection and use? (Consider circumstances of material deprivation, inequality, institutional and structural discrimination, and maldistribution of resources and social goods.)

- What are the histories of these injustices? Have they developed or become entrenched across generations? Which groups within impacted communities have they affected the most?

- How, if at all, do these conditions inform disparate access to the benefits of data collection and processing?

- How, if at all, does the distribution of benefits from data processing, in turn lead to furthering material conditions of injustice for some groups?
- How can our data practices support societal efforts to recognise and rectify these injustices?
- How can I ensure that our data innovation agendas and data practices address injustice through a transformation of these material conditions?

## Equitably open access to data through responsible data sharing

- Are my data practices (and those of my firm or organisation) currently supporting and advancing responsible data sharing?
- How can our policies and practices equitably open access to data by promoting responsible data sharing so that it supports the reusability of research, the improvement of datasets, broadened access to benefits, as well as oversight and scrutiny? How can such practices and policies accordingly:
  - o Protect the privacy, rights, and freedoms of affected data subjects and communities from where the data comes from by being privacy optimised, impact aware, and security-compliant?
  - o Implement sufficient and transparently reported processes throughout the project's lifecycle to ensure that all data used in producing the system and that shared is accurate, reliable, relevant, appropriate, up-to-date, and of adequate quantity and quality for the use case, domain, function, and purpose of the system?
  - o Ensure the implementation of governance protocols that safeguard data integrity across the lifecycle, promoting trustworthy and responsible data management?
  - o Ensure the implementation of specialised protocols that support data integrity, validity, and veracity in safety-critical environments, where appropriate?
  - o Ensure the implementation of specialised protocols that support data security and confidentiality where data is particularly sensitive or otherwise confidential, where appropriate?
  - o Support proposals for communities becoming monetary and material beneficiaries of their aggregate data?

## Equitably advancing access to research and innovation capacity

- How can our data policies and practices advance just access to research and innovation capacities that enable societally beneficial insights, discoveries, and innovations to be equitably produced, replicated, and enjoyed? How can such policies and practices accordingly:
  - o Address asymmetrical dynamics of sharing between more and less well-resourced research collaborators including those from high-income countries (HICs) and those from low-/middle-income countries (LMICs)?
  - o Promote the redress of asymmetries between HICs and LMICs in know-how, education, training, and research and innovation capacitation?
  - o Promote equitable access to the benefits of data work to overcome digital divides both within HICs and between HICs and LMICs?

| |
|---|
|     o   Promote international research collaboration that incorporates asymmetry-aware practices and enables participatory parity? |

### Equitably advance access to the capabilities of individuals, communities, and the biosphere to flourish

- How can our policies and practices ensure that data collection and use increase the scope of impacted communities' possible opportunities to realise their capabilities for well-being, flourishing, and the actualisation of their potential:

  o through the direct benefits of data systems?

  o through the improvement of the personal, socioeconomic, and environmental conditions required for realisation of capabilities in practice?

- How can our policies and practices prevent the data systems we build and use from creating or exacerbating existing obstacles to impacted communities for realising their capabilities?

- How can I advance data policies and practices—and data innovation agendas, more generally—that prioritise individual, community, and biospheric well-being?

- How can I advance data policies and practices—and data innovation agendas, more generally—which demand that data collection and use be considered in terms of the affordances they provide for the ascertainment of well-being, flourishing, and the actualisation of individual and communal potential for these?

- What educational and engagement mechanisms could be put in place through the innovation governance policies my project team creates and follows to encourage an inclusive understanding of human, societal, and biospheric well-being that incorporates Indigenous notions of the fullness, creativity, harmony, and flourishing of human and biospheric life (like the Maori commitment to *Manaakitanga* or well-being nourished through communal relationships, the African commitment to *Ubuntu*, and the commitment of the Abya Yala Indigenous traditions of Bolivia and Ecuador to 'living well' or *sumak kawsay* in Quechua, *suma qamaña* in Aymara, or *buen vivir* in Spanish)? (See Annex 1 for more details on these concepts)

### Confronting questions of equitable access involves four dimensions of data justice

- How can I advance data policies and practices—and data innovation agendas, more generally—that ensure individuals and communities impacted by data collection and use realise all four dimensions of data justice? Specifically, how can I advance data innovation agendas that:

  o Ensure the equitable distribution of the social goods and obligations, burdens and opportunities, risks and benefits, and rights and privileges that emerge from data collection and use? (*distributive justice*)

  o Ensure the material preconditions necessary for the universal realisation of the potential for human flourishing? (*capabilities-centred social justice*)

  o Establish the equal dignity and autonomy, and the equal moral status, of every person through the affirmation of reciprocal moral, political, legal, and cultural regard? (*representational and recognitional justice*)

  o Ensure that past wrongs are rectified through reparation, reconciliation, and meaningful dialogue? (*restorative and reparational justice*)

### Promote the airing and sharing of data injustices across communities through data witnessing

- In what ways can our practices of data collection, processing, and use make visible potential injustices and harms done to its members? (For instance, abusive behaviour captured by a social media platform making online harm visible; or data collected by a social service agency making discriminatory practices of racial targeting or profiling visible)
- How can we support impacted communities to draw on these forms of data witnessing to expose and challenge injustices where these arise?
- What support and empowerment mechanisms could we put in place to encourage communities impacted by our data practices to share experiences of injustice that are captured by data witnessing?
- How can we facilitate the sharing of experiences of injustice captured by data witnessing so that other, wider communities have access to this information?

### Promote the airing and sharing of data injustices across communities through transparency

- Are our current practices of collecting, processing, and using data sufficiently transparent to ensure that impacted communities have access to the information needed to understand and challenge any injustices that could emerge from these practices?
- If not, how can we ensure that our practices of data collection, processing, and use are sufficiently transparent?
- Do we use complex or opaque approaches to algorithmic modelling or application development? If so, could these approaches interfere with the ability of members of impacted communities to access, understand, and challenge outcomes of our data systems that adversely affect their rights?
- Is 'explainability' taken into account when we develop our systems, and are considerations of how information about algorithmic inputs, outputs, and operation will be delivered to stakeholders in a clear, non-technical manner part of our design practices?
- Do we have sufficient independent and transparent review, auditing, and oversight processes in place so that there can be actionable recourse and effective remedy where sought by adversely impacted individuals and communities?

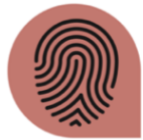 **Identity**

## Interrogate, understand, and critique harmful categorisations

- Do our data aggregation, categorisation, and labelling practices ensure that they accurately reflect the ways in which members of impacted communities self-identify? Do such practices (and those of reviewing automated labelling processes, where present) include the perspectives of members of impacted communities—especially those who are marginalised, vulnerable, or historically discriminated against?

- How might our categorisations of members of impacted communities harm their identity claims (i.e. the ways they self-identify) or limit/negatively impact their access to goods, services, or public benefits?

- How can we institute policies and processes that prevent data practices where categorisations of sensitive identity characteristics (such as race, gender, sex, or religious affiliation) are harmful, racialising, misgendering, or otherwise discriminatory? Where such harms do occur, how can we safeguard their rectification as well as effective remedy for those impacted?

- How can we institute policies and processes that ensure members of communities impacted by our data practices have opportunities to contest or correct data relating to aspects of their identities?

- How can we institute policies and processes that ensure transparency and accountability relating to the ways that data are used to classify members of communities impacted by our data practices based on aspects of their identity?

## Challenge erasure

- What data policies and processes can we put in place to help us prevent and/or mitigate instances of data collection, processing, and use where categorisations or the grouping of categories erase elements of the identities of members of impacted communities that they value and demand to be recognised?

    o For instance, the designers of a data system may group together a variety of non-majority racial identities under the category of "non-white", or a data system may record gender only in terms of binary classification and erase the identity claims of non-binary and trans people.

- What policies and processes can we put in place to help us prevent and/or mitigate instances of data collection, processing, and use that disparately injure people who possess intersectional characteristics of identity which render them vulnerable to harm?

    o For instance, a facial recognition system could be trained on a dataset that is primarily populated by images of white males, thereby causing the trained system to systematically perform poorly for darker skinned females. If the designers of this system have not taken into account the vulnerable intersectional identity (in this case, darker skinned females) in their bias mitigation and performance testing activities, this identity group becomes invisible and so too do injuries done to its members.

- What policies and processes can we put in place to help us prevent and/or mitigate instances where missing data in the datasets we use are concentrated within specific groups in ways that could harm or disparately impact them? Where this does occur, how can we ensure that any such harm is avoided?

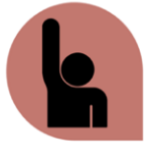

# Participation

## Democratise data work and govern data democratically

- To what extent are the design, development, and deployment processes behind our production and use of data systems democratically governed and socially licensed?
- Are impacted communities, domain experts, and other relevant stakeholders included and meaningfully consulted in the design, development, and deployment stages of our data projects?
- How can we initiate policies and processes to ensure that impacted communities possess appropriate agenda-setting and decision-making agency around our practices of data collection, processing, and use?
- How can we initiate policies and processes to ensure that impacted communities are able to participate in articulating collective visions for the direction that our data innovation agendas should take?
- How can we initiate policies and processes to ensure that impacted communities participate in the assessment and determination of which sorts of data practices are to be deemed as unacceptable and which sorts are to be deemed as permissible or desirable?

## Challenge existing, domination-preserving modes of participation

- How do current logics and justifications of data practices reinforce or institutionalise prevailing power structures and hierarchies and how can I (and others in my firm or organisation) engage in the interrogation of these structures and hierarchies? (Refer to the power pillar for further direction)
- In what ways could the way we approach community participation and involvement in our data innovation practices and their governance operate to normalise or support existing power imbalances and the harmful data practices that could follow from them?

## Ensure transformational inclusiveness rather than power-preserving inclusion

- How can I (and others in my firm or organisation) ensure that, where opportunities arise for the inclusion of members of impacted communities in our data innovation practices, that the terms of inclusion are equitable, symmetrical, and equality-promoting?
- How can I (and others in my firm or organisation) ensure, in these instances, that the inclusion of members of impacted communities is not normalising or supporting existing power imbalances in ways that could perpetuate data injustices and fortify unequal relationships?

- How can I (and others in my firm or organisation) develop critical approaches to the term "inclusion" that ensure its use does not reproduce power hierarchies and that detect where its use may represent "virtue signalling", insincerity, or duplicity?

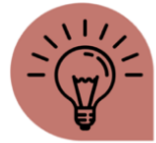 **Knowledge**

### Embrace the pluralism of knowledges

- How can I, and my firm or organisation, initiate policies and practices which ensure that the distinctive knowledges of impacted communities (i.e., their unique ways of seeing, understanding, and being in the world—especially in their lived experience of data innovation) inform and are respected in our practices of data collection, processing and use that impact them?
- To what extent do we value non-technical, socially-situated knowledge in our work?
- How can we centre and actively draw on non-technical, socially-situated knowledges in our data practices?

### Challenge the assumed or unquestioned authority of technical, professional, or "expert" knowledge across scientific and political structures

- What data policies and practices can I initiate to ensure that members of my firm or organisation recognise that our processes of knowledge creation in data science and innovation are social processes that require rational scrutiny by external stakeholders and wider public engagement?
- How can I initiate data policies and practices which ensure that the "expertise" behind this knowledge creation is held to account and aligns with wider societal values?
- How can I initiate data policies and practices which ensure the clear and accessible public communication of research and innovation purposes/goals and data analytic and scientific results, so that impacted communities and relevant stakeholders can interrogate the claims and arguments we put forward to justify data-driven decision-making and data innovation agendas?
- What kinds of upskilling, knowledge development, and resources do members of communities impacted by our data practices need to be prepared to receive, understand, and rationally scrutinise the public communication of our research and innovation purposes and our data analytic and scientific results?
    - o How can my firm or organisation support necessary upskilling and knowledge development?

## Prioritise interdisciplinarity and pursue a reflexive and positionally aware objectivity that amplifies marginalised voices

- How can I, and members of my firm or organisation, pursue understandings of data innovation environments—and of the sociotechnical processes and practices behind them—that are informed by plurality of methods and perspectives (which draw on insights from many credible sources and academic disciplines)?

- How can our approach to producing and using data systems integrate the lived experience of impacted communities with a wide range of academic and specialised knowledges, enabling an appreciation and incorporation of a wide range of insights, framings, and understandings?

- How can I, and members of my firm or organisation, approach our understandings of data innovation environments—and of the sociotechnical processes and practices behind them—with the kind of objectivity, impartiality, and neutrality that actively takes into account the voices of the marginalised, vulnerable, and oppressed, which have previously been excluded from considerations?

- How can we question claims of objectivity, impartiality, and neutrality that mask privilege and the privileged interests of dominant groups?

## Cultivate intercultural sharing, learning, and wisdom

- In what ways can I, and members of my firm or organisation, incorporate insights, learning, and wisdom from a diverse and inclusive range of sociocultural groups—especially as these insights, learning, and wisdom might inform the values, beliefs, and purposes behind data research and innovation agendas and practices?

- How can we set up or tap into networks of communication and collaboration with a diverse and inclusive range of communities and sociocultural groups, so that we can come together to cultivate shared understandings and constructively explore differences?

- In what ways can I, and members of my firm or organisation, incorporate the insights, learning, and wisdom of other communities and sociocultural groups as these insights, learning, and wisdom might inform the values, beliefs, and purposes behind data research and innovation agendas and practices?

- How can we draw on the principles and priorities of data justice to find commonality and build solidarity with impacted communities and sociocultural groups?

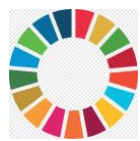 **Sustainable Development Goals**

- Are there opportunities for my firm or organisation to engage in data innovation practices which reduce the negative impacts of issues such as poverty, climate change, gender inequality, amongst others outlined in the SDGs to address their root causes?

- What measures is my firm or organisation taking to make its products or services more affordable for lower income groups and accessible for marginalised or vulnerable populations?

*For developers and practitioners involved in data innovation directly related the achievement of the SDGs:*
- Are my data innovation practices informed by engagement with the communities and individuals impacted by the issues I am trying to address, specifically as these relate to solving one of more of the SDGs?
- Could the implementation and use of the model/system I am building to promote the achievement of one or more SDGs negatively impact the well-being and livelihood of local markets?
- If the SDG-supportive model/system that my firm or organisation is using has been developed externally (outside of the community or country context in which it will be implemented), are we imposing any norms, standards, or beliefs on impacted communities?
- Are we using data that was acquired or collected without consent to achieve an SDG?
- How does this data system we are deploying to achieve the SDG fit into existing infrastructures, practices, institutions, and norms in the country or local setting?
- Are the benefits of new data systems or tools built to achieve the SDGs distributed equitably across the groups and communities they affect?
- Does our development team consist of a diverse backgrounds with respect to gender, race, and other protected characteristics?
- Do our data practices, which are aimed at achieving one or more SDGs, operate, in fact, to produce negative impacts for already marginalised groups, exacerbate inequality, or hinder the progress towards other SDGs?
- Have we considered the environmental impacts of designing, developing, and deploying our data system to achieve one or more of the SDGs (including hosting of the model)?
  - Have we considered using a simpler model where possible to decrease the negative impacts to the environment?

# Guiding Questions – The Safe-D Principles

| *Data Collection and Use Lifecycle* | DESIGN<br>• Agenda Setting, Commissioning, and Project Planning<br>• Problem Formulation<br>• Data Extraction and Procurement<br>• Preliminary Data Analysis | DEVELOP<br>• Data Pre-Processing and Feature Engineering<br>• Model Selection and Training<br>• Software Engineering<br>• Interface Design<br>• Testing and Validation<br>• Reporting | DEPLOY<br>• Productionalisation<br>• Implementer and User Training<br>• System Use and Monitoring<br>• Maintenance, Updating, or Deprovisioning | |
|---|---|---|---|---|
| **Safety and Sustainability**<br><br>• Robustness<br>• Security<br>• Accuracy and Performance Metrics<br>• Reliability | Will the prospective system serve an essential or primary function in a high impact or safety critical sector (e.g., transport, social care, healthcare, other divisions of the public sector)? If so, were previous safety cases or other forms of assurance documentation for similar technologies consulted during the planning of the system to anticipate and identify possible risks?<br><br>Will we carry out an evaluation as to whether building the system is the right approach given available resources and data, existing technologies and processes, the complexity of the use-contexts involved, and the nature of the policy or social problem that needs to be solved?<br><br>Will we carry out a human rights or ethical and social impact assessment for the prospective system? Has an appropriate degree of stakeholder engagement been incorporated into the impact assessment process? | If the algorithmic model(s) or technique(s) used by the AI system have a non-deterministic, probabilistic, evolving, or dynamic character that prevents or hinders the system's intended functionality from being formalised into specific and checkable design-time requirements (or that impairs commonly accepted methods of formal verification and validation), how will we ensure the system's accuracy, reliability, and robustness?<br><br>Will the system's design be based on well-understood techniques that have previously been in operation and externally validated for a similar purpose and in the same sector? If not, how will we ensure that diligent processes of testing, verifying, and externally validating the performance of the system occurs? How will we establish system monitoring and performance evaluation protocols that are proportionate to the system's technological maturity? | Will an independent security analyst carry out extensive penetration testing on our systems to ensure that sensitive data will not be revealed to non-trusted parties?<br><br>Where appropriate, will model reliability be evaluated and optimised using sensitivity analyses and perturbations to training data to minimise the risk performance failure when encountering novel data in the runtime environment?<br><br>How will we incorporate sufficient processes to ensure that the deployment of the system does not harm the physical, psychological, or moral integrity of implementers or adversely impact their dignity, autonomy, and ability to make free, independent, and well-informed judgements?<br><br>How will we ensure that processes of monitoring the system during its operation involve regular re-evaluations of performance that are sufficient to keep pace with real world changes that may cause concept drifts and shifts in underlying data distributions? | |

| | | | |
|---|---|---|---|
| | When we assess the potential impacts of the system and its social sustainability, will a wide range of harms that could result from the system be explored, including those direct or indirect, physical, emotional, mental, economic, social and others, as they relate to individuals and groups?<br><br>Does statute or regulation in the sector or domain in which the system will operate require any other types of impact assessment for the specific use-case of the system we are planning to develop (e.g., data protection impact assessment, equality impact assessment, human rights impact assessment, etc.)? If so, have these been carried out?<br><br>Could the prospective system be repurposed or used in unintended ways that may cause harm to impacted stakeholders?<br><br>If the prospective system is replacing a human, technical, or hybrid system that serves the same or similar function, will we fully consider the costs and benefits of replacement and whether the new system introduces new performance and safety risks?<br><br>If the design, development, and deployment of the AI system will have potentially significant impacts on the environment, will sufficient and transparently reported processes be | When performance metrics for the AI system are considered, determined, and reported, will the prioritisation of error types (e.g., false positives/ negatives) be:<br>• Informed by the specific context of the use case and by the potential effects of differential error rates on affected sub-populations (in particular, on vulnerable or protected groups).<br>• Clearly and accessibly presented, so that the rationale behind the chosen metrics is made explicit and understandable in plain, non-technical language?<br><br>When we consider, determine, and present the performance metrics for the system, will the prioritisation and reporting of metrics beyond accuracy (e.g., sensitivity, precision, specificity) be informed by the specific context of the use case and its performance needs (e.g., a system whose effective identification of rare events is more critical than its overall accuracy rate)?<br><br>Will our model training methods be oriented by the controllability of the system's runtime environment and the degree of certainty and understanding possessed about that environment? In cases where the runtime environment is less controllable and more unpredictable, how will we adjust our model training, testing, and validating methods to endure its accuracy, reliability, and robustness? | Does the system have ongoing monitoring from a human-in-the-loop to minimise the physical harm that could arise from its operations? |
| | implemented throughout the project's lifecycle to ensure that the system, in both its production and use, complies with applicable environmental protection standards and supports the sustainability of the planet?<br><br>Could the system present motivations or opportunities for malicious parties to hack or corrupt it to achieve substantial financial gains, political goals, or other perceived benefits? If so, how are these risks being anticipated, avoided, and mitigated?<br><br>How are we planning to implement sufficient and transparently reported processes throughout the project's lifecycle to ensure that measures put in place to safeguard the system's safety, security, and robustness are appropriately proportional to potential risks of hacking, adversarial attack, data poisoning, model inversion, or other cybersecurity threats?<br><br>How are we planning to implement sufficient and transparently reported processes throughout the project's lifecycle to stress test the AI system for cybersecurity vulnerabilities and resilience? | Will the trained model be externally validated? If so, will external validation be carried out across an appropriately wide range of environments to ensure the system is robust?<br><br>Has the model been evaluated and developed to minimise its vulnerability to inversion attacks (NB: model inversion attacks attempt to reconstruct training data from model parameters)?<br><br>Will the team consider model development options that contribute to reduction of the energy consumption required for model-building and system use? | |
| **Accountability**<br><br>• Traceability<br>• Auditability | Will all identified stakeholders be consulted prior to the development of our system to help critically evaluate our project plans and ensure they are intelligible? | Are members of the development team free and empowered to object to potentially harmful design decisions without fear of reprisal? | Where we are developing algorithmic models for external use, will we make available comprehensive documentation to any individual or organisation who wishes to deploy our model in their own system, including suggestions about appropriate/ inappropriate use cases? |
| • Clear Data Provenance and Data Lineage<br>• Accessibility<br>• Reproducibility<br>• Responsible Governance<br>• Answerability | Will processes of formulating the problem to be solved by the system and of defining its target variable (or measurable proxy) be opened to input from stakeholder engagement and public scrutiny? If not, how are we ensuring sufficient social license, traceability, and auditability?<br><br>Will we implement sufficient and transparently reported processes throughout the project's lifecycle to ensure end-to-end accountability across the production and use of the system? Namely, will we:<br>  • Ensure that the system is auditable by design, allowing for the end-to-end traceability and oversight of its processes of production and use.<br>  • Establish a continuous chain of human responsibility for all roles involved in the project lifecycle to allow for end-to-end answerability in the event that the human rights or fundamental freedoms of affected individuals have been negatively impacted.<br>  • Transparently report all relationships to external vendors, service providers, and other relevant business organisations across the entire supply chain and demonstrate that we have carried out due diligence regarding the ethical practices of these parties.<br>  • Enable designated public authorities and third parties, | Will we include sufficient and transparently reported processes of external peer review and evaluation by independent domain and technical experts in the evaluation, verification, and validation of the AI model?<br><br>Will we transparently report how the model has been trained and tested, including which parts of the data have been used to train the model, which have been used to test it, and which have formed the holdout data?<br><br>Will all relevant details of our model training, testing, and validation be recorded on an accessible team repository (e.g., predictor selection processes, baselines for model comparison, performance metrics)? | Where our system is deployed in public spaces, will clear messaging be established to ensure that all individuals are aware that the system is in operation?<br><br>Will we retain system logs store them securely to support any internal or external audit or review?<br><br>Will a member of our organisation be tasked with monitoring the use of the system for compliance with relevant standards, regulations, and statutes and will this person be authorised to revoke use if the system fails to comply?<br><br>Will we keep a record of every model or system update, how each version has changed, and how this affects the model/system's performance and impacts? |
| | where appropriate, to assess its compliance with existing legislation, regulation, and standards instruments across the entire project lifecycle?<br><br>Will the origin of our data sources and collection processes be sufficiently documented to support reproducibility?<br><br>Where personal data is used, will a data protection officer identify and confirm the appropriate controllers and processors of the data as well as ensure compliance to the principles of data protection?<br><br>Will a full and end-to-end record of the data lifecycle be maintained which includes a comprehensive recording of data provenance, procurement, pre-processing, lineage, storage, and security as well as qualitative input from team members about determinations made with regard to data representativeness, data sufficiency, source integrity, data timeliness, data relevance, and unforeseen data issues encountered across the workflow? | | |
| **Fairness**<br><br>• Bias mitigation<br>• Diversity and Inclusiveness<br>• Non-discrimination<br>• Equality | Do the sector(s) or domain(s) in which the prospective system will operate, and from which the data used to train it are drawn, contain historical legacies and patterns of discrimination, inequality, bias, racism, or unfair treatment of minority, marginalised, or vulnerable groups? If so, could be these legacies and patterns be replicated or augmented in the functioning of the | Where feature engineering, whether automated or carried out by humans, involves the grouping, disaggregating, or excluding of input features related to protected or potentially sensitive characteristics (e.g., decisions about combining or separating categories of gender or ethnic groups) or proxies for these, how will we incorporate processes | How will we train implementers and users of the system to prevent automation biases (overreliance or overcompliance) and other implementation biases that can arise from distrust or dismissal of automated decision support systems?<br><br>How will we monitor potential discriminatory impacts of the system once it is in operation? |

system or in its outputs and short- and long-term impacts?

Will members of identity and demographic groups that are most at risk of harm by the prospective system be consulted about design intentions to help identify and remove potential discriminatory effects?

Will an equality impact assessment be performed to identify and mitigate potential discriminatory impacts that could arise through the deployment of this system?

Will the problem to be addressed by the system be formulated though a multistakeholder process to ensure the inclusion of a broad range of perspectives and potential concerns and to stem biases that may arise from the positionality limitations of the project team?

How will we go about identifying any underlying structural biases that may play a role in translating our objectives into target variables and measurable proxies?

Where appropriate, will independent domain experts analyse the data to identify and mitigate selection biases and other bias that may be baked into the dataset?

How are we planning to implement sufficient and transparently reported processes throughout the project's in the production of the system to mitigate emergent forms of bias and to make the rationale behind these decisions transparent and accessible to impacted individuals and other relevant stakeholders?

During pre-processing, how will we take steps to balance the distribution of the data sets' classes and mitigate other possible biases?

How will we implement sufficient and transparently reported processes throughout the project's lifecycle to ensure that the inferences generated from the model's learning mechanisms are reasonable, fair, equitable, and do not contain discriminatory correlations or influences of lurking or hidden proxies for discriminatory features that may act as significant factors in the generation of its output?

If the system will use unstructured data or a combination of structured and unstructured data, how will we incorporate mechanisms and processes in the project lifecycle to ensure that the inferences generated from that data by the system are reasonable, fair, and do not contain lurking proxies or correlations that are discriminatory or inequitable?

Will we make explicit the formal definition(s) of fairness we have chosen to mathematically govern the system's

Will the system be tested with a diverse set of end users to ensure that its outputs are accessible to all at the time they are needed and deliver relevant and interpretable information?

Is the infrastructure in place to allow for testing and monitoring of fairness/bias issues, given that demographics, ground truth and thus model applicability and relevance can change over time?

| | | | |
|---|---|---|---|
| | lifecycle to ensure that all data used in producing the system are sufficiently balanced and representative of the individuals and groups it is affecting?<br><br>Where there is human involvement in the data lifecycle, how are we planning to implement transparent and publicly accessible measures to ensure mitigation of potential measurement errors or biases in collection, measurement, and recording processes? | allocation of errors and outcomes and why we have chosen these fairness criteria?<br><br>Where appropriate, will we use post-processing techniques to minimise the classifier's correlation with protected attributes?<br><br>If data labelling or annotation is partly or fully automated, how will we ensure that sufficient and transparently reported processes of human oversight are implemented to mitigate the negative impact of biases generated by automated labelling or annotation, especially in cases where the dataset includes social and demographic categories that can import patterns of discrimination and proxies for protected characteristics? | |
| **Explainability**<br><br>• Interpretability<br>• Explanation-Aware Design<br>• Responsible Model Selection<br>• Accessible Rationale Explanation<br>• Responsible Implementation and User Training | Will we approach our agenda setting and project planning activities in an explanation-aware manner that considers:<br>• The kinds of explanation that impacted people would expect given the use-case and domain context (e.g., explanations that deliver meaningful information about safety and performance in a medical AI/ML system use case or explanations that deliver meaningful information about fairness and non-discrimination in a criminal justice AI/ML system use case). | If the algorithmic model(s) or technique(s) to be used by the system have a complex, high-dimensional, or non-linear character that impairs or prevents the interpretability and explainability of the system, how will we ensure that rationale behind the system's outputs has an appropriate degree of accessibility given the sector, domain context, and risks that its use may pose to impacted individuals and communities?<br><br>Where complex or potentially opaque models are under consideration, will processes of model selection include appropriate and transparent considerations | How will we ensure that implementers of the system are sufficiently trained so that they are able to fully understand:<br>• the strengths and limitations of the system and its outputs<br>• the potential conditions of situational complexity, uncertainty, anomaly, or system failure that may dictate the need for the exercise of human judgment, common sense, and practical intervention?<br><br>How will we train our implementers to interpret which correlations are consequential for providing a meaningful explanation by drawing on their domain knowledge or the decision recipient's specific circumstances? |

- The resources needed to build a complex, non-linear system that is sufficiently explainable (for instance, by the use of surrogate explainability techniques that identify relative feature importance).
- The need for both process-based explanation (i.e. information on the governance of your AI system across its design and deployment) and outcome-based explanation (i.e. information that makes the reasoning behind the system-generated outcome clear, understandable, and in plain language)?

Will we hold stakeholder engagements to facilitate input on what impacted people expect from an accessible explanation about the outcome of system-supported decision-making?

Have we put in place and documented processes that optimise the end-to-end transparency and accountability of our AI model/system to facilitated process-based explanation?

Are my project team and I aware of what the interpretability and transparency expectations and requirements are in our sector or domain?

Have my project team and I considered the setting and the sector in which our AI

of the system's explainability by taking into account:

- The normal expectations of intelligibility and accessibility that accompany the function the system will fulfil in the sector or domain in which it will operate.
- The availability of more interpretable algorithmic models or techniques in cases where the selection of an opaque model poses risks to the physical, psychological, or moral integrity of affected individuals or to their human rights and fundamental freedoms.
- The availability of the resources and capacity that will be needed to responsibly provide supplementary methods of explanation (e.g., simpler surrogate models, sensitivity analysis, or relative feature importance) in cases where an opaque model is deemed appropriate and selected?

In choosing our algorithmic model—and determining its needed degree of interpretability—we will consider the domain context, setting, and sector in which the system will be used, its level of impact (e.g., safety critical or high stakes) and how these factors affect the depth and types of explanation we should provide?

Will the interface through which users interact with our system be developed to meet universal design standards and promote accessibility for all users, including those with visual or cognitive impairments?

Will user testing be conducted prior to the full development of the system? Will this testing involve a representative sample of users who are asked to ensure they could satisfactorily interpret the outputs of the system and provide suitable explanations about its functions to affected persons?

Where stakeholder engagement is incorporated into system design, will previously identified stakeholders be consulted again towards the end of the project lifecycle to evaluate whether they were happy with the accessibility and comprehensiveness of the explanations being offered by the system?

How will we ensure that sufficient and meaningful information is provided that indicates to affected individuals when the system is being used and how and where to complain in the event of an adverse impact or harm?

| | | | |
|---|---|---|---|
| | model will be used and how this affects the explanations we provide? | How will we select a range of features that optimise for both interpretability and predictive power?<br><br>Where the system processes social or demographic data, how will we ensure that our selected model is sufficiently interpretable so that we can demonstrate that its architecture contains no discriminatory proxies, correlations, or inferences that have been determined by its learning mechanisms?<br><br>Where the system processes unstructured or high-dimensional data, will we be clear about why we are doing this and what impact this will have on the system's interpretability and explainability? | |
| **Data Quality, Integrity, Protection, and Privacy**<br><br>• Source Integrity and Measurement Accuracy<br>• Timeliness and Recency<br>• Relevance, Appropriateness, and Domain Knowledge | Did our project team complete tailored training on data governance prior to the start of this project?<br><br>Were domain experts and stakeholders consulted to identify the minimum level and kinds of data that need to be collected to ensure satisfactory performance of the system?<br><br>How are we planning to implement sufficient and transparently reported processes throughout the project's lifecycle to ensure that:<br>• all data used in producing the system are accurate, reliable, relevant, appropriate, up-to-date, and of adequate quantity and quality for the use case, domain, | If collected or procured datasets have missing or unusable data, which methods will we use for addressing these deficiencies, and will these methods be transparent and made accessible to relevant stakeholders?<br><br>How will we ensure that processes of labelling and annotating the data used to produce the system is transparently reported and made accessible for audit, oversight, and review by appropriate authorities and relevant stakeholders?<br><br>Was personal data for each of the records collected used to verify that none of the extracted features contained hidden proxies? After this assessment was carried out, was the personal data destroyed? | If the system uses dynamic data, collected and processed in real time (or near real time), for continuous learning, how will we ensure sufficient data quality, integrity, and measurement accuracy? What measures will we put into place to mitigate associated risks?<br><br>If there is a possibility of de-anonymising or identifying data subjects through data linkage with existing data, publicly available datasets, or data that could be easily obtained, how will we ensure the privacy of impacted individuals across the lifetime of the system and beyond?<br><br>For any new data that are collected in the runtime environment of the system, will these be gathered in a structured format with accompanying metadata? If not, how will we ensure appropriate data quality and annotation? |
| • Adequacy of Quality and Quantity<br>• Balance and Representativeness<br>• Attributable<br>• Consistent, Legible, and Accurate<br>• Complete<br>• Contemporaneous<br>• Traceable and auditable<br>• Consent<br>• Data Security<br>• Data Minimisation<br>• Transparency<br>• Proportionality<br>• Purpose Limitation<br>• Accountability<br>• Lawfulness, fairness, and transparency<br>• Respect for the rights of data subjects | function, and purpose of the system<br>• all data used in producing the system are attributable, consistent, complete, and contemporaneous with collection<br>• there is proper recording, traceability, and auditability of the provenance and lineage of all data used in producing the system, and any other data involved in the dynamic learning, tuning, or re-training of the system across its lifecycle?<br><br>Where personal data are used in the production of the prospective system, will we make information available to impacted rights-holders and other relevant stakeholders about the consent or legitimate basis to use that data for the purpose of the system?<br><br>Where personal data are used in the production of the AI system, will we make information available to impacted rights-holders and other relevant stakeholders about the consent or legitimate basis to use that data for the purpose of the system?<br><br>If consent or the legitimate basis to use personal data is implied, will we consult rights-holders and other relevant stakeholders to identify acceptability of the data use or concerns that need to be addressed? | Have training and testing splits been fully documented?<br><br>Was model comparison undertaken to determine that all personal data that are collected were necessary for the purposes of the task by iteratively removing different types and recording the decrease in model performance?<br><br>Was the model selection process constrained by the requirement that any model must be interpretable to ensure that individuals' right to be informed is respected?<br><br>Has external validation of the model been carried out prior to full deployment to verify whether the training data were adequate stand-in for data encountered in novel settings?<br><br>Was cross-validation carried out to determine the generalisability of training data samples and models? | Will data be stored in an interoperable and reusable format to promote replicability and transparency?<br><br>Where appropriate, will there be a document available on our website that explains how our model works and also provides details of the relevant input features and outputs?<br><br>Will mechanisms be established that allow users to request full data erasure if they no longer want to use the system?<br><br>Before using the system, will users be provided with a readable and accessible privacy policy outlining the purposes for all data that is collected and requiring acknowledgement and informed consent to proceed?<br><br>Will automated or human oversight triggers be set up to check with appropriate frequency whether the model is still representative of the original data distribution?<br><br>Will you establish a process for maintaining awareness of new sources of risk (e.g., privacy risks, security vulnerabilities) that can arise from or need to be incorporated into the system's operation? |

Has the system been designed to function without the need to collect any personal data?

Has metadata been stored alongside the datasets to help aid and improve data analysis?

Have multiple copies of the dataset been stored securely and used to verify consistency throughout the project?

Has an automated script been established that automatically flags for human checking any unexpected deviations, missing data, or unexpected formatting?

Have domain experts been consulted about the findings from the exploratory data analysis to verify the relevance of the input variables?

# Annex 1: 12 Principles and Priorities of Responsible Data Innovation

The information contained below serves as background material to provide you with a means of accessing and understanding some of the existing human rights, fundamental freedoms, and value priorities that could be impacted by the use of AI technologies. A thorough review of this table and an engagement of the links to the relevant Charters, Conventions, Declarations, and elaborations it contains is a critical first step that will help you identify the salient rights, freedoms, and values that could be affected by your project. You should also explore whether your organisation has engaged in any previous impact assessments (data protection impact assessment, equality impact assessment, ethical and social impact assessment, environmental impact assessment, etc.)—and review these where they are present.

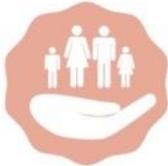

| Principles and Priorities | Corresponding Rights and Freedoms with Selected Elaborations | Resources for Principles and Priorities and Corresponding Rights and Freedoms |
|---|---|---|
| **Respect for and protection of human dignity** | *All individuals are inherently and inviolably worthy of respect by mere virtue of their status as human beings. Humans should be treated as moral subjects, and not as objects to be algorithmically scored or manipulated.*<br><br>~<br><br>**-The right to human dignity, the right to life and the right to physical, mental and moral integrity**<br><br>**-The right to be informed of the fact that one is interacting with an AI system rather than with a human being**<br><br>**-The right to refuse interaction with an AI system whenever this could adversely impact human dignity** | **Universal Declaration of Human Rights:**<br>-Preamble, Universal Declaration of Human Rights – *Dignity*<br><br>**International Covenant on Civil and Political Rights:**<br>-Article 6, International Covenant on Civil and Political Rights – *Right to life*<br><br>**European Convention on Human Rights (ECHR):**<br>-Article 2, European Convention on Human Rights – *Right to life*<br><br>-Article 2, 'Guide on Article 2 of the European Convention on Human Rights', Council of Europe – *Right to life*<br><br>**African Commission on Human and Peoples' Rights**<br>473 Resolution on the need to undertake a Study on human and peoples' rights and artificial intelligence (AI), robotics and other new and emerging technologies in Africa - ACHPR/Res. 473 |
| **Interconnectivity, solidarity, and intergenerational reciprocity** | *All humans are interconnected to a greater whole, which transcends time and thrives when all its constituent parts are enabled to thrive. This unbounded bond of solidarity extends from the closest relationship between kin to the* | **UNESCO:**<br>-III.1 Values, Recommendation on the Ethics of Artificial Intelligence, *Living in peaceful, just and interconnected societies* |

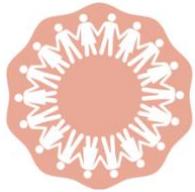

*living totality of the biospheric whole. Membership in this greater community also places a responsibility on the present generation to take account of the well-being and flourishing of future generations. Intergenerational reciprocity involves looking backward in considering the wisdom and learning of past generations and looking forward in considering the rights and well-being of lives not yet lived (two, four, seven, or more generations in the future).*

~

**-The right of future generations to due moral regard and consideration**

**- *Kaitiakitanga* (Maori): The responsibility to ensure sustainable futures for the biosphere and for people, families, communities, and humanity**

**- *Manaakitanga* (Maori): The responsibility to extend care, compassion, hospitality, and generosity to all others including strangers and the environment. Shared *Manaakitanga* supports well-being, dignity, and the stewardship of healthful and spiritual living.**

**-The Seventh Generation Principle (Haudenosaunee Confederacy, Iroquois): Give regard to the well-being of the seventh generation ahead of you in your practices, works, actions, and deliberations and draw on the experience and wisdom of the seventh generation that came before**

**-The values of *Ubuntu* (Sub-Saharan Africa): Ethical life is measured by the meaningful relationships formed by each individual with an interconnected and interdependent whole of people, community, and environment. One's humanity is affirmed by connecting with and taking care of others and by recognising their dignity in works, deliberations, and deeds.**

**Other resources:**

The Maori Report**, Independent Maori Statutory Body**

Treaty of Waitangi/Te Tiriti and Māori Ethics Guidelines for: AI, Algorithms, Data and IOT**, 2020**

The World People's Conference on Climate Change and the Rights of Mother Earth**, Bolivia 2010**

The Constitution of the Iroquois Nations**, 1916**

What is Ubuntu?**, Desmond Tutu 2013**

I am because you are**, Michael Onyebuchi Eze, UNESCO 2011**

| Environmental flourishing, sustainability, and the rights of the biosphere 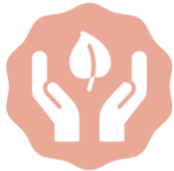 | *All humans draw oxygen from the Earth's air, draw nourishment from its soil, and live as interconnected parts of a living biospheric community. The interrelated organisms of this unbounded community share a common origin, a common history, and a common ecological fate. Members of humanity, as benefactors and inheritors of such a circle of life and of the life-giving gifts of the earth, should seek practices of living that secure environmental flourishing, sustainability, and the rights of the biosphere. These practices of living should aim for a harmony and balance with the interdependent ecologies of the biosphere in solidarity with it. They should also respect nature's right to flourish, to endure, and to regenerate life without harmful anthropogenic influence. All people involved in AI and data innovation lifecycles should prioritise environmental flourishing, sustainability, and the rights of the biosphere, ensuring that they use the affordances of technology to do battle with climate change and biodiversity drain rather than contribute to them.*<br><br>*~*<br><br>**-The right of** *Pachamama***: 'Nature or** *Pachamama***, where life is reproduced and exists, has the right to exist, persist, maintain and regenerate its vital cycles, structure, functions and its processes of evolution'. (Article 1, Constitution of Ecuador)**<br><br>**-***Sumak kawsay* **(Quechua),** *suma qamaña* **(Aymara),** *buen vivir* **(Spanish): "living well" or "collective well-being" but also the priority of a shared pursuit of the fullness, creativity, harmony, and flourishing of human and biospheric life.**<br><br>**-** *Kaitiakitanga* **(Maori): The responsibility to ensure sustainable futures for the biosphere and for people, families, communities, and humanity** | **UNESCO:**<br>**-**III.1 Values, Recommendation on the Ethics of Artificial Intelligence, *Environment and ecosystem flourishing*<br><br>**Other resources:**<br><br>The Constitution of Ecuador, **2008**<br><br>17 Principles of Environmental Justice, **First National People of Colour Environmental Leadership Summit 1991**<br><br>Bali Principles of Climate Justice**, 2002**<br><br>The Maori Report**, Independent Maori Statutory Body**<br><br>Treaty of Waitangi/Te Tiriti and Māori Ethics Guidelines for: AI, Algorithms, Data and IOT, **2020**<br><br>The World People's Conference on Climate Change and the Rights of Mother Earth, **Bolivia 2010**<br><br>The Albuquerque Declaration, **Native People-Native Homelands Climate Change Workshop-Summit, Albuquerque, New Mexico, 1998** |

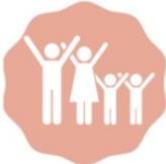

| | | |
|---|---|---|
| | **- 'Environmental Justice affirms the sacredness of Mother Earth, ecological unity and the interdependence of all species, and the right to be free from ecological destruction'. (First National People of Colour Environmental Leadership Summit)** | |
| **Protection of human freedom and autonomy** | *Humans should be empowered to determine in an informed and autonomous manner if, when, and how AI and data-intensive systems are to be used. These systems should not be employed to condition or control humans, but should rather enrich their capabilities.*<br><br>*~*<br><br>**-The right to liberty and security**<br><br>**-The right to human autonomy and self-determination**<br><br>**-The right not to be subject to a decision based solely on automated processing when this produces legal effects on groups or similarly significantly affects individuals**<br><br>**-The right to effectively contest and challenge decisions informed and/or made by an AI system and to demand that such decisions be reviewed by a person**<br><br>**-The right to freely decide to be excluded from AI-enabled manipulation, individualised profiling, and predictions. This also applies to cases of non-personal data processing**<br><br>**-The right to have the opportunity, when it is not overridden by competing legitimate grounds, to choose to have contact with a human being rather than a robot** | **Universal Declaration of Human Rights:**<br> -Article 3, Universal Declaration of Human Rights – *Right to life, liberty, and the security of person*<br><br>-Article 18, Universal Declaration of Human Rights – *Right to freedom of thought, conscience, and religion*<br><br>-Article 19, Universal Declaration of Human Rights – *Right to freedom of opinion and expression*<br><br>**African Commission on Human and Peoples' Rights**<br>473 Resolution on the need to undertake a Study on human and peoples' rights and artificial intelligence (AI), robotics and other new and emerging technologies in Africa - ACHPR/Res. 473<br><br>**International Covenant on Civil and Political Rights:**<br> -Article 9, International Covenant on Civil and Political Rights – *Right to liberty and security of person*<br><br>-Article 18, International Covenant on Civil and Political Rights – *Right to freedom of thought, conscience, and religion*<br><br>-Article 19, International Covenant on Civil and Political Rights – *Freedom of expression*<br><br>**European Convention on Human Rights (ECHR):**<br> -Article 5, European Convention on Human Rights – Right to liberty and security |
| | | -Article 5, 'Guide on Article 5 of the European Convention on Human Rights', Council of Europe – *Right to liberty and security*<br><br>-Article 9, European Convention on Human Rights – *Freedom of thought, conscience, and religion*<br><br>-Article 9, 'Guide on Article 9 of the European Convention on Human Rights', Council of Europe – *Freedom of thought, conscience, and religion*<br><br>-Article 10, European Convention on Human Rights – *Freedom of expression*<br><br>*-Article 10,* 'Guide on Article 10 of the European Convention on Human Rights', Council of Europe – *Freedom of expression* |
| **Prevention of harm and protection of the right to life and physical, psychological, and moral integrity**<br><br>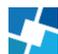 | *The physical and mental integrity of humans and the sustainability of the biosphere must be protected, and additional safeguards must be put in place to protect the vulnerable. AI and data-intensive systems must not be permitted to adversely impact human well-being or planetary health.*<br><br>*~*<br><br>**-The right to life and the right to physical and mental integrity**<br><br>**-The right to the protection of the environment**<br><br>**-The right to sustainability of the community and biosphere** | **European Convention on Human Rights (ECHR):**<br> -Article 2, European Convention on Human Rights – *Right to life*<br><br>-Article 2, 'Guide on Article 2 of the European Convention on Human Rights', Council of Europe – *Right to life* |
| **Non-discrimination,** | *All humans possess the right to non-discrimination and the right to equality and equal treatment under the law. AI and* | **Universal Declaration of Human Rights:**<br> -Article 7, Universal Declaration of Human Rights – *Equality before the law* |

| fairness, and equality 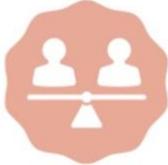 | data-intensive systems must be designed to be fair, equitable, and inclusive in their beneficial impacts and in the distribution of their risks.<br><br>~<br><br>**-The right to non-discrimination, including intersectional discrimination**<br><br>**-The right to non-discrimination and the right to equal treatment. This right must be ensured in relation to the entire lifecycle of an AI system (design, development, implementation, and use), as well as to the human choices concerning AI design, adoption, and use, whether used in the public or private sector.** | **African Commission on Human and Peoples' Rights**<br>473 Resolution on the need to undertake a Study on human and peoples' rights and artificial intelligence (AI), robotics and other new and emerging technologies in Africa - ACHPR/Res. 473<br><br>**International Covenant on Civil and Political Rights:**<br> -Article 6, International Covenant on Civil and Political Rights – *Right to life*<br><br>-Article 26, International Covenant on Civil and Political Rights – *Non-discrimination*<br><br>**European Convention on Human Rights (ECHR):**<br> -Protocol No. 12, European Convention on Human Rights<br><br>-Article 14, European Convention on Human Rights – *Prohibition of discrimination*<br><br>-Article 14 and Article 12 of Protocol No. 12, 'Guide on Article 14 of the European Convention on Human Right and on Article 1 of Protocol No. 12 to the Convention', Council of Europe – *Prohibition of discrimination*<br><br>**Office of the United Nations High Commissioner for Human Rights:**<br> -OHCHR, International Convention on the Elimination of All Forms of Racial Discrimination<br><br>-OHCHR, Convention on the Elimination of All Forms of Discrimination against Women |
| **Rights of Indigenous Peoples and** | *Indigenous peoples have a right to self-determination, to recognition of equal standing, and to remedy and reparation for the historical and systemic denial of their rights. These rights should be contextualised in accordance* | **The United Nations**<br>- United Nations Declaration on the Rights of Indigenous Peoples<br><br>The Maori Report**, Independent Maori Statutory Body, 2016** |
| **Indigenous Data Sovereignty**<br><br>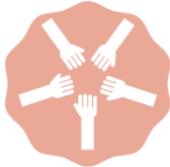 | *with the unique sociocultural histories and lived experience of the Indigenous people to whom such rights apply. Indigenous peoples also have a right to control data from and about their communities, activities, and lands and to shape the way these are collected and used. This encompasses both collective rights of benefit, access, ownership, and control and individual data-related rights and freedoms like rights to privacy and dignity.*<br><br>~<br><br>**-The rights to the restoration of equality, reparation, and self-determination**<br><br>**-** ***Rangatiratanga* (Maori): The empowering unity of a self-determining and sovereign community that is bound together by the reciprocal involvement of leadership and community members in collective governance, problem solving, and the articulation of shared goals and visions**<br><br>**-** ***Makarrata* (Aboriginal and Torres Strait Islander): The coming together after a struggle, confronting harms done, truth telling, righting the wrongs of the past, and restoring peace, solidarity, and community** | Treaty of Waitangi/Te Tiriti and Māori Ethics Guidelines for: AI, Algorithms, Data and IOT, **2020**<br><br>Compendium of Māori Data Sovereignty, **2022**<br><br>Barunga Statement**, Aboriginal and Torres Strait Islander Peoples 1988**<br><br>Uluru Statement from the Heart**, Aboriginal and Torres Strait Islander Peoples, National Constitutional Convention 2017**<br><br>Idle No More Movement, **First Nations of Canada 2012**<br><br>The CARE Principles for Indigenous Data Governance, **2020** |
| **Data protection and the right to** | *The design and use of AI and data-intensive systems that rely on the processing of personal data must secure a person's right to respect for private and family life, including the individual's right to control their own data. Informed, freely given, and unambiguous consent must play a role in this.*<br><br>~<br><br>**-The right to respect for private and family life and the protection of personal data** | **Universal Declaration of Human Rights:**<br> -Article 12, Universal Declaration of Human Rights – *Right to respect for privacy, family, home, or correspondence*<br><br>**African Commission on Human and Peoples' Rights**<br>473 Resolution on the need to undertake a Study on human and peoples' rights and artificial intelligence (AI), robotics and other new and emerging technologies in Africa - ACHPR/Res. 473<br><br>**African Union** |

| respect of private and family life 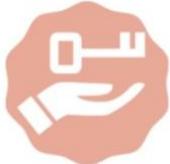 | -The right to physical, psychological, and moral integrity in light of AI-based profiling and emotion/personality recognition<br><br>-All the rights enshrined in Convention 108+ of the Council of Europe and in its modernised version, and in particular with regard to AI-based profiling and location tracking | -African Union Convention on Cyber Security and Personal Data Protection, 2014<br><br>**European Convention on Human Rights (ECHR):**<br> -Article 8, European Convention on Human Rights – *Right to respect for private and family life*<br><br>-Article 8, 'Guide on Article 8 of the European Convention on Human Rights. Right to respect for private and family life, home and correspondence', Council of Europe – *Right to respect for private and family life* |
|---|---|---|
| **Economic and social rights** 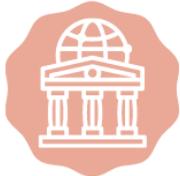 | *Individuals must have access to the material means needed to participate fully in work life, social life, and creative life, and in the conduct of public affairs, through the provision of proper education, adequate living and working standards, health, safety, and social security. This means that AI and data-intensive systems should not infringe upon individuals' rights to work, to just, safe, and healthy working conditions, to social security, to the protection of health, and to social and medical assistance.*<br><br>~<br><br>-The right to just working conditions, the right to safe and healthy working conditions, the right to organise, the right to social security, and the rights to the protection of health and to social and medical assistance | **African Union**<br>Digital Transformation Strategy for Africa (2020-2030)<br><br>**Universal Declaration of Human Rights:**<br> -Article 3, Universal Declaration of Human Rights – *Right to life, liberty, and the security of person*<br><br>-Article 12, Universal Declaration of Human Rights – *Right to private home life*<br><br>-Article 22, Universal Declaration of Human Rights – *Right to social security*<br><br>-Article 22, Universal Declaration of Human Rights – *Workers' rights*<br><br>**International Covenant on Economic, Social and Cultural Rights:**<br> -Article 6, International Covenant on Economic, Social, and Cultural Rights – *The right to work*<br><br>-Article 7, International Covenant on Economic, Social, and Cultural Rights – *Right to just and favourable conditions of work*<br><br>-Article 8, International Covenant on Economic, Social, and Cultural Rights – *Right to organise* |
| | | -Article 9, <u>International Covenant on Economic, Social, and Cultural Rights</u> – *Right to social security* |
| **Accountability and effective remedy**<br><br>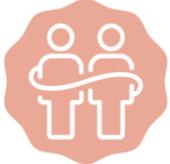 | *Accountability demands that the onus of justifying outcomes that have been influenced by data-driven and AI systems be placed on the shoulders of the human creators and users of those systems. This means that it is essential to establish a continuous chain of human responsibility across the whole data innovation lifecycle. Making sure that accountability is effective from end to end necessitates that no gaps be permitted in the answerability of responsible human authorities from first steps of the design of a system to its deprovisioning. Accountability also entails that every step of the process of designing and implementing the system is accessible for audit, oversight, and review. Where a system harms people, they have a right to actionable recourse and effective remedy, so that responsible parties can be held accountable.*<br><br>*~*<br><br>**-The right to an effective remedy for violation of rights and freedoms. This should also include the right to effective and accessible remedies whenever the development or use of AI and data-intensive systems by private or public entities causes unjust harm or breaches an individual's legally protected rights.** | **Universal Declaration of Human Rights:**<br>-Article 8, <u>Universal Declaration of Human Rights</u> – *Right to an effective remedy*<br><br>**International Covenant on Civil and Political Rights:**<br>-Article 2, <u>International Covenant on Civil and Political Rights</u> – *Right to effective remedy*<br><br>**European Convention on Human Rights (ECHR):**<br>-Article 13, <u>European Convention on Human Rights</u> – *Right to an effective remedy*<br><br>-Article 13, '<u>Guide on Article 13 of the European Convention on Human Rights.</u>', Council of Europe – *Right to an effective remedy* |

| Democracy 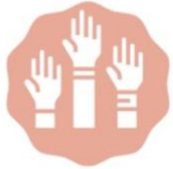 | *Individuals should enjoy the ability to freely form bonds of social cohesion, human connection, and solidarity through inclusive and regular democratic participation, whether in political life, work life, or social life. This requires informational plurality, the free and equitable flow of the legitimate and valid forms of information, and the protection of freedoms of expression, assembly, and association.*<br><br>~<br><br>**-The right to freedoms of expression, assembly, and association**<br><br>**-The right to vote and to be elected, the right to free and fair elections, and in particular universal, equal and free suffrage, including equality of opportunities and the freedom of voters to form an opinion. In this regard, individuals should not be subjected to any deception or manipulation.**<br><br>**-The right to (diverse) information, free discourse, and access to plurality of ideas and perspectives**<br><br>**-The right to good governance** | **Universal Declaration of Human Rights:**<br>-Article 19, <u>Universal Declaration of Human Rights</u> – *Right to freedom of opinion and expression*<br><br>-Article 20, <u>Universal Declaration of Human Rights</u> – *Right to freedom of peaceful assembly and association*<br><br>**International Covenant on Civil and Political Rights:**<br>-Article 19, <u>International Covenant on Civil and Political Rights</u> – *Freedom of expression*<br><br>-Article 21, <u>International Covenant on Civil and Political Rights</u> – *Freedom of assembly*<br><br>-Article 22, <u>International Covenant on Civil and Political Rights</u> – *Freedom of association*<br><br>-Article 25, <u>International Covenant on Civil and Political Rights</u> – *Right to participate in public affairs, good governance, and elections*<br><br>**European Convention on Human Rights (ECHR):**<br>-Article 3 of Protocol No.1, <u>European Convention on Human Rights</u> – *Right to free elections*<br><br>- Article 3 of Protocol No. 1, <u>Guide on Article 3 of Protocol No. 1 to the European Convention of Human Rights</u> – *Right to free elections*<br><br>-Article 10, <u>European Convention on Human Rights</u> – *Freedom of expression*<br><br>-*Article 10,* <u>'Guide on Article 10 of the European Convention on Human Rights'</u>, Council of Europe – *Freedom of expression*<br><br>-Article 11, <u>European Convention on Human Rights</u> – *Freedom of assembly and association* |
| | | -Article 11, 'Guide on Article 11 of the European Convention on Human Rights', Council of Europe – *Freedom of assembly and association* |
| **Rule of law**<br>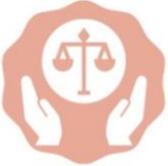 | *AI and data-intensive systems must not undermine judicial independence, effective remedy, the right to a fair trial, due process, or impartiality. To ensure this, the transparency, integrity, and fairness of the data and data processing methods must be secured.*<br><br>*~*<br><br>**-The right to a fair trial and due process. This should also include the possibility of receiving insight into and challenging AI-informed decisions in the context of law enforcement or justice, including the right to review of such decisions by a human. The essential requirements that secure impacted individuals' access to the right of a fair trial must also be met equality of arms, right to a natural judge established by law, the right to an independent and impartial tribunal, and respect for the adversarial process.**<br><br>**-The right to judicial independence and impartiality, and the right to legal assistance**<br><br>**-The right to an effective remedy, also in cases of unlawful harm or breach an individual's human rights in the context of AI and data-intensive systems** | **Universal Declaration of Human Rights:**<br>-Article 8, Universal Declaration of Human Rights – *Right to an effective remedy*<br><br>- Article 10, Universal Declaration of Human Rights – *Right to a fair trial*<br><br>**International Covenant on Civil and Political Rights:**<br>-Article 2, International Covenant on Civil and Political Rights – *Right to effective remedy*<br><br>-Article 14, International Covenant on Civil and Political Rights – *Right to fair trial*<br><br>**European Convention on Human Rights (ECHR):**<br>-Article 6, European Convention on Human Rights – *Right to a fair trial*<br><br>-Article 6, 'Guide on Article 6 of the European Convention on Human Rights.', Council of Europe – *Right to a fair trial*<br><br>-Article 13, European Convention on Human Rights – *Right to an effective remedy*<br><br>-Article 13, 'Guide on Article 13 of the European Convention on Human Rights.', Council of Europe – *Right to an effective remedy* |

## Annex 2: Sustainable Development Goals

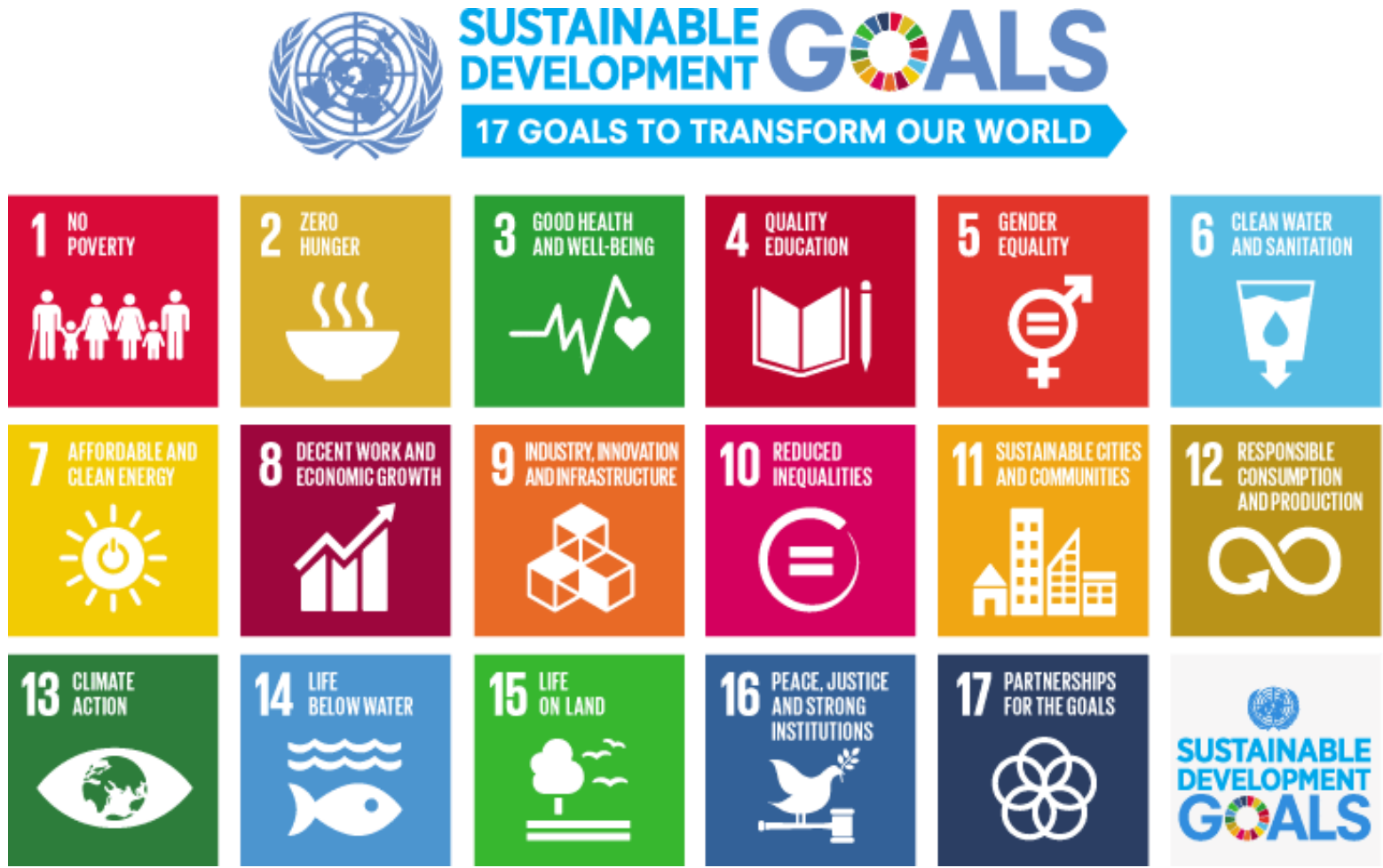

*Image is from the United Nations Sustainable Development Goals blog post[20]*

---

[20] United Nations, 2015

# Annex 3: Insights from the Policy Pilot Partner Reports

A central aspect of the Advancing Data Justice Research and Practice project (ADJRP) is the project's collaboration with 12 partner organisations from across Africa, the Americas, Asia, and Oceania to enhance our understanding of data justice with a broad spectrum of regional, national, and local perspectives. We asked the partner organisations to engage with their communities on the meaningfulness of the data justice pillars and other components of this guide while it was in draft form. This annex summarises the feedback from these partner organisations derived from surveys, interviews, and workshops with policymakers, developers, and impacted community members in more than a dozen countries.

The partners whose insights inform this annex are:
- AfroLeadership (Cameroon)
- CIPESA - Collaboration on International ICT Policy for East and Southern Africa (Uganda)
- CIPIT - Centre for Intellectual Property and Information Technology Law (Kenya)
- Digital Empowerment Foundation (India)
- Digital Natives Academy (Aotearoa/New Zealand)
- Digital Rights Foundation (Pakistan)
- Engage Media (Indonesia/Philippines)
- Gob_Lab - Universidad Adolfo Ibáñez (Chile)
- Internet Bolivia (Bolivia)
- ITS Rio - Institute of Technology and Society (Brazil)
- Open Data China (PRC)
- WOUGNET - Women of Uganda Network (Uganda)

## Prominent feedback and recommendations

**Data Justice:** The concept of data justice was novel for many audiences, and our partner organisations found that it was an unfamiliar term to many —though not all—of the respondents. In addition to conceptual unfamiliarity, in some cases, the term data justice did not easily translate into local languages. For example, there is no word for "data" in Urdu, which complicates linking the concept to narratives about justice (Digital Rights Foundation).[21]

While in many cases, respondents identified data justice with related concepts, such as fairness and dignity, in at least a few other cases, data justice was equated with *legal* justice (i.e., the work of courts and law enforcement). As a result, in some contexts "justice" did not conjure a positive valence because of local histories of state violence and oppression employed by officials claiming to be on the side of justice (Digital Natives Academy). Such concerns are exacerbated by the potential for AI/ML to be employed oppressively using the legitimising claims of public safety and national security to carry out inequitable or authoritarian agendas. This insight motivates us to employ particular nuance and care in our work to define data justice to ensure that its meaning is equated with the broader goals of fairness and emancipation rather than within the constraints of any particular legal structure or oppressive programmes of social control.

Even where data justice is not conceived of purely in legalistic terms, we cannot assume that it will be universally understood as emancipatory or located in a human rights framework. How data justice is conceptualised and operationalised is likely to reflect variances in the needs, values, and cultural and political climate of a given society. In contexts with a tradition of resistance to hegemonic authority (governmental, corporate, or both), data justice is understood as a move towards resisting or reforming systems of social control and violence (Digital Natives Academy, WOUGNET). Where the authority and control of governments and/or business are accepted by a large share of the population, data justice may be viewed more narrowly in economic terms, as affecting consumer rights, labour relations, and access to innovation (Open Data China). It may be incumbent upon the ADJRP project to reflect on strategies to either "meet audiences where they are" or to do additional work to develop shared understandings of data justice that promotes an emancipatory and respectful vision that functions across societal differences. Beyond this, the results of the Policy Pilot Partner collaborations and our desk-based research recommend the view that the concept of data justice is contextually bound and plural. We have tried to integrate this understanding that data justice is both pluralistic and situated in the guides.

Another challenge for conceptualising and operationalising data justice are the social and economic conditions in which a significant portion of marginalised persons currently live. Partner organisations frequently mention "digital divide" issues such as digital literacy and lack of access to infrastructure, but they also point out that other factors interfere with attempts to develop an inclusive account of data justice which could combat such digital inequalities. In many locales of interest to data justice discourse, large population segments struggle even to meet their basic needs and face obstacles including poor sanitation, low reading literacy, military conflict, poor health, and hunger. For these populations, awareness of data justice issues may be very low even while data extraction and intervention by data-intensive technologies (for instance, in the provision of social services and international aid) may impact their lives. Data justice related issues are, in any case, challenging to prioritise over basic needs to a degree that enables the involvement of a full complement of voices (Digital Empowerment Foundation, ITS Rio). Furthermore, where digital technologies have improved otherwise desperate conditions, some are hesitant to adopt a critical stance towards technology, a stance that appears to be implied by the data justice discourse (Engage Media).

- *Work to develop shared understandings of data justice that overcomes language barriers and supports the emancipatory aspirations of those facing injustice in both material and societal forms. Encourage reflective engagement of the contextually situated and pluralistic character of data justice.*

**Positionality:** Partner organisations drew attention to the perspective from which this project emerges. Questions were raised about the data justice implications of the project itself; respondents expressed scepticism about the potentially extractive desire of a UK institution to acquire knowledge from an historically colonised people (Digital Natives Academy). Further evidence of this appears in, among other places, the project's move to shift attention away from data protection as a prominent data justice aim. In countries where state violence and repression is enabled by the collection of and access to data about populations, data protection remains a centrally important element in struggles for justice (Digital Natives Academy). Similarly, we are cautioned against broad characterisations and assumptions of disadvantage; cultures outside of the Global North are multi-faceted. We are cautioned, for example, from implying that all people living in a particular region are poor. Such a presumption is common amongst Global North perspectives and is potentially exacerbated by data collection practices by Western NGOs that focus on poor populations (WOUGNET). These insights elide with other concerns raised about the positionality of this work being Eurocentric (despite our claims and efforts to the contrary) and at risk of being out of touch with non-Western experiences of coloniality and modernity. We welcome and accept this critique. We are reminded that the ADJRP project is an opportunity for the project team to learn from others as we simultaneously provide resources for learning.

- *The project team should commit to the additional, necessary work of consultation, inclusion, and reflective self-development to produce work that is viewed as relevant, legitimate, and offered in service of meaningful and holistic intercultural justice.*

**Accessibility of the material:** Some partner organisations offered criticisms of choices of language in the materials. Some respondents suggested that the pillars overgeneralise populations rather than accounting for cultural uniqueness. These respondents also questioned the term "pillars" as reflecting a Western perspective (Digital Natives Academy). Others observed accessibility challenges along two dimensions. First, it was felt that some of the descriptive material supporting the pillars was framed in academic and technical language that some audiences (e.g., policymakers) may find dense and alienating (CIPESA, Engage Media, Gob_Lab). Second, aspects of the project appear to assume a readership that accepts that data processing can be a source of material inequity, and the associated analysis of power relations in technology production and regulation frames some parties as oppressors, implicating some readers who are unlikely to identify as such (Gob_Lab). While the project team has worked to make the language of its materials more accessible in subsequent drafts, there is always more work to be done, including in following recommendations to include more concrete examples to illustrate abstractions. In anticipation of this need for examples, a track of work was initiated early in the ADJRP project to build a repository of use cases from around the world that tell stories both of challenges to data justice and of transformative data justice practices that illustrate the pillars. This piece, *Data Justice Stories: A Repository of Case Studies*, will be published alongside this guide. As far as displeasing some readers who may feel implicated as creating data inequities, it is likely to be more challenging to reframe data justice in terms that do not cause discomfort for some readers.

- *Ensure that the material is based on a foundation of sound, well-reasoned arguments and inclusive language to ensure that intended audiences see themselves as partners in data justice.*

### Other insights and recommendations (in no particular order)

**Accountability and Recourse:** A holistic conception of data justice should include means to hold those responsible for data injustice accountable. Overlapping this concern, people who experience harm from data collection and use should have avenues of recourse available to them to seek remedies and hold those responsible accountable (Engage Media).

- *Our work could do more to address accountability and recourse as a feature of data justice.*

**Business transparency:** in addition to making data-driven systems more explainable and transparent to those who use or are otherwise affected by them, the details of data and technology procurement by governments and business-to-business data sharing should also be considered as targets for data justice transparency efforts (WOUGNET).

- *Broaden the scope of transparency to include business practices and agreements*

**Domestic violence:** Data-driven technologies can play a role in the enablement of domestic abuse. This is a specific and impactful data injustice case to consider (WOUGNET).

- *Be attentive to identity-related harms from 'unintended' uses of data*

**Disability justice**: The identity and access pillars are likely to be strengthened by making explicit reference to abledness and disability as data justice issues (WOUGNET).

- *Account for disability rights*

**Audience diversity:** It was suggested there may be value in differentiating between 'impacted' stakeholders (i.e., potentially harmed or disadvantaged) and general consumers (i.e., potentially affected but do not express concerns about direct harm) to make the work more relatable to more recipients (Open Data China). It was also suggested that our audience distinctions overgeneralise and fail to account for the diversity of experiences. E.g., indigenous developers are likely to have unique perspectives and needs (Digital Natives Academy).

- *Be mindful of audience, including those who do not fit easily into the three categories of 'developer', 'policymaker', and 'impacted communities'.*

**Rule of law:** In many countries, existing laws governing data justice issues (e.g., data protection and privacy) are routinely unenforced or circumvented by both state and non-state actors. (WOUGNET).

- *Data protection should be considered a component of data justice.*

**Regulatory power and abuse:** In some national contexts, the strengthening of regulatory agencies and associated laws can aid the cause of data justice, while in others it provides oppressive power to authoritarians and crony governments.

- *Be attentive to how data justice might be enacted in particular contexts—and the roles and responsibilities of those who are entrusted to be promote data justice.*

### Feedback specifically related to the pillars

**Power:** Some respondents were concerned that the power pillar may not account for the full nuance of power and the difficulty to recognise data and technological power everywhere it resides. Where most people may see such power residing in governments and large companies, it may be harder to see when it is a feature of local and small business interests. Other respondents were concerned that the project's portrayal of power is binary, where there are oppressors and oppressed, when the actual landscape of power cuts across obvious categories. For example, we should consider the nuanced power relations of Global South governments in which they hold power over their constituents but are themselves frequently made subservient to Global North governments and companies (Gob_Lab). Furthermore, the interplay of power and influence should be recognised to account for cases in which they do not manifest together (CIPESA).

- *Attend to the nuance of power – degrees of power held by different stakeholders and spectrums of power.*

**Equity:** This was a challenging concept for some partner organisations and their local communities because of the term's inexact translation into local languages (Digital Rights Foundation, ITS Rio). In other contexts, the concept was more readily understood as a feature of social and economic hierarchies. For these groups, the meaning of technological progress varies significantly based on one's geography (e.g., urban vs. rural) and social position (e.g., young tech enthusiast vs. precarious already vulnerable) (Engage Media).

- *Work on developing a shared understanding of equity that functions in multiple cultural and social contexts.*

**Access:** There was some variance in how this pillar was understood. For some respondents, access was portrayed as an issue of access to data and barriers to that access. However, for others, access was primarily framed in terms of a digital divide, with a particular focus on infrastructure and connectivity being salient. There were multiple accounts of large population segments without assured connectivity. Digital literacy was also mentioned as essential to consider. At least one respondent group emphasised the importance of these notions of access as fundamental to human rights given their role in participation in contemporary civic and commercial life.

- *Work on developing a shared understanding of this pillar. Be attentive to questions of infrastructure as a feature of this pillar.*

**Participation:** For some respondents, the element of participation was portrayed as a tension. They argued that, on the one hand, there is a need for technology providers and regulators to do more to make their work inclusive, aware, and potentially simplified in order to meet affected persons and communities where they are. On the other hand, there is a need for investing in the work of developing more expertise in society so as not to impede technological progress with process but rather to enable more forward movement in technological development and uptake (Gob_Lab). This tension points to an underlying strain in approaches to data innovation between more horizontal and participatory technology practices and more vertical strategies of technological governance. Mediating between these should be approached cautiously so as not to contribute to further epistemic injustice and the denigration of local knowledge.

Where participation was described as engagement between decision makers and affected persons, some respondents argued that increasing the diversity of those involved in data and technological practices was important, while others were distrustful of public institutions and cynical about participatory work being easily co-opted and corrupted by political operators and other powerful interests (Gob_Lab). Furthermore, there were concerns that some members of society would unlikely be invited as participants in any collaborative processes owing to power relations and status assignments that treat some as 'unworthy'. Participation was also understood by some as the difference between opting in and opting out of technology use. Arguments were offered, on the one hand, that opting out can be a form of resistance, while others argued that allowing some to opt out creates a drag on the society as a whole.

- *Be attentive to barriers to meaningful participation as well the potential burden on relevant stakeholders as a form of injustice.*

**Knowledge:** Concerns were raised about how public officials, civic entrepreneurs, and technology companies discount existing bodies of knowledge and seem to actively unlearn or leave behind what is known about societal issues as they charge forward towards the goal of digital transformation. An additional point for the project team to consider is the framing of this pillar for societies with a rich oral tradition and limited written one. Oral knowledge is less easily datafied and risks erasure by digital systems. Furthermore, there are concerns about the risks of acquiring knowledge from indigenous communities in ways that threaten data sovereignty. Well-meaning inclusion efforts may be seen as colonial and extractive (Digital Natives Academy).

- *Recognise the "unlearning" of knowledge as a challenge for this pillar.*
- *Broaden the understanding of knowledge to account for oral traditions.*
- *Recognise the issue of data sovereignty in relation to the goals of the knowledge pillar.*

**Identity:** In relatively homogenous societies and societies where individualism is deemphasised, the identity pillar may not be immediately salient without being linked directly with the power pillar. Identitarian concerns may become more legible and relatable when examined as an aspect of power and hierarchy (Open Data China).

- *Consider the identity pillar from the perspective of cultures that are non-individualistic.*

**Other issues of note**

**Power and agency:** There were concerns as well about the feasibility of putting the pillars and reflections into practice when the majority of technological power resides outside of the national context where they operate. This was expressed across all target audiences: marginalised people lack the resources to mobilise on issues of data justice; developers may be forced to compromise when faced with market conditions; policy experts are constrained by lack of jurisdiction over the actions of major companies sited outside their national boundaries.

**Representation:** In addition to concerns about the representation of non-Western people and concepts in data, there were also concerns raised about the fit of technologies to local contexts. Too often "adaptation" stands in for context-aware development, resulting in a sense of exclusion. For indigenous populations whose very existence is threatened and whose visibility is muted in many societies, there is a tension between the benefits of being made visible by representation in data and concerns about data sovereignty, cultural exploitation, and digital abuse (Digital Natives Academy).

**Conceptual novelty and awareness:** Concerns were raised about the lack of a conceptual basis among many affected individuals and communities creating barriers to even starting a conversation about data justice. Literature on social justice issues may not be available in many languages (e.g., indigenous, regional languages) making it difficult for advocates to join data justice to similar narratives. This was reflected by respondents who struggled to articulate a meaning of data justice that corresponds with what is used in the materials provided.

**Techno-optimism and inevitability:** A key challenge noted by one partner organisation is the prevailing attitude that technology should play a steering role in progressing their society towards economic and other improvements. There are some lessons in this perspective, particularly in national contexts in which non-technical support infrastructures are weak and digital technologies, however flawed, offer improved conditions that might otherwise remain elusive (Digital Empowerment Foundation). Consequently, some respondents resisted emphasising the risks and social issues raised by data and technologies, favouring perspectives that emphasise potential benefits (ITS Bolivia). Others were more critical. They emphasised that, where digital technologies were elevated as means to improvement (i.e., as a saving force), they could be uncritically taken to embody progress in and of themselves. Such an idealisation could lead to downsides being largely ignored and other efforts to achieve social equity being set aside (Digital Empowerment Foundation, Engage Media).

**Stakeholder engagement:** At least one partner organisation noted challenges working with policymakers, who they found resistant to engaging on the topic and/or requiring significant advance work to engage (Digital Empowerment Foundation). In some cases, people involved in policy chose to participate in providing feedback as individuals rather than from their professional perspectives. It was not made clear the source of this resistance, but it is something the project team should consider. Perhaps this signals that the concept of 'data justice' is seen as threatening to those in political positions and therefore must be approached with particular care for some audiences.

# Annex 4: ADJRP Positionality Statement

As researchers and individuals, we are committed to social justice and to revealing the systemic bases of intersectional discrimination in our research practices and life choices. We represent various communities including LGBTQ+ identities, ethnicities, women in STEM, migrants, and citizens of LMICs. For this reason, some members of our team relate to marginalised stakeholders from both a position of kinship and one of solidarity, while others confront their privilege with reflexivity and critical self-awareness. Our team participates in research activities that promote justice in pursuit of a pluralistic, anti-racist, gender-equitable, and accessible society. A key argument that motivates our research is that artificial intelligence and associated technologies are potential sites of production and reproduction of systemic advantages for people in positions historically associated with social power. This valence for AI is not inevitable and we seek to combat it through the work of explication, illumination, and alternative framings. We recognise and interrogate our own positionalities of power and privilege and see opportunities to use these advantages to lift up others, and to promote justice, equity, and liberation.

We also direct our expertise and labour to social justice causes in our communities. Members of our team support prison education programs, have advised government institutions in efforts to lower barriers to legal justice for marginalised communities, develop modes of participatory community engagement to bolster the voices of marginalised voices in decision-making processes and research governance, lobby local governments in technology civil rights matters, develop digital security capacity-building and tools for harassed social and political leaders and activists, develop AI tools that are inclusive in design and practice, and use human-in-the-loop data science methodologies to combat issues like food insecurity, amongst others. In short, we are collectively committed to the work of justice and to revealing the systemic basis of intersectional discrimination in our research and our lives.

In collaboratively formulating this team positionality statement, each of us contributed an individual positionality statement, which was aggregated in our team positionality statement shared here. Members of our team have roots in or come from regions and countries across the world, from South Asia and Australia to Argentina, Venezuela, Great Britain and the United States. Some of us identify as cisgender, others as trans persons, and others as neither of these. While some of us identify as socially privileged and relatively affluent, others have faced poverty and gained a formal education despite financial and familial barriers.

By engaging in practices of critical self- awareness, we endeavour to draw on each of these unique social and cultural positions to bring about progressive social change and to gain insights and analytical leverage about data justice. As one of us puts it, 'I am committed to promoting a pluralistic, anti-racist, gender equitable, and accessible society through my research, activism, and other life activities. I seek to reveal and combat the sources of systemic and intersectional oppression and hierarchical domination in my own society and within the multi-stakeholder communities in which I participate'.

Another of us emphasises how they draw directly on their identity in framing their research: 'I have developed a programme of research activities that places the law, human rights, diversity, and inclusion at the core of responsible data, data flows, and AI research, innovation, and governance. In my projects, I draw on my own diversity to inform on equality and inclusion issues, playing particular focus on (a) improving and informing on data capture, representativeness, language and identity labels of hidden and marginalised populations, and (b) fostering multi-disciplinary, multi-sector, stakeholder and community engagement in the design of data capture, flows and interventions to address societal challenges (e.g. slavery and migration, use of biometrics and digital traces), working with international colleagues and organisations that can best inform and engage the population who would be impacted'.

Some of us navigate lived experience, confronting intersectional discrimination, and managing the adversities of code-switching, while others reflexively acknowledge their inheritance of legacies of unquestioned privilege along with the limited mindsets that derive therefrom. Some of us experience both of these, coping with harms that are rooted in deep-seated discrimination while simultaneously inhabiting other socially privileged strata. All of our team members who identify as socially privileged have pursued a career defining 'commitment to facilitating and amplifying the voices of people and communities in less privileged positions'. However, we also consider the potential for illocutionary disablement from securitising or speaking on behalf of others and from speaking from a space where we may not have the authority. Nevertheless, from such a critical self-acknowledgement of privilege, comes a deep sense of responsibility namely, the responsibility to marshal the advantages of carrying out research in power centres of the Global North and at well-funded research institutions in order to serve the interests of those on our planet who are all too- often marginalised, de-prioritised, and exploited in the global data innovation ecosystem.

# A Note on Sources

This guide is intended to be a companion to three other pieces of research that have been published contemporaneously: *Advancing Data Justice Research and Practice: An Integrated Literature Review, Advancing Data Justice Research and Practice: Annotated Bibliography and Table of Organisations*, and *Data Justice Stories: A Repository of Case Studies*. Expansions on the ideas presented here and references for source material can be found in the *Integrated Literature Review*. All of these documents are located [here](#).

**For sections of this guide related to technical background, stakeholder engagement, and practical guidance, we have drawn on:**

Esteves, A. M., Factor, G., Vanclay, F., Götzmann, N., & Moreira, S. (2017). Adapting social impact assessment to address a project's human rights impacts and risks. *Environmental Impact Assessment Review*, *67*. https://doi.org/10.1016/j.eiar.2017.07.001

Götzmann, N., Bansal, T., Wrzoncki, E., Veiberg, C. B., Tedaldi, J., & Høvsgaard, R. (2020). Human rights impact assessment guidance and toolbox. *The Danish Institute for Human Rights.* https://www.humanrights.dk/tools/human-rights-impact-assessment-guidance-toolbox

Kernell, E. L., Veiberg, C. B., & Jacquot, C. (2020). Guidance on Human Rights Impact Assessment of Digital Activities: Introduction. *The Danish Institute for Human Rights.* https://www.humanrights.dk/sites/humanrights.dk/files/media/document/A%20 HRIA%20of%20Digital%20Activities%20-%20Introduction_ENG_accessible.pdf

Leslie, D., Burr, C., Aitken, M., Katell, M., Briggs, M., Rincón, C. (2021) *Human rights, democracy, and the rule of law assurance framework: A proposal.* The Alan Turing Institute. https://doi.org/10.5281/zenodo.5981676

Leslie, D., Burr, C., Aitken, M., Cowls, J., Briggs, M. (2021). *Artificial intelligence, human rights, democracy, and the rule of law: A primer.* The Council of Europe. https://rm.coe.int/cahai-feasibility-study-primer-final/1680a1eac8

Leslie, D., Rincón, C., Burr, C., Aitken, M., Katell, M., & Briggs, M. (2022a). *AI Sustainability in Practice: Part I.* The Alan Turing Institute.

Leslie, D., Rincón, C., Burr, C., Aitken, M., Katell, M., & Briggs, M. (2022b). *AI Sustainability in Practice: Part II.* The Alan Turing Institute.

Leslie, D. (2019). *Understanding artificial intelligence ethics and safety: A guide for the responsible design and implementation of AI systems in the public sector.* The Alan Turing Institute. https://doi.org/10.5281/ZENODO.3240529

**Other excellent resources on community and stakeholder engagement which we have drawn on here include:**

https://www.thersa.org/globalassets/reports/2020/IIDP-citizens-assembly.pdf

https://www.local.gov.uk/sites/default/files/documents/New%20Conversations%20 Guide%2012.pdf

https://datajusticelab.org/wp-content/uploads/2021/06/PublicSectorToolkit_english.pdf

https://www.communityplanningtoolkit.org/sites/default/files/Engagement.pdf

# References

Akbari, A. (2019). Spatial|Data Justice: Mapping and digitised strolling against moral police in Iran. *Development Informatics Working Paper*, University of Manchester. https://hummedia.manchester.ac.uk/institutes/gdi/publications/workingpapers/ di/di_wp76.pdf

Ashmore, R., Calinescu, R., & Paterson, C. (2019). Assuring the Machine Learning Lifecycle: Desiderata, Methods, and Challenges. *ArXiv.* https://arxiv.org/pdf/1905.04223.pdf

Burton, S., Habli, I., Lawton, T., McDermid, J.A., Morgan, P., & Porter, Z. (2020). Mind the gaps: Assuring the safety of autonomous systems from an engineering, ethical, and legal perspective. *Artificial Intelligence, 279.* https://doi.org/10.1016/j.artint.2019

Burr, C., & Leslie, D. (2021). Ethical Assurance: A Practical Approach to the Responsible Design, Development, and Deployment of Data-Driven Technologies.

Cinnamon, J. (2019). Data inequalities and why they matter for development. *Information Technology for Development*, *26*(2), 214–233. https://doi.org/10.1080/02681102.2019.1650244

Dagne, T. (2020). Embracing the Data Revolution for Development: A Data Justice Framework for Farm Data in the Context of African Indigenous Farmers. *The Journal of Law, Social Justice and Global Development*, *25*. https://doi.org/10.31273/LGD.2019.2502

Dencik, L., Hintz, A., & Cable, J. (2016). Towards data justice? The ambiguity of anti-surveillance resistance in political activism. *Big Data & Society*, *3*(2), https://journals.sagepub.com/doi/pdf/10.1177/2053951716679678

Esteves, A. M., Factor, G., Vanclay, F., Götzmann, N., & Moreira, S. (2017). Adapting social impact assessment to address a project's human rights impacts and risks. *Environmental Impact Assessment Review*, *67*. https://doi.org/10.1016/j.eiar.2017.07.001

Götzmann, N., Bansal, T., Wrzoncki, E., Veiberg, C. B., Tedaldi, J., & Høvsgaard, R. (2020). Human rights impact assessment guidance and toolbox. *The Danish Institute for Human Rights*. https://www.humanrights.dk/tools/human-rights-impact-assessment-guidance-toolbox

Haraway, D. (1988). Situated Knowledges: The Science Question in Feminism and the Privilege of Partial Perspective. *Feminist Studies, Inc*, 14(3), 575-599. https://doi.org/10.2307/3178066

Heeks, R., & Renken, J. (2016). Data justice for development: What would it mean? *Information Development*, *34*(1), 90–102. https://doi.org/10.1177/0266666916678282

Information Commissioner's Office (ICO) and The Alan Turing Institute (ATI). (2020). *Explaining Decisions Made with AI*. https://ico.org.uk/for-organisations/guide-to-data-protection/key-data-protection-themes/explaining-decisions-made-with-ai/

Johnson, J. A. (2014). From open data to information justice. *Ethics and Information Technology*, *16*(4), 263–274. https://doi.org/10.1007/s10676-014-9351-8

Kennedy, L., Sood, A., Chakraborty, D. & Chitta, R.M. (2019). Data justice through the prism of Information politics and resource injustice: A case study form Hyderabad's urban frontier. (Working paper 78). *Centre for developmental informatics, global development institute SEED.* https://hummedia.manchester.ac.uk/institutes/gdi/publications/workingpapers/di/di_wp78.pdf

Kernell, E. L., Veiberg, C. B., & Jacquot, C. (2020). Guidance on Human Rights Impact Assessment of Digital Activities: Introduction. *The Danish Institute for Human Rights*. https://www.humanrights.dk/sites/humanrights.dk/files/media/document/A%20HRIA%20of%20Digital%20Activities%20-%20Introduction_ENG_accessible.pdf

Kidd, D. (2019). Extra-activism: Counter-mapping and data justice. *Information, Communication & Society*, *22*(7), 954–970. https://doi.org/10.1080/1369118X.2019.1581243

Leslie, D., Rincon, C., Burr, C., Aitken, M., Katell, M., & Briggs, M. (2022a). AI Sustainability in Practice: Part I. *The Alan Turing Institute and the UK Office for AI.*

Leslie, D., Rincon, C., Burr, C., Aitken, M., Katell, M., & Briggs, M. (2022b). AI Sustainability in Practice: Part II. *The Alan Turing Institute and the UK Office for AI.*

Mulder, F. (2020). Humanitarian data justice: A structural data justice lens on civic technologies in post-earthquake Nepal. *Journal of Contingencies and Crisis Management*, *28*(4), 432–445. https://doi.org/10.1111/1468-5973.12335

Punathambekar, A., & Mohan, S. (2019). *Global digital cultures: Perspectives from South Asia*. University of Michigan Press. https://library.oapen.org/handle/20.500.12657/23985

Nussbaum, M. (2006). "Education and Democratic Citizenship: Capabilities and Quality Education." *Journal of Human Development and Capabilities*, 7(3), 385-395. https://doi.org/10.1080/14649880600815974

Sen, A. (1999). *Development as Freedom*. Alfred Knopf.

SL Controls. (n.d.). What is ALCOA+ and Why Is It Important to Validation and Data Integrity. https://slcontrols.com/en/what-is-alcoa-and-why-is-it-important-to-validation-and-data-integrity/

Sweenor, D., Hillion, S., Rope, D., Kannabiran, D., Hill, T., & O'Connell, M. (2020). *ML OPS: Operationalizing Data Science*. O'Reilly Media, Inc.

Taylor, L. (2017). What is data justice? The case for connecting digital rights and freedoms globally. *Big Data & Society*, *4*(2), 1-14. https://doi.org/10.1177/2053951717736335

Taylor, L. (2019). Global data justice. *Communications of the ACM*, *62*(6), 22-24. https://doi.org/10.1145/3325279

United Nations. (2015). "Sustainable Development Goals kick off with start of year." https://www.un.org/sustainabledevelopment/blog/2015/12/sustainable-development-goals-kick-off-with-start-of-new-year/